\documentclass{aa}
\usepackage[colorlinks,linkcolor=blue,citecolor=blue]{hyperref}

\usepackage{graphicx} 
\usepackage{placeins}
\usepackage{txfonts}
\usepackage{natbib}
\usepackage[normalem]{ulem}

\begin{document}

\title{The 700 ks {\sl Chandra} Spiderweb Field I: Evidence for widespread nuclear activity in the
protocluster}

\author{
P. Tozzi\inst{1}, L. Pentericci\inst{2}, R. Gilli\inst{3}, M. Pannella\inst{4}, 
F. Fiore\inst{5}, G. Miley\inst{6}, M. Nonino\inst{5}, H.J.A. R\"ottgering\inst{6}, 
V. Strazzullo\inst{5}, C. S. Anderson\inst{7}, S. Borgani\inst{4,5,8,9}, 
A. Calabrò\inst{2}, C. Carilli\inst{10}, H. Dannerbauer\inst{11,12}, 
L. Di Mascolo\inst{4}, C. Feruglio\inst{5}, R. Gobat\inst{13},  S. Jin\inst{11}, 
A. Liu\inst{14}, T. Mroczkowski\inst{15}, C. Norman\inst{16,17}, E. Rasia\inst{5,8}, 
P. Rosati\inst{18}, A. Saro\inst{4,5,8,9}
% B. Venemans\inst{11}, 
}

\institute{
INAF - Osservatorio Astrofisico di Arcetri, Largo E. Fermi, I-50122 Firenze, Italy 
\email{paolo.tozzi@inaf.it} 
\and
INAF - Osservatorio Astronomico di Roma, Via Frascati 33, I-00040 Monteporzio (RM), Italy 
\and
INAF - Osservatorio di Astrofisica e Scienza dello Spazio, via Piero Gobetti 93/3, 40129 Bologna, Italy
\and
Università di Trieste, Dipartimento di Fisica, Sezione di Astronomia, via Tiepolo 11, 
34143 Trieste, Italy
\and
INAF-Osservatorio Astronomico Trieste, via Tiepolo 11, 34123 Trieste, Italy
\and
Leiden Observatory, PO Box 9513, 2300 RA Leiden, The Netherlands
\and
Jansky Fellow of the National Radio Astronomy Observatory, P. O. Box 0, Socorro, NM 87801, USA
\and
INFN - Sezione di Trieste, via Valerio 2 - 34127 Trieste, Italy
\and
Institute of Fundamental Physics of the Universe, via Beirut 2, 34151 Grignano, Trieste, Italy
\and
National Radio Astronomy Observatory, P. O. Box 0, Socorro, NM 87801, USA
\and 
Instituto de Astrofísica de Canarias (IAC), 38205 La Laguna, Tenerife, Spain
\and
Universidad de La Laguna, Dpto. Astrofísica, 38206 La Laguna, Tenerife, Spain
\and 
Instituto de Física, Pontificia Universidad Católica de Valparaíso, Casilla 4059, Valparaíso, Chile
\and Max Planck Institute for Extraterrestrial Physics, Giessenbachstrasse
1, 85748 Garching, Germany
\and
European Southern Observatory (ESO), Karl-Schwarzschild-Str. 2, D-85748 Garching, Germany
\and
Space Telescope Science Institute, 3700 San Martin Dr., Baltimore, MD 21210, USA
\and
Johns Hopkins University, 3400 N. Charles Street, Baltimore, MD 21218, USA
\and
Dipartimento di Fisica e Scienze della Terra, Universit\'a degli Studi 
di Ferrara, via Saragat 1, I-44122 Ferrara, Italy
% Kapteyn Astronomical Institute, University of Groningen, Postbus
% 800, 9700 AV Groningen, The Netherlands
% \and
% INFN, Sezione di Bologna, viale Berti Pichat 6/2, I-40127 Bologna, Italy
% \and
% INAF - Osservatorio Astronomico di Brera, Via E. Bianchi, 46, I-23807 Merate (LC), Italy
% \and
% INAF - Istituto di Astrofisica Spaziale e Fisica cosmica di Milano, Italy
% \and
}

\titlerunning{{\sl Chandra} Spiderweb Field}
\authorrunning{Tozzi et al.}

\abstract
{}
{We present an analysis of the 700 ks {\sl Chandra} ACIS-S observation of the field around the  
radio galaxy J1140-2629 (the Spiderweb Galaxy) at $z=2.156$, focusing on the 
nuclear activity in the associated large-scale environment. }
{We identified unresolved X-ray sources in the field down to 
flux limits of $1.3\times  10^{-16} $ and $3.9 \times 10^{-16}$ erg/s/cm$^2$ 
in the soft (0.5-2.0 keV) and hard (2-10 keV) band, respectively.  We searched for counterparts
in the optical, near-infrared (NIR), and submillimeter catalogs available in the literature to 
identify X-ray sources belonging to the protocluster and derived their
X-ray properties. }
{We detect 107 X-ray unresolved sources within 5 arcmin (corresponding to 2.5 Mpc) 
of J1140-2629, among which 13 have optical counterparts
with spectroscopic redshift $2.11<z<2.20$, and 1 source
has a photometric redshift consistent with this range. 
The X-ray-emitting protocluster members are distributed approximately over a 
$\sim 3.2 \times 1.3$ Mpc$^2$ rectangular region.
An X-ray spectral analysis for all the sources within the protocluster
shows that their intrinsic spectral slope is
consistent with an average $\langle \Gamma \rangle \sim 1.84\pm 0.04$. 
Excluding the Spiderweb Galaxy, the best-fit intrinsic absorption for 
five protocluster X-ray members is 
$N_H>10^{23}$ cm$^{-2}$, while another six have upper limits of the order of 
a few times 10$^{22}$ cm$^{-2}$. Two sources can only be 
fitted with very flat $\Gamma\leq 1$, and are therefore considered Compton-thick candidates.
The 0.5-10 keV rest-frame luminosities of the 11 Compton-thin protocluster members
corrected for intrinsic absorption are greater than $2\times 10^{43}$ erg/s. 
These values are typical for the bright end 
of a Seyfert-like distribution and significantly greater than X-ray luminosities 
expected from star formation activity. The X-ray luminosity function of the AGN in the 
volume associated to the Spiderweb protocluster in the range $10^{43}<L_X<10^{44.5}$ erg/s 
is at least ten times higher than that in the field at the same redshift and significantly flatter, 
implying an increasing excess at the bright end. 
The X-ray AGN fraction is measured to be  $ 25.5 \pm 4.5 $\% 
of the spectroscopically confirmed members
in the stellar mass range $\log(M_*/M_\odot)>10.5$. This value
corresponds to an enhancement factor of $6.0^{+9.0}_{-3.0}$ for the nuclear activity
with $L_{0.5-10 keV}> 4\times 10^{43}$ erg/s with
respect to the COSMOS field at comparable redshifts and stellar mass range.}
{We conclude that the galaxy population in the Spiderweb protocluster 
is characterized by enhanced X-ray nuclear activity triggered by environmental 
effects on megaparsec scales.}

\maketitle

\section{Introduction}

% PROTOCLUSTERS: WHAT THEY ARE, WHY THEY ARE IMPORTANT
Protoclusters are defined as overdense regions in the high-$z$ Universe and are expected to evolve into massive, virialized clusters of galaxies at the 
present epoch \citep[see][for a review]{2016Overzier}. Finding 
and characterizing protoclusters is key to studying the large-scale structure of the 
Universe and the transformational processes that affect star formation and nuclear 
activity in the member galaxies, such as powerful mergers, gas cooling, and feedback 
effects.  In practice, 
although a protocluster is observationally identified as 
a high-z structure that is overdense compared to its surroundings, 
the dynamical state of the structure, and therefore whether it will evolve 
into a low-z cluster, is usually highly uncertain \citep[see, e.g.,][]{2015Muldrew}. 
At present, there are no standard methods to systematically search for 
{\sl bona fide} protoclusters, and this is why biased-tracer techniques are often used to identify interesting 
targets.  One of these consists in searching around high-redshift, powerful radio 
galaxies, which are often found to be a beacon for overdense regions
\citep[see][]{2006Miley}. Independently of the selection method, in order to confirm 
their nature and to trace the activity of several processes occurring during 
their rapid evolution, protocluster candidates must be followed up intensively with 
multiwavelength campaigns.

% THE ROLE OF X-RAY OBSERVATIONS
In this respect, X-ray observations play a key role, particularly for studying: (1)  
unresolved emission from active galactic nuclei (AGN) indicative of accretion
onto nuclear supermassive black holes (SMBHs); (2) the less intense emission from 
strongly star forming galaxies; and (3)  diffuse emission from hot gas and/or relativistic plasma.
Unfortunately, $z\gtrsim 2$ protoclusters are usually very faint in the X-rays 
and therefore only deep, high-resolution observations can provide
useful information in this band.
%  A QUICK REVIEW OF PROTOCLUSTER STUDIES WITH X-RAY
Until now only a few deep ($t_{\rm exp}\geq 200$ ks) 
observations of protoclusters have been made with {\sl Chandra} or XMM-Newton. 
In a few cases a faint, extended emission has been interpreted as the signature 
of a low-mass virialized halo 
\citep[][]{2011Gobat,2016Valentino,2016Wang}.
% \citep[as in the case of CLJ1449;][]{2011Gobat,2016Valentino,2016Wang,2019Gobat}.  
In some other studies, the thermal or relativistic nature of the
diffuse emission is still unclear \citep{2019Gilli,2021Champagne}.  In addition to the fact that X-ray data on protoclusters are so few, another clear requirement to search for and characterize
the diffuse emission in high-z protoclusters and at the same time 
the unresolved emission from AGN members is data of sufficiently high sensitivity and  
angular resolution ($\sim 1$ arcsec). 

Regarding the X-ray nuclear activity, there is increasing evidence  
that this is enhanced in overdense, protocluster regions.
\citet{2009Lehmer} and \citet{2010Digby-North} found, respectively, 12 and 4 X-ray sources with 
$L_{2-10keV}\geq 3-5\times 10^{43}$ erg s$^{-1}$ probably associated to the protoclusters 
identified in the SSA22 and HS1700+64 fields. These numbers exceed the 
X-ray source counts in deep fields at the same redshifts  by a factor
of 10-20, although the uncertainties are large
\citep[see also][]{2017Krishnan}. The enhancement of X-ray AGN in high-z protoclusters 
is particularly impressive when compared with the 
overdensity of X-ray sources in evolved clusters at $z\sim 0.5$, which is only a factor of 
$\sim 3$ \citep{2006Martini,2007Branchesi,2009Kocevski,2015Ehlert}. However, we note that the 
physical quantity that should be measured is not the X-ray source density, but 
the AGN fraction among protocluster members, 
and therefore the ratio of this fraction to the same quantity measured in the field, defined 
as the enhancement factor $f_{\rm enh}$.  To be meaningful, this quantity
should be evaluated in a well-defined stellar mass range, and possibly differentiated
between star-forming and passive galaxies, in order to 
compare the full distribution of AGN accretion rates in the protocluster
members to those in field galaxies at comparable redshift.
% \citep[see][]{2018Aird}.  
A robust characterization of $f_{\rm enh}$ in protoclusters that accounts 
for the dependence on redshift and on the mass scale is still not available.
Furthermore, the lack of quantitative estimates on the luminosity function of AGN in 
protoclusters and on their obscuration implies large uncertainties on 
an important, related  quantity, namely the energy budget released in the surrounding 
medium by the nuclear activity. 

% Other $z> 2$ clusters and protoclusters have shallower X-ray coverage 
% and a firm statistic cannot be derived. 
Finally, another relevant missing piece of information in protoclusters is
the measurement of the hot component in the surrounding diffuse baryons (the 
forming intracluster medium, ICM), which can be used to constrain the virialization status 
of the halo.
% In addition to environmental effects on  triggering nuclear activity in
% protocluster galaxies, it is possible to investigate the effect of protocluster 
% AGN on the surrounding medium.  A
Characterization of the proto-ICM, including its morphology and its 
thermodynamical and chemical properties, can provide an independent and direct measure
of the heating efficiency associated to the feedback and, at the same time,
of the gas accretion into the protocluster potential well.  In addition, 
feedback processes not only affect the ICM, but can also couple directly
with the member galaxies.  An interesting example was 
recently found in the field of the high-redshift quasar SDSSJ1030 
(observed by {\sl Chandra} with a 500 ks exposure),
where a protocluster at $z\sim 1.7$ centered on a bright FRII radio galaxy shows 
star-forming member galaxies apparently triggered by the AGN radio jet \citep{2019Gilli}.  
If confirmed, this phenomenon may shed new light on the 
strong interplay between the member galaxies and the surrounding medium in the first stages of 
formation of large structures.

%  ON THE NEED OF X-RAY OBSERVATIONS IN PROTOCLUSTERS
% WHY WE ARE HERE: THE DEEP CHANDRA DATA ON THE SPIDERWEB
As of today, the available data are providing tantalizing evidence for 
the complex and intricate phenomena that occur in such high-density, 
rapidly evolving regions at cosmic noon.  A comprehensive physical picture
of the accretion, aggregation, and phase transformation of baryons in protoclusters
will require large sample statistics (still limited by the number of currently known protoclusters) and a multiwavelength approach, including 
high-resolution X-ray observations. 
Here we address these issues using deep  {\sl Chandra} 
ACIS-S observations ($\sim 700 $ ks) of the remarkable Spiderweb Galaxy (J1140-2629) 
complex at $z=2.156$ \citep[redshift obtained from low-resolution spectroscopy by ][]{1995vanOjik}. 
The central, powerful radio galaxy, which is embedded in a giant Ly$\alpha$ 
halo \citep{1997Pentericci,2006Miley}, is surrounded by a $\gtrsim 2$ Mpc-sized 
overdensity (corresponding to a radius of 4 arcmin) of star forming galaxies 
(Ly$\alpha$ and H$\alpha$ emitters), dusty starbursts, and red galaxies likely 
in the process of shaping a nascent red sequence \citep{2008Zirm}.  
The complex around the Spiderweb Galaxy is considered to be a typical 
protocluster region and is expected to evolve 
into a massive cluster in less than 1.5 Gyr, with the radio galaxy itself showing characteristics of a cD progenitor \citep{2006Miley}. 
% (Miley et al.  2009). 
% The combination of the deep exposure and the unparalleled angular resolution of {\sl Chandra} 
% allows us to address the most compelling science issues concerning the protocluster environment
% in this key object. 

% PAPER ORGANIZATION
In the present paper, we focus on the nuclear activity in the protocluster members, while 
the physics associated to the diffuse emission surrounding the Spiderweb Galaxy
is investigated in a companion paper based on X-ray and 
SZ data (Tozzi et al. in preparation; DiMascolo et al. in preparation).
This paper is organized as follows. In Section \ref{review} we summarize 
the properties of the Spiderweb 
Galaxy complex on the basis of previous observations. In Section \ref{multiwav_section}
we briefly describe the multiwavelength data set used in this work.  
In Section \ref{datareduction}
we outline the X-ray data acquisition and reduction techniques.  
Section \ref{xraysources} deals with the identification all the X-ray 
sources in the field and their number counts.  
In Section \ref{identification} we list the galaxies associated to 
the large-scale structure (protocluster members) and characterize their nuclear activity in 
terms of X-ray luminosity and spectral properties.  
% down to luminosities $\sim 4 \times 10^{42}$ erg s$^{-1}$) in the rest-frame 2-10 keV band. 
In Sections \ref{function} and \ref{fraction}, we present the X-ray luminosity 
function and the AGN fraction of the protocluster members, respectively.
In Section \ref{enhancement}, we discuss the enhancement factor implied 
by our new measurements and discuss our results in comparison to previous
X-ray studies of protoclusters.  Possible extensions of this work in 
promising directions are briefly discussed in Section \ref{perspectives}.
Finally, our conclusions are summarized in Section \ref{conclusions}.
For the sake of clarity, we refer to the radio galaxy and its immediate
surroundings as the Spiderweb Galaxy, or J1140-2629, 
while "Spiderweb Complex" indicates the 
large-scale structure potentially associated to the radio 
galaxy in a broad redshift range ($2.0<z<2.3$).
% to a large extent, 
% not necessarily doomed to collapse in a single virialized halo by $z=0$. 
The term "protocluster" is instead used to include only the 
galaxies in the narrower redshift range $2.11<z<2.20$ \citep[see also][]{2021Jin}, assuming that 
all the protocluster members will end up in a massive ($\sim$few times  10$^{14} M_\odot$)
cluster today.  Throughout this paper, we adopt the seven-year WMAP cosmology 
with $\Omega_{\Lambda} =0.73 $, $\Omega_m =0.27$, and $H_0 = 70.4$ km s$^{-1}$ 
Mpc$^{-1}$ \citep{2011Komatsu}. In this cosmology, at $z=2.156$, 1 arcsec 
corresponds to 8.473 kpc, the Universe is  3.13 Gyr old, 
and the lookback time is 77\% of the age of the Universe.  Quoted error 
bars correspond to a 1 $\sigma$ confidence level unless noted otherwise.

%%%%%%%%%%%%%%%%%%%%%%%%%%%%%%%%%%%%%%%%%%%%%%%%%%%%%%%%%%%%
% REVIEW OF LITERATURE ON THE SPIDERWEB
% A SHORT SUMMARY OF OBSERVATIONAL PAPERS ON THE SPIDERWEB
% CLASSIFIED ACCORDING TO SCIENCE TOPICS
% HERE WE FOCUS ONLY ON THE PROTOCLUSTER STRUCTURE
%%%%%%%%%%%%%%%%%%%%%%%%%%%%%%%%%%%%%%%%%%%%%%%%%%%%%%%%%%%%

\section{The Spiderweb Complex: previous observational campaigns 
and main results\label{review}}

Since its discovery in 1994, the Spiderweb Galaxy (J1140-2629) 
and the Spiderweb Complex have become an 
intensively studied region. In the last 25 years, 54 papers have been published 
about this field including observations in radio, sub-millimeter (submm), infrared (IR), 
optical, and X-ray wavelengths.  In this section, we summarize the main 
results that have been obtained 
so far, focusing on the protocluster environment.

% The results on the central 
% Spiderweb galaxy (J1140-2629) and its immediate surroundings 
% will be reviewed in a companion paper (Tozzi et al. 2021b).

\subsection{Discovery and peculiarities}

The Spiderweb Galaxy was discovered during the implementation of an ESO Key Program
designed to find high-z radio galaxies (HzRG) by targeting ultra-steep spectrum radio 
sources \citep{1994Roettgering}.  The object was noticed as peculiar 
because of its unusually clumpy and bent radio morphology 
coupled with an exceptionally high rotation measure ($\sim 6200$ rad/m), 
indicating that the radio synchrotron jet is surrounded and its 
propagation constrained by a dense ($10^{-1}- 10^{-2}$cm$^{-3}$) external medium 
\citep{1997Carilli,1997Pentericci,1998Athreya}.  Optical follow-up classified 
it as a narrow emission-line galaxy at $z=2.156$ \citep[see][]{1997Rottgering}, and
revealed a clumpy optical continuum morphology and a spectacular Ly$\alpha$ halo 
(still one of the largest known in the Universe)
with a luminosity of $42.5 \times 10^{44}$ erg/s
elongated along the jet.  The total stellar mass inferred
from the K-band luminosity is $10^{12} M_\odot$ \citep{1997Pentericci}, 
confirmed by Spitzer data \citep{2007Seymour}, making the Spiderweb one of the 
most massive high-z galaxies known.  Using the Hubble Space Telescope (HST) Advanced Camera for Surveys (ACS) observations, 
the clumpy features, embedded in the 200 kpc Ly$\alpha$ halo, whose distribution 
resembles that of flies trapped in a spiderweb, have been shown to 
correspond to approximately ten star-forming satellite galaxies moving with peculiar 
velocities of several hundred km/s and presumed to be merging with the central radio 
galaxy \citep{2006Miley}.  

Eventually, \citet{2011Nesvadba} found that the optical and near-infrared (NIR) spectrum of the nucleus 
of J1140-2629 shows a broad, spatially unresolved H$\alpha$ line
(blended with [NII]) with a FWHM of $\sim 14900$ km/s.
The presence of broad nuclear lines is rare in optical
counterparts of high-redshift radio sources typically classified as TypeII AGN. 
% Also, several emission lines have been 
% detected \citep[see][]{2008Humphrey} including [OII], [NeIII], H$_\beta$ narrow, [OIII], 
% [OI], [NII], H$_\alpha$ 
% narrow and broad, [NII], and [SII], and a strong CIV and HeII 
% emission with equivalent widths of $101 \AA$ and $87$ \AA, respectively. 
% Such an optical spectrum is consistent 
% with photoionization from the AGN, with an estimated average electron temperature of 
% $T_e=14100^{+1000}_{-600}$ K, and a roughly solar abundance as derived from the $N/H$ ratio.
The black hole mass is estimated to be $M_{\Large \bullet} \sim 2\times 10^{10}M_\odot$, 
putting J1140-2629 a factor of $\sim 2$ above the local $M_{\Large \bullet}-\sigma$ relation, 
in agreement with its expected evolution. % \citep{2011Nesvadba}. 
Taken together, the observed characteristics suggest that the 
Spiderweb is indeed a massive radio galaxy forming at the center of a dynamically evolving 
protocluster region and is likely to evolve into a brightest cluster galaxy.
The Spiderweb is therefore one of the most revealing laboratories 
for studying the evolution of a protocluster in the early Universe,
and may represent a typical phase in the 
formation of massive galaxy clusters with prominent central galaxies.

\subsection{Identification of protocluster members and their properties}

The search for protocluster members of the Spiderweb Complex was based on a series of
works with spectroscopic follow-up and narrow band imaging in the field. 
A first sample of 50 candidate Ly$\alpha$ emitters in a $6\times 6$ arcmin$^2$ field 
\citep{2000Kurk} provided support to the existence of several star forming 
galaxies in the protocluster. Of these candidates, 15 were 
eventually confirmed to have redshift 
in the range $2.16\pm 0.02$ with VLT-FORS
spectroscopy by \citet{2000Pentericci}. One of these confirmed sources shows 
clear signs of AGN activity with very broad Ly$\alpha$ and CIV emission lines,
while all the others simply show narrow Ly$\alpha$ emission.  
More extensive VLT imaging in  NB0.38, B, R, and I bands with  FORS2 and in the 
$J_s$, H, $K_{\rm S}$, and NB2.07 bands with ISAAC identified 40 H$\alpha$ candidate emitters 
and 44 extremely red objects (EROs) consistent with $z\sim 2$, whose density 
increases towards the central galaxy suggesting a physical 
association \citep{2004aKurk}. 
% (within a radius of 40 arcsec)
% EROs and H$\alpha$ emitter candidates are more concentrated 
% towards PKS 1138-262 than the candidate $Ly\alpha$ emitters.  
Estimated star formation rates (SFRs) without dust correction are in the 
range 2-50 $M_\odot$/yr and 0.4-4.3 
$M_\odot$/yr for the H$\alpha$- and Ly$\alpha$-emitter candidates, respectively.
% The range becomes $\sim 1-60 M_\odot$/year when derived from the UV continuum for both source 
% category, resulting in an total value of $\sim 500 M_\odot$/yr.  
The SFR per unit mass, estimated from K-band magnitudes, 
are $2.5$ times higher than in local massive clusters.  

% The total stellar masses of the non-AGN 
% Ly$\alpha$ and H$\alpha$ emitter candidates are $3.14$ and $8.9\times 10^{11} M_\odot$, 
% respectively. Overall, the difference in values between the Ly$\alpha$ and H$\alpha$
% emitters indicates that the Ly$\alpha$ emitters are still undergoing
% accretion, while the more massive
% H$\alpha$ emitters are older and have retained more metal-rich gas and dust. 
% that depress the $Ly\alpha/H\alpha$ ratio. The general conclusion is 
% that J1140-2629 is in a density peak that
% will most probably evolve into a massive galaxy cluster. 

In a spectroscopic follow-up in the IR band with ISAAC on the VLT, an additional 
nine H$\alpha$ emitter candidates within 0.6 Mpc from J1140-2629 
were confirmed as protocluster members \citep{2004bKurk}. 
% with a lower velocity dispersion (360 km/s). 
The narrowness of the emission lines 
of eight of the nine confirmed sources indicates that they 
are powered by star formation dominated by a very young ($<100$ Myr) 
stellar population with moderately 
high metallicity.  Only in one of the emitters does the 
H$\alpha$ appear extremely broad with a FWHM exceeding 5000 km/s.  
The contribution of these nine candidates to the SFR density in the protocluster shows it to be 
at least an order of magnitude larger than that found at 
$\langle z \rangle =2.23$ in the HDF-N.

Near-infrared spectroscopy of narrow-band-selected
H$\alpha$ emitters with MOIRCS on Subaru presented by  \citet{2014Shimakawa,2018Shimakawa}
% The SFR limit set by the limiting flux in the H$\alpha$ line corresponds to 20-25 $M_\odot$/yr.  
% extended to 27 
significantly increased the number of spectroscopically confirmed protocluster members.   
An additional 4 protocluster members were identified among 13 H$\alpha$ emitter candidates
using ALMA CO(3–2) ($\nu_{rest}=345.796$ GHz) observations \citep{2019Tadaki},
all of them also detected in CO(4–3) by \citet{2018Emonts}.
Recently, \citet{2021Jin} obtained 46 CO(1-0) detections in the range $2.09<z<2.22$, using
13 individual pointings with ATCA.  
% Finally, Perez-Martinez et al. (in preparation) 
% spectroscopically identified several sources with KMOS in the context of the MAHALO survey.

% \citet{2010Tanaka}, find that galaxies in the Spiderweb comples have similar 
% ages but shorter star-formation time scales, lower star formation rates and, 
% consistently, weaker dust extinction compared to GOODS galaxies at $z\sim 2$, 
% and assemble their mass $\sim 1 $ Gyr earlier than field galaxies. In addition, they 
% stress the role of frequent mergers during the assembly stage of clusters, that 
% can be the main reason for a shorter SF time scale and a more intense initial 
% star formation.  They conclude that these may be the hint of the environmental 
% dependence of galaxy  properties in the Spiderweb.  However, the presence of $ 5$ 
% sources with a $SFR>1000 M_\odot$/yr is better explained as a to nature rather than 
% nurture, since the structure is young.  
Photometric studies also helped to characterize the protocluster population.
Infrared Surveys indicated an overdensity of 3-4 $\sigma$ with respect to 
the field at $24\mu$m \citep{2012Mayo}, with Spitzer IRAC 
colors \citep{2012Galametz}, and 
% at $250\mu$m 
with Herschel SPIRE \citep{2013Valtchanov,2014Rigby}.  
The distribution of IR protocluster galaxy candidates confirms
the approximate filamentary shape of the protocluster, 
previously obtained using optical narrowband selection.
% ($RA=11:41:04.44, DEC=-26:35:08.6$), within which 76 SPIRE sources are detected at $250\mu$m.  
% The same region is not overdense in the other SPIRE maps at 350 and 500 $\mu$m. \citet{2013Valtchanov}
% also derived photo-z using a modified black body. The values of $z_{BB}$, however, are not 
% reliable taken individually, but the overall distribution has a peak at $\langle z_{BB}\rangle \sim 1.6$
% in the overdense region, despite with a strong dependence on the assumed dust temperature
% (for example $\langle z_{BB}\rangle \sim 2.2$ if $T=38 $ K instead of $30$ K, making it a structure
% possibly linked to the protocluster). 
% \citet{2014Rigby} surveyed 19 HzRGs surveyed with {\sl Herschel} SPIRE in the three bands, 
% \and found that only a few of them appear to be overdense, with J1140-2629 being one of them. 
% \The general conclusion, however, is that identifying protocluster with Herschel is not as efficient 
% \compared to other wavelengths, and that FIR data typically identify structures different from 
% \those probed with narrow-band and mid-IR selection. 
In the submm band, an overdensity of a factor of four is observed with APEX LABOCA at 870 
$\mu$m \citep{2014Dannerbauer}. The submm flux distribution 
at the protocluster redshift would imply an SFR that is
higher by about an order of magnitude with respect to the H$\alpha$-derived values.
The sample of narrow-band-selected candidates 
was extended to 68 H$\alpha$ emitters with the 
MAHALO Deep Cluster Survey \citep{2018Shimakawa}  with the Subaru Telescope,
17 of which are newly discovered.
% , plus 13 narrow band emitter 
% that are classified as new H$\alpha$ emitter candidate. 
In addition, $Bz'K_s$ selection identified 34 distant red galaxies, 
half of which have photo-z consistent with being protocluster members.
% (and 3 of them are already spectroscopically confirmed members).  
% Their primary goal is to determine
% the stellar mass function of protocluster members and quantify passive and AGN fraction as a 
% function of $M_*$. 

Overall, the galaxy population of the Spiderweb Complex appears to be a mix of
active and star forming galaxies, and already massive, passive galaxies 
that constitute a nascent red sequence \citep{2007Kodama,2008Zirm,2012Zirm},
% mostly traced by the 7 best-candidates elliptical galaxies, 
% with a slope steeper than the local one. This work was extended by \citet{2012Zirm}, 
% that measured a lower stellar mass
% density in protocluster members with respect to the field at the 
% same redshift, that can be interpreted as the cumulative effect of the enhanced merger rate in the
% protocluster, that makes their stellar mass profile more extended. 
% The nascent red sequence has also been observed also by  \citet{2007Kodama} with 
% MOIRCS on Subaru.  The spatial distribution of the elliptical protocluster candidates 
and that approximately follow an elongated structure extending $\sim 5$ Mpc in diameter.
% previously observed 
% for the Ly$\alpha$ and H$\alpha$ emitters. 
\citet{2013Tanaka} found that the weak red sequence is 
consistent with a formation redshift in the range $3<z<4$. 
In addition, \citet{2013Koyama} found that the properties of the 
H$\alpha$ emitters are consistent with them having formed a large part of 
their stellar mass at $z> 2$, but observed in a still intense star-forming 
phase.  The presence of both star forming and red galaxies may be related 
to a typical phase associated to environmental effects driven by the protocluster collapse 
that accelerates their formation and evolution. 
% typical  for protoclusters at $z\sim 2$, 
% However, the two populations are clearly distinct.
% A possible reason for that is that the emission line members are likely to be star forming, 
% with masses too small ($M<10^{10}M_\odot$) to be detected 
% in the NIR imaging. 

Finally, we note that four out of six ($60$\%) massive $H_\alpha$ emitters (HAEs)
($M_* > 10^{11} M_\odot$) were previously identified as bright X-ray 
sources \citep[see][]{2002Pentericci,2005Croft}.  Also, \citet{2005Croft} 
confirmed two X-ray sources as protocluster members
using LRIS Keck spectroscopy, and classified both as AGN due to the presence 
of NV, CIV, and [CIII] in the spectrum, bringing the total number of 
X-ray-emitting members to five.
% In addition, one of the previously confirmed 
% Ly$\alpha$ emitters also showed narrow NV in the new spectrum. 
This finding suggests that protocluster members that host X-ray AGNs
may be in the transition stage from dusty star-forming galaxies to
passive populations. 

\subsection{Role of the new X-ray observations}

The nature of the galaxy population of the Spiderweb Protocluster 
will be further investigated in this work thanks to the addition of our deep X-ray 
data. Our study relies on a collection of 252 unique sources in the field that 
have spectroscopic or narrow-band/photometric
redshift in the range $2.0<z<2.3$ (see Section 3).
A detailed list of all the spectroscopic members and color- or narrow-band-selected candidates, 
and the identification of the Spiderweb Complex candidates, is shown in Section \ref{multiwav_section}.

% The Spiderweb Complex shows a clear course overdensity 
% in all previously observed bands, from the sub-mm to the X-rays, ranging from 
% factors of $2-6$ to $\sim 100$, depending on the source selection and 
% the specific region considered within the complex.  
Using our deep {\sl Chandra} observation, we explore the following 
aspects: (i) the frequency of X-ray nuclear activity among
the protocluster members; (ii) the enhancement of nuclear
activity in the protocluster; (iii) the distribution of phases 
in the diffuse baryons in the halo of the Spiderweb Galaxy, 
and the relation with its strong nuclear activity; 
and (iv) the presence of diffuse thermal emission possibly associated to
the virialization of a central halo.  In this paper, we focus on the 
AGN population.  The presence of hot, diffuse baryons around the 
Spiderweb Galaxy and their relation to its nuclear activity 
is discussed in two papers based on X-rays and the SZ signal 
(Tozzi et al. in preparation; Di Mascolo et al. in preparation).  
The main results and the data products of the present paper 
can be found on the project webpage\footnote{\tt http://www.arcetri.inaf.it/spiderweb/}.

% Answering these specific questions may help us in solving more general aspects 
% concerning galaxy evolution, such as: is the halo of hot and cold gas a typical 
% phase of the formation of BCG progenitors? Can we constrain the nature and time scale of 
% the environmental effects that drive the galaxies from dusty-star forming ones to 
% red passive ellipticals in protoclusters? 
% Can we answer the "nature or nurture" question 
% for the evolution of cluster ellipticals?  Is the Spiderweb Complex an 
% archetypical protocluster (in other words, is it evolving into a massive 
% virialized structure by $z=0$)?

% Which is the role of nuclear activity in the evolution of the protocluster galaxies? 
% Do we see and increase in the fraction of X-ray emitting sources among the H$\alpha$ emitters?  
% What is the dominant population 
% of the protocluster galaxies (dusty starburst and old ellipticals?), and how many 
% can still be discovered?  

%%%%%%%%%%%%%%%%%%%%%%%%%%%%%%%%%%%%%%%%%%%%%%%%%%%%%%%%%%%%%%%%%%%%%%
% DETAILED DESCRIPTION OF MULTIWAVELENGTH DATA USED IN THIS WORK
%%%%%%%%%%%%%%%%%%%%%%%%%%%%%%%%%%%%%%%%%%%%%%%%%%%%%%%%%%%%%%%%%%%%%%

\section{Multiwavelength data used in this work\label{multiwav_section}}

In this work, we exploited part of the large set of multiwavelength data 
publicly available for the Spiderweb Complex field.  
% Optical and NIR images have been used to identify the position 
% of the counterparts of the X-ray sources.  
We used the HST images for the central regions and the 
Subaru images ---which cover the entire {\sl Chandra} field of view (FOV)--- 
to identify the optical counterparts of the X-ray sources.  In particular, 
we used a 23 ks exposure in the F814W band with ACS (HST Proposal 10327, PI H. Ford), 
and B, $r'$, $z'$ exposures obtained with the Suprime camera at the Subaru telescope 
and covering a $\approx $ $30\times 30$ arcmin$^2$ field.
%and a stack of 1h15m from  exposures obtained with the Suprime camera at the Subaru 
%elescope  in $z'$ filter and covering a $30\times 30$ arcmin$^2$ field.
%of a $30\times 30$ arcmin field observed with Suprime at the Subaru telescope in the $z$ band.
We also used %VLT-FORS1 images obtained in the [OII] observer filter \footnote{ESO 63.O-0477, P.I. Miley}, 
VLT-VIMOS imaging data in the U band, and VLT-HAWKI images in Y, H, and K bands.  
The Suprime B, $r'$, and $z'$ raw data were retrieved from the SMOKA archive \citep{2002Baba}. 
The images were reduced with a custom pipeline because of the four-ports-readout layout 
of the camera chips. Images have been {\em overscan}, {\em master bias,} and {\em master flats} 
corrected. We then used an improved version of  own  pipeline as described in \citep{2009Nonino}, 
which currently uses Scamp \citep{2006Bertin} with GAIA-EDR3 as astrometric reference, 
and Swarp \citep{2002Bertin} for the final coaddition. A similar approach was 
used for the VLT-VIMOS and VLT-HAWKI images, with the raw data retrieved from the 
ESO Archive. Removal of the instrumental signatures was performed using 
IRAF\footnote{IRAF was distributed by the National Optical Astronomy Observatory, 
which was managed by the Association of Universities for Research in Astronomy (AURA) 
under a cooperative agreement with the National Science Foundation.}.
For the NIR HAWKI images, the appropriate darks were retrieved from the 
ESO Archive, stacked and applied to the raw data. For the optical data, 
the bias and the flats were used to create static weights which account 
for chips defects. For the NIR data, darks and flats were instead used.

%Suprime and VIMOS data have 
%been reduced in the standard manner, astrometrized using Scamp %\citep{2006Bertin} 
%with GAIA-EDR3 as reference. Final stacks have been obtained with Swarp \citep{2002Bertin}.
%We also use VLT-FORS data.
% ADD DETAILS AND DATA REDUCTION OF THE HST IMAGE BY BRAMMER AND THE SUBARU IMAGE
% Subaru images in $B$, $rp$ nd $zp$ bands.  

We used optical and NIR photometry to estimate host galaxy masses and rest-frame luminosities of 
the X-ray AGNs in our sample. Detailed descriptions of the photometric measurements, 
the multi-wavelength catalog, the photometric redshift, and our stellar mass estimation 
will be  provided elsewhere (Pannella et al., in preparation). In the following, for the 
sake of completeness we briefly summarize the data and procedures relevant here.

\begin{table*}
\centering
\caption{Optical and NIR data.}
\begin{tabular}{lcccc}
\hline
Filter        & Instrument    & FWHM     & 5$\sigma$ &  Science Archive\\
              & /Telescope    & (arcsec) & (AB)      &  /Project ID(s)\\
\hline
$U$           & VIMOS/VLT     & 0.86     & 26.4 &  ESO - 383.A-0891\\
$B$           & S-Cam/Subaru  & 1.12    & 25.8 &  SMOKA - o17409\\
$r'$          & S-Cam/Subaru  & 0.73     & 25.2 &  SMOKA - o17409\\
$z'$          & S-Cam/Subaru  & 0.65     & 24.4&  SMOKA - o10144\\
$Y$           & HAWK-I/VLT    & 0.46     & 24.5 &  ESO - 088.A-0754, 091.A-0106\\
$H$           & HAWK-I/VLT    & 0.58     & 23.2 &  ESO - 088.A-0754, 091.A-0106\\
$K_{\rm S}$   & HAWK-I/VLT    & 0.40     & 22.8 &  ESO - 088.A-0754, 091.A-0106\\
\hline
\end{tabular}
\label{tab_data}
\tablefoot{The first to fourth columns indicate filter name, 
instrument/telescope, seeing FWHM, and $5\sigma$ limiting magnitude in a 2 arcsec diameter 
aperture, respectively. In the fifth column we list the science archive and 
proposal ID of the raw dataset.}
\end{table*}

We used Very Large Telescope (VLT) 
and Subaru stacked images to create a multi-wavelength 
catalog with seven passbands (U,B,$r'$,$z'$,Y,H,Ks) from U to K. 
In Table~\ref{tab_data} we summarize the main properties of the dataset used.
We used SExtractor in dual image mode to measure photometry. The $K_{\rm S}$-band 
HAWK-I image was adopted as the primary detection image,
because it represents the best compromise ---among all available bands--- between
the need for a robust tracer of galaxy stellar mass and sufficient angular resolution
($\simeq$0.4"), which simplifies catalog assembly and photometry measurements. 
The whole catalog contains 2 279 objects over the $K_{\rm S}$ image field of 
approximately 84 arcmin$^2$ and down to an AB magnitude of 22.8 
(i.e., the image 5$\sigma$ limiting magnitude).
The images used here have very different resolutions. Rather than 
convolving all images to the lower resolution, which would result in 
a significant loss of information, we account for this in the estimate of 
aperture colors by applying PSF-matching corrections based on the growth curve 
of point-like sources. To limit uncertainties in such corrections we use 2" 
diameter apertures to sample the galaxy spectral energy distributions (SEDs).

Photometric redshifts are computed using EAZY (Brammer et al. 2008). Global 
photometric zero points are adjusted iteratively by minimizing the difference between 
the various photo-z values 
and the available spectroscopic redshifts 
\citep[e.g.,][]{2004Gabasch,2010Ilbert,2015Pannella}.  By comparison with 
the available spectroscopic sample, 
the rms of $(z_{phot}-z_{spec})/(1+z_{spec})$ is $\sim$6\%.  
Stellar masses are derived using a modified 
version\footnote{https://github.com/cschreib/fastpp} of FAST \citep{2009Kriek},
adopting the \citet{1955Salpeter} IMF, \citet{2003Bruzual} stellar population 
synthesis models, delayed exponentially declining star formation histories (SFHs), 
the \citet{2000Calzetti} dust extinction law (with A$_V$ = 0 to 4), and solar metallicity. 
In deriving the stellar mass, we use the photo-z obtained using the FAST code, 
while we use the spectroscopic redshifts when available.

To compile a comprehensive list of spectroscopic redshifts, 
we use a collection of catalogs of all the spectroscopically 
confirmed protocluster members, member candidates selected on the basis of color or narrow-band 
photometry, and random sources with redshift in the field.  
The 17 catalogs, based on 12 different publications, are listed in Table \ref{catalogs}.
The catalogs have a significant overlap, because several of them 
include and extend previous results. The total number of unique sources with any redshift
information in the field is 252, out of which 161
have spectroscopic redshifts.  We also compile a complementary list of 
91 protocluster member candidates that were identified thanks to narrow-band or color selection
in previous works.
Most (90\%) of the sources (both confirmed and candidate members)
are included in a box of $6.3\times 3.7$ arcmin$^2$ 
(corresponding to $3.13 \times 1.84$ Mpc at $z=2.156$) roughly centered
on the Spiderweb Galaxy and elongated along the E-W direction. Only nine sources with 
redshift information are outside a circle of 5 arcmin centered on the Spiderweb Galaxy, 
and none of them are in the range $2.0<z<2.3$.

\begin{figure}
\begin{center}
\includegraphics[width=0.49\textwidth]{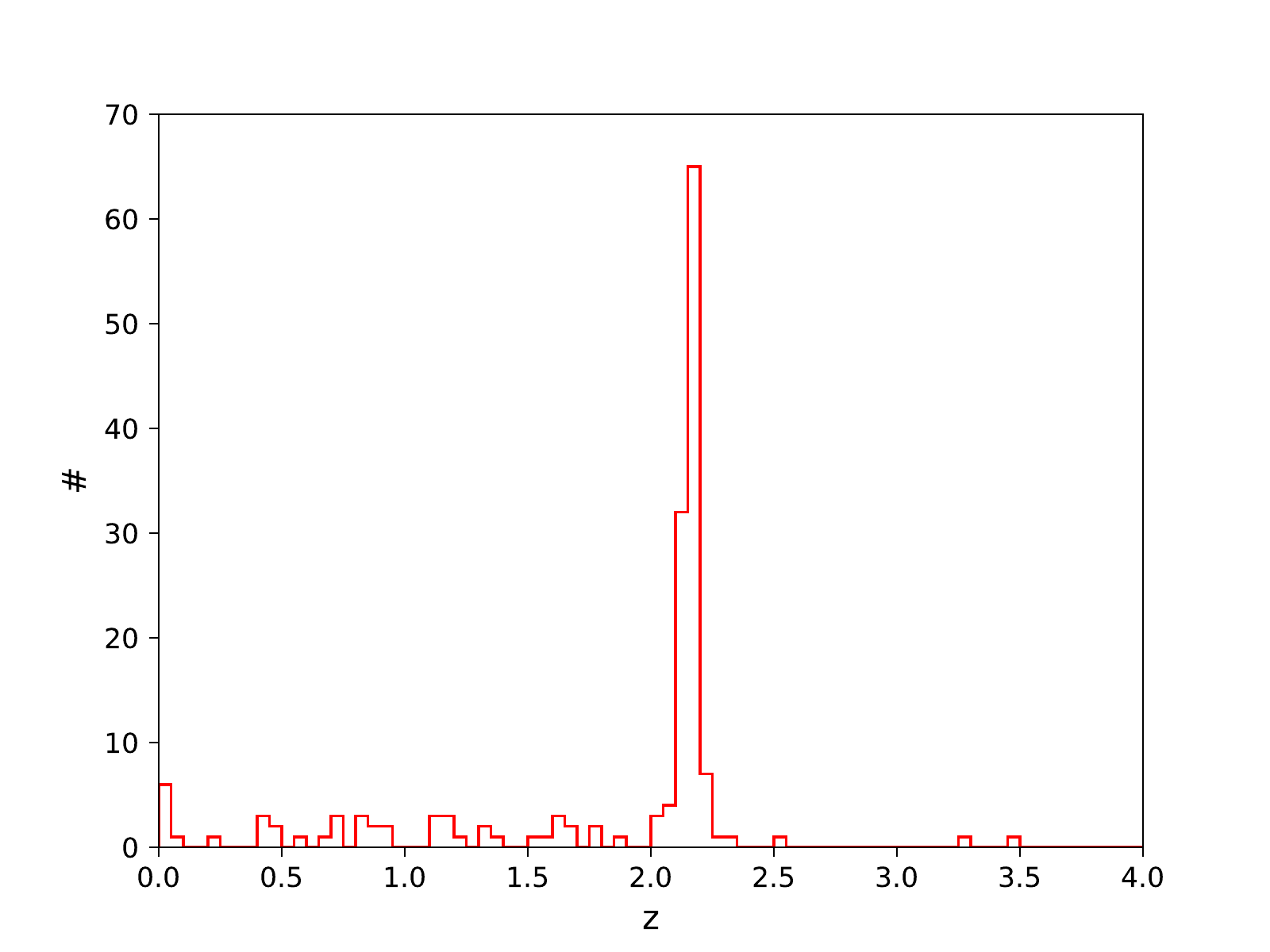}

\includegraphics[width=0.49\textwidth]{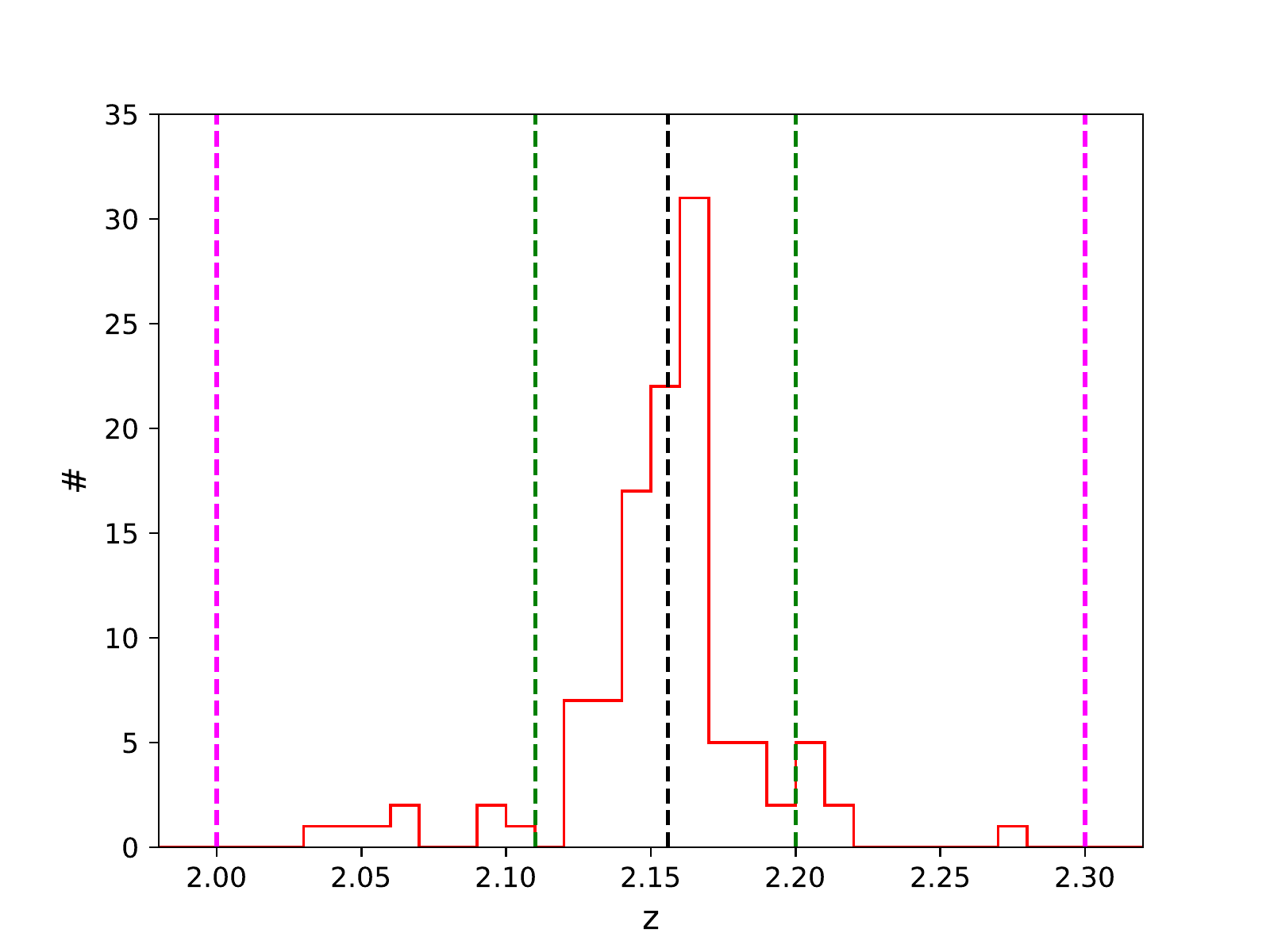}
\caption{Redshift distribution in the Spiderweb field.   
Upper panel: Distribution of all the 161 spectroscopic redshifts measured in the field. 
Lower panel: redshift distribution in the range $2.0<z<2.3$.
Vertical dashed lines show the redshift of the Spiderweb Galaxy as measured with
low resolution spectroscopy by \citet{1995vanOjik} (black line), 
the range adopted to define the Spiderweb Complex (magenta lines)
and the protocluster (green lines), as discussed in the text. }
\label{zdist}
\end{center}
\end{figure}

The distribution of the 161
available spectroscopic redshifts is shown in Figure 
\ref{zdist}.  Clearly the majority of the redshift values are close to the 
Spiderweb redshift $z=2.156$, because mostly protocluster member candidates have been
targeted.  We consider a redshift interval $\Delta z=0.3$ roughly 
symmetric with respect to the Spiderweb redshift, corresponding to a 
velocity difference of $\Delta v\sim c\times \Delta z/(1+z)$ associated to 
the Hubble flow.  This interval spans all the sources possibly associated to 
the Spiderweb large-scale structure.  If we consider the range $2.0<z<2.3$, 
corresponding to a maximum $\Delta v \sim 28500$ km/s, 
we find 112 sources.  We do not expect to see a normal distribution for the redshift, and therefore
we adopt the biweight estimator for the central value and the standard deviation, 
as suggested by \citet{1990Beers} for sample size $n\geq 100$\footnote{See equations 5, 6 and 9 in
\citet{1990Beers}}.  We find $z_{BI}=2.15754$ and $\sigma_{BI}=0.0208612$.  Therefore, 
two- and three-sigma intervals centered on $z_{BI}$ are $2.116<z <2.1996$ and
$2.095<z<2.220$, respectively.  We decided to assume the $2\sigma$ intervals (approximated as
$2.11<z<2.20$) as the protocluster redshift range.

In the range  $2.11<z<2.20$, corresponding to 
a maximum $\Delta v \sim 8500$ km/s, we find 96 sources.  We compute the
formal velocity dispersion $\sigma_v$ in this range and find
% $\sigma_v\sim 3100$ km/s in the range  $2.0<z<2.3$, and 
$\sigma_v\sim 1490\pm 150$ km/s. 
This value should be compared to that expected for a flat distribution 
in the same redshift range as expected for sources following the Hubble flow, 
% in a given redshift interval, which 
% are $\sim 8200$ in the $2.0<z<2.3$ 
which is $\sim 2700$ km/s.  
We conclude that the sources are clearly detached from the Hubble flow, 
and show a global velocity dispersion that would correspond to a very-high-mass ($M>10^{15}\, M_\odot$), virialized single halo. Given the mass scale and the distribution of sources in the sky, 
the virialized, single-halo hypothesis is clearly unrealistic.  In addition, 
we notice that excluding the new spectroscopic sources identified
by \citet{2021Jin} with the CO(1-0) line, the velocity dispersion 
drops to $\sigma_v=980\pm 100$ km/s (that would correspond to a temperature
of $\sim 6$ keV for a virialized halo). 
As already shown in \citet{2021Jin}, this points out that the CO sources show a different dynamical behavior, and are probably
distributed along a large-scale filament or a galaxy superprotocluster. 
In addition, half of the CO emitters have measured velocities
larger than the escape velocities of any realistic virialized 
halo hosting the Spiderweb Galaxy, suggesting that the
outer regions traced by the CO sources are still far from 
being accreted onto the main halo.
In this work we do not attempt to further investigate the dynamical structure of the Spiderweb 
complex. However, the number of spectroscopically confirmed members in the Spiderweb
Complex is now the highest for a single structure at $z>2$ and a detailed, 
spatially resolved dynamical analysis will be a relevant piece of information to 
constrain the current status and the future evolution of the structure. 

% If we exclude the sources by Jn 2021, we find a dispersion of ~960 km/s, 
% closer to that of a large halo.
In summary, in this paper we consider the 96 sources included in the 
narrow redshift bin $2.11<z<2.20$ as belonging to the protocluster, 
while the additional 16 sources included in the range $2.0<z<2.3$ 
are considered part of the large-scale structure associated to the Spiderweb Complex, 
and are most likely not dynamically bound. % linked to it, but may be eventually included in the 
% protocluster as the gravitational collapse goes on.  
The corresponding redshift intervals are
shown in Figure \ref{zdist} with vertical dashed lines.

\begin{table}
\caption{Catalogs of spectroscopically confirmed protocluster members and member 
candidates used in this work. }
\begin{center}
\begin{tabular}[width=0.5\textwidth]{lcc}
\hline
%ER: With "Nr. of Sources" the table went over the boundaries
%ER: Alternatively you could use two lines for
%ER: "Color \\ 
%ER: or narrow-band selected"
Reference & Selection method & Members \\
\hline
\hline
Spectroscopic & &  \\
\hline
\hline
\citet{2000Pentericci} & Ly$\alpha$ emission & 15\\
\citet{2004bKurk} & H$\alpha$ emission & 9 \\
\citet{2005Croft} & Ly$\alpha$ emission   & 7 [17]\\
\citet{2010Doherty} & H$\alpha$ emission & 2 [2] \\
\citet{2011Kuiper} & OII emission & 15 [1]\\
\citet{2013Tanaka} & OII/OIII emission & 4 [10] \\
\citet{2014Dannerbauer} & Specz & 11 [4] \\
% THIS WAS WRONG \citet{2018Shimakawa} & H$\alpha$ emission & 29 \\
\citet{2014Shimakawa} & H$\alpha$ emission & 23 \\
\citet{2019Tadaki} &  CO emission & 4 \\
\citet{2021Jin} &   CO emission & 46 \\
\hline
\hline
Color or narrow-band selected & & \\
\hline
\hline
\citet{2002Pentericci} & photo-z &  3 \\ 
\citet{2004aKurk} & Ly$\alpha$ emitter & 40\\
\citet{2004aKurk} &  H$\alpha$ emitter & 40  \\
\citet{2004aKurk} &  EROs & 44 \\
\citet{2014Dannerbauer} & FIR photo-z & 2 [4] \\
\citet{2018Shimakawa} &  H$\alpha$ emitter & 68 \\
\citet{2019Tadaki} &   H$\alpha$ emitter & 9 \\
% \citet{2021Perez-Martinez} &   Specz & 15? \\
% ATTENTION, NOT ALL TANAKA2013 ARE SPECZ
% MAGENTA CIRC = Z FROM NED WITHIN 14 ARCMIN Field_Spiderweb_redshift_NED.reg
\hline
\end{tabular}
\end{center}
\label{catalogs}
\tablefoot{In the second column, we list the identification/selection method, 
while in the third column we report the number of 
confirmed or candidate members of the Spiderweb complex ($2.0<z<2.3$).  
Occasionally, the number of random sources in the Spiderweb field
outside the range $2.0<z<2.3$ is reported in square brackets.  
The catalogs were  obtained independently in different works, but often overlap significantly
because they are often derived from the same datasets. The total number of unique sources is 252. 
We find 161 sources with spectroscopic redshifts, of which  112 are in the range $2.0<z<2.3$.}
\end{table}

\section{X-ray data reduction\label{datareduction}}

\begin{table}
\caption{{\sl Chandra} observations of the Spiderweb Galaxy Complex. }
\label{obsid}
\begin{center}
\begin{tabular}[width=0.5\textwidth]{lcc}
\hline
Obsid & date & Exptime (s) \\
\hline
898         &   2000-06-06       &      27\,378 \\
21483    &      2019-11-18       &      49\,312 \\
22924    &      2020-02-26       &      33\,475 \\
22923    &      2020-02-28       &      34\,598 \\
23175    &      2020-03-01       &      29\,938 \\
22925    &      2020-03-06       &      39\,541 \\
22926    &      2020-03-07       &      43\,375 \\
22921    &      2020-03-17       &      24\,642 \\
23190    &      2020-03-17       &      24\,736 \\
23069    &      2020-03-23       &      19\,648 \\
23201    &      2020-03-26       &      27\,613 \\
21482    &      2020-04-02 &    34\,492 \\
22927    &      2020-04-02       &      32\,534 \\
23205    &      2020-04-03       &      23\,514 \\
22928    &      2020-04-10       &      29\,572 \\
23215    &      2020-04-13       &      19\,803 \\
21484    &      2020-04-22       &      46\,504 \\
22929    &      2020-04-23       &      51\,231 \\
23186    &      2020-06-13       &      14\,770 \\
22922    &      2020-07-22       &      18\,487 \\
21481    &      2020-08-13 &    40\,383 \\
22905    &      2020-08-04 &    49\,291 \\
\hline
\end{tabular}
\end{center}
\tablefoot{In the last column we list the exposure time after data reduction.}
\end{table}

The Spiderweb Galaxy was observed with a Chandra Large Program observation of 700 ks with ACIS-S 
granted in Cycle 20 (PI P. Tozzi). The observations were completed in the period November 2019
--August 2020, split into 21 separate pointings.  To this data set, we add the first X-ray observation 
with ACIS-S, dating back to June 2000 for a total of 39.5 ks.  
All the 22 Obsid used in this work are listed in Table \ref{obsid}.

Data reduction is performed starting from the level = 1 event files with {\tt CIAO 4.13}, with the 
latest release of the {\sl Chandra} Calibration Database at the time of writing {\tt (CALDB 4.9.4)}. 
We run the task {\tt destreak} to flag and remove spurious events (with moderate to small pulse 
heights) along single rows, which is particularly important for ACIS-S4.
We run the tool {\tt acis\_detect\_afterglow} to flag residual charge from cosmic rays in CCD 
pixels. Finally, 
as all our new observations are taken in the VFAINT mode we run the task 
{\tt acis\_process\_events} with the parameter {\tt apply\_cti=yes} 
to flag background events that are most likely associated to cosmic 
rays and reject them.  With this procedure, the ACIS
particle background can be significantly reduced compared to
the standard grade selection. The data are then filtered to include
only the standard event grades 0, 2, 3, 4, and 6. The {\sl level 2} event 
files generated in this way 
are visually inspected for flickering pixels or hot columns left after the standard reduction
but we find none after filtering out bad events. We also carefully inspect
the image of the removed photons, particularly to verify whether we have pile-up 
effects\footnote{https://cxc.cfa.harvard.edu/ciao/ahelp/acis\_pileup.html}.
We find that slightly more than 100 net counts in the total band are removed at the position of 
the nucleus of J1140-2629 across the 22 exposures.  This indicates that 
the source, which is by far the brightest in the field, suffers some amount of pile-up. However, 
we verify {\sl a posteriori} that this effect does not impact the spectral analysis of the
nucleus and therefore it is not worth giving up the VFAINT cleaning to recover less 
than 1\% of the flux in the nucleus. In particular, the VFAINT cleaning is very important
for characterization of the diffuse emission, which is presented in detail in 
further papers from the collaboration focused on the inverse Compton emission associated
to the radio jets \citep{2022Carilli,2022Anderson} and on the thermal component
(Tozzi et al. in preparation; see also Di Mascolo et al. in preparation). 

Finally, time intervals with high background are filtered by performing a 3$\sigma$ 
clipping of the background level. The light curves are extracted in the 2.3--7.3 keV band, 
and binned with a time interval of 200 s.  The time intervals when the background exceeds the average value  by 
$3\sigma$ are removed with the tool {\tt deflate}.  
The final total exposure time after data reduction and 
excluding the dead-time correction amounts to 715 ks (corresponding to the 
LIVETIME keyword in the header of {\sl Chandra} fits files), including the
first observation. The average removed time interval for the new observations 
amounts to only $\sim 0.3$\% of the observing time, except for Obsid 898, where 
$\sim 30$\% of the exposure time has been removed due to high-background time intervals. 
The monochromatic exposure maps are created at the energies of 1.5 and 4.5 keV for the 
soft and hard band respectively by running the tools {\tt mkinstmap} and {\tt mkexpmap}. 

\begin{figure}
\begin{center}
\includegraphics[angle=0,width=0.49\textwidth]{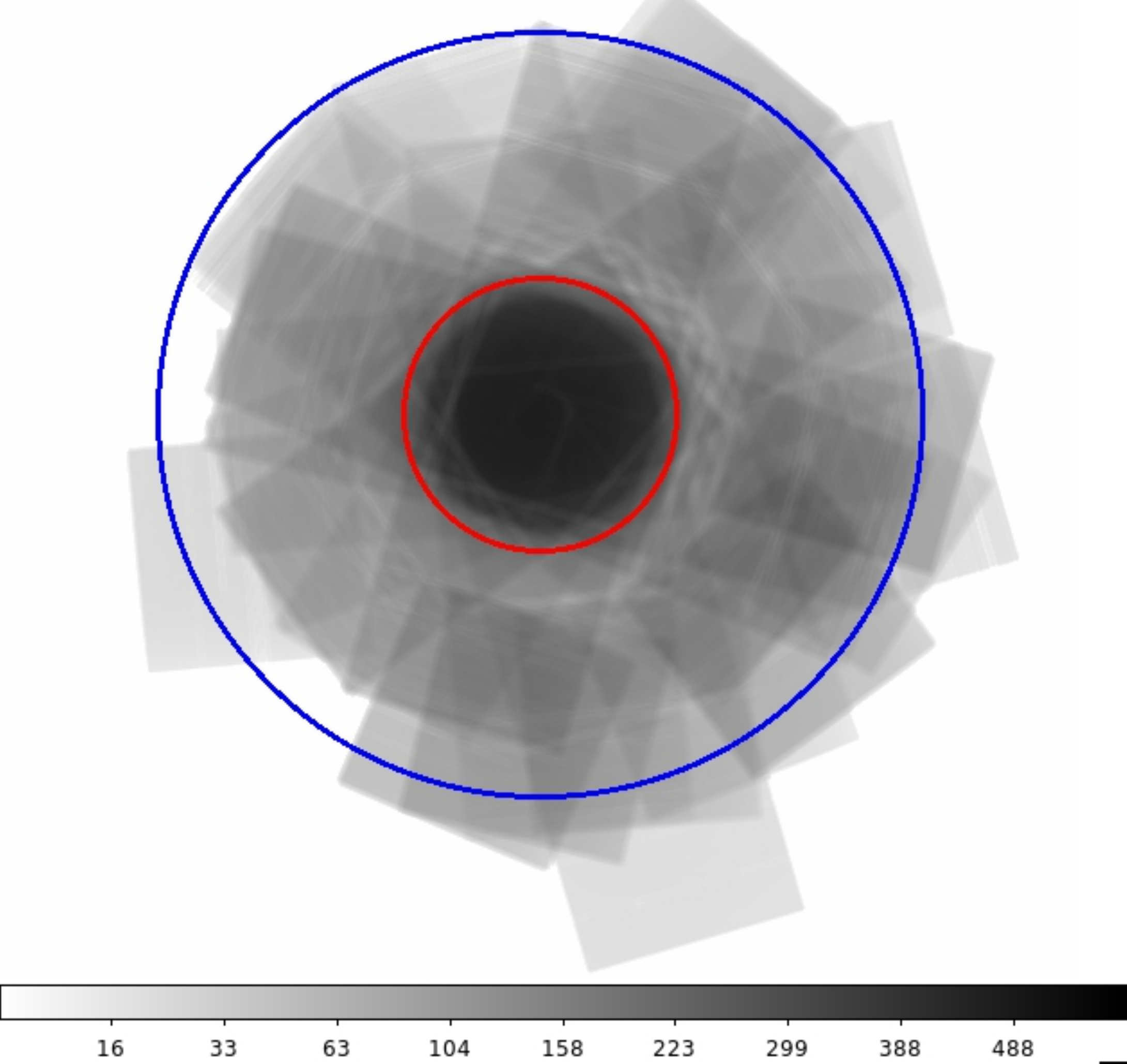}
\caption{Time-weighted, monochromatic exposure map at 1.5 keV 
of the Spiderweb Complex field.  The maximum value is 470 cm$^2$.
The innermost (red) circle shows the circular region investigated in this work, with a radius of 5 arcmin 
($\sim 2.5$ Mpc at $z=2.156$). All the points inside this circle 
have an effective area of larger than 240 cm$^2$. The outer annulus, 
with radii of 5 and 14 arcmin, has 
a much lower effective exposure, and a degraded angular resolution, and is 
not included in the present analysis.}
\label{FOV}
\end{center}
\end{figure}

The 22 {\sl level 2} event files are merged together with the tool {\tt reproject\_obs}, 
using the reference coordinates of Obsid 21483.  The exposure maps are reprojected 
onto the reference frame of Obsid 21483 with the tool {\tt reproject\_image}, and 
are then added together weighting each map by the fractional exposure time of the 
corresponding Obsid. Images for source detection are created in the 0.5-2 keV, 2-7 keV 
and total (0.5-7 keV) bands, with no binning to preserve the full angular resolution 
of $\sim 1$ arcsec (FWHM) at the aimpoint (1 pixel corresponds to 0.492 arcsec). 
The images include all the CCDs that were "on" while observing, namely ACIS-S3 
(where the aimpoint is always located), ACIS-S2, ACIS-S4, and ACIS-I3 (except Obsid=898, in which 
also ACIS-S5 and ACIS-I2 were on).  As the aimpoint is approximately the same for 
the 22 Obsid, and considering the geometry of the CCD positions, the effective 
exposure is optimal within a radius of few arcmin from the J1140-2629 position, 
while it is significantly lower in a wide annulus covered by the flanking CCDs.  
We identify a region with a radius of 5 arcmin centered on 
J1140-2629, roughly coincident with all the points with an effective 
area larger than 240 cm$^2$. This radius corresponds to $\sim 2.5$ Mpc
at $z=2.156$.  Within this region, the sensitivity 
falls by a factor of about two from the center to the border.  Beyond this radius, 
the flanking CCDs cover a larger area up to a radius of $\sim 14 $ arcmin, 
but with a much lower sensitivity and a significantly degraded angular resolution.  
The two regions drawn on the monochromatic 1.5 keV exposure map of the
field are shown in Figure \ref{FOV}.  The region at off-axis angle between 5 
and 14 arcmin is not studied in this work.
% will be studied in a future work.

\begin{figure*}
\begin{center}
\includegraphics[width=0.98\textwidth]{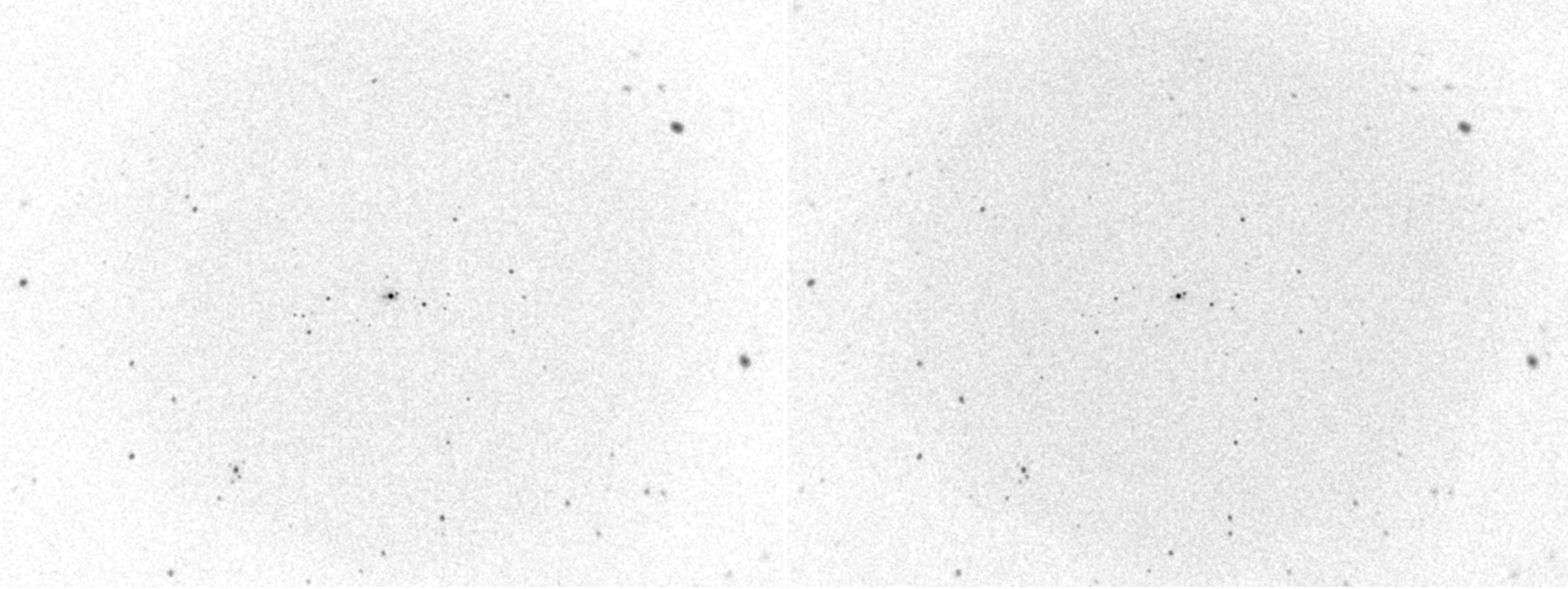}
\includegraphics[width=0.98\textwidth]{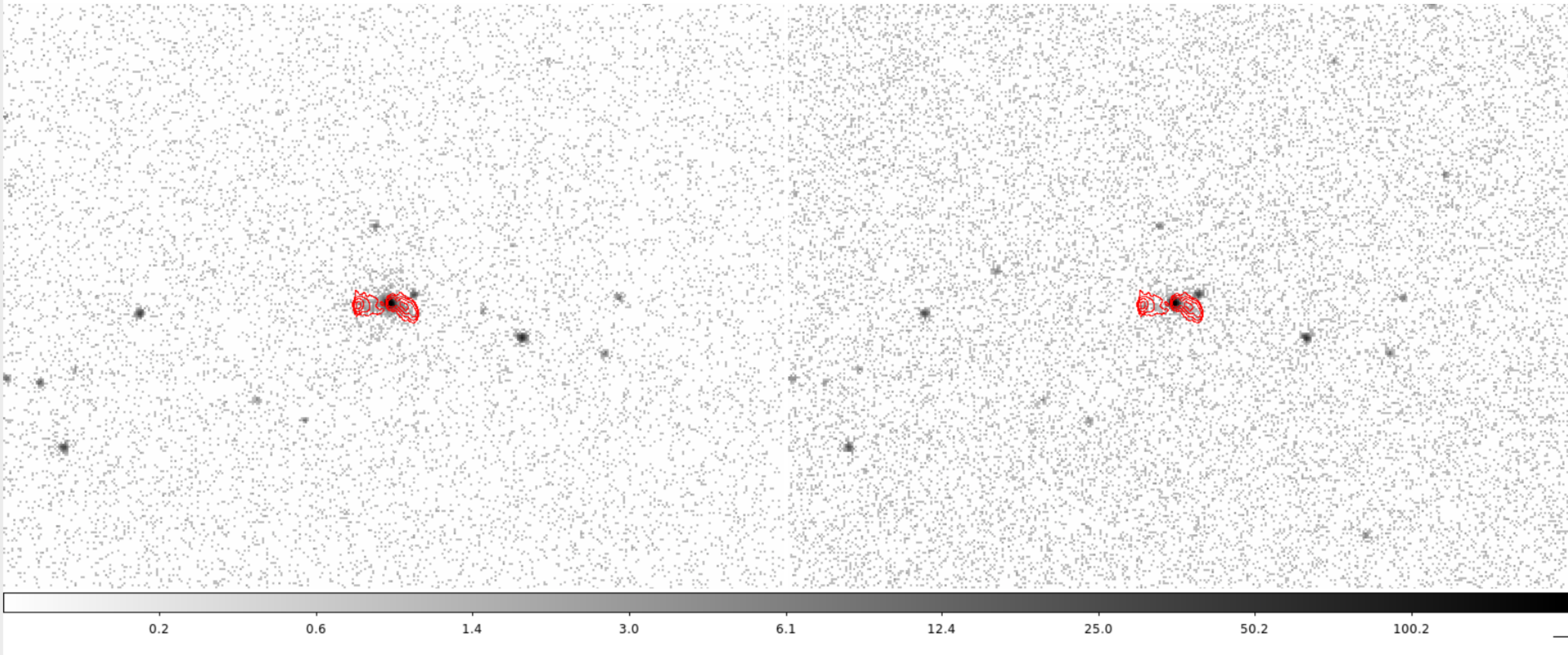}
\caption{ {\sl Chandra} X-ray images of the Spiderweb Galaxy field.
Upper panels: {\sl Chandra} ACIS-S images
in the soft 0.5-2 keV (left) and hard 2-7 keV band (right).  
A box of $12\times 9 $ arcmin by side centered on J1140-2629 is shown.
The images have been smoothed with a $\sim 1 $ arcsec 
Gaussian kernel to emphasize the unresolved sources.  Sources
at the outskirts of the image appear larger due to the effect of the 
increase of the PSF.  
% An excess in the density of X-ray sources
% is visible close to Spiderweb Galaxy, following an approximate filamentary structure.  
Lower panels: Close-up of the Spiderweb Galaxy ($3\times 2$ arcmin by side)
at full angular resolution, in the soft (0.5-2 keV, left) 
and hard (2-7 keV, right) band.  Extended emission is clearly visible 
in both bands.  Red contours show radio emission observed in the 10 GHz band with the JVLA
\citep[see][]{2022Carilli} at levels of $0.03$, $0.2$, $2$ and $20$ mJy/beam.}
\label{Ximage}
\end{center}
\end{figure*}

The upper panels of Figure \ref{Ximage} show an image of the central part of the field
(a box $12\times 9$ arcmin by side) in the 
soft and hard band.  The images have been smoothed with a $\sim 1 $ arcsec 
kernel to emphasize the unresolved sources.  The color scales are the same 
in both bands (max 200 counts in log scale) and therefore the larger background in the 2-7 keV band with respect
to the soft 0.5 -2 keV band is noticeable.  
% At a first glance, unresolved X-ray sources 
% appear to be more concentrated towards the Spiderweb Galaxy.  
In the lower panels, close-up images of the Spiderweb Galaxy (a box 
of $3\times 2$ arcmin by side, no smoothing applied) 
are shown in the soft and hard band.  The radio contours in the 
10 GHz band obtained with JVLA \citep{2022Carilli} are
shown in red.  Upon visual inspection, we see that the unresolved sources surrounding 
the Spiderweb Galaxy are distributed approximately along the jet directions, 
as already noticed for the spatial distribution of the 
narrow-band selected sources.  Finally, we note that the 
extended emission in the soft band appears not only along the jets but also in the 
direction perpendicular to the jet axis.
% , while in the hard band it essentially follows the jets.  
% The detailed analysis of the diffuse emission
% will be presented in two companion papers (Carilli et al. 2021, Tozzi et al. 2021b). 

To perform spectral analyses, we extract the spectra and 
compute the ancillary response file (ARF) and redistribution matrix file (RMF) 
at the source position
for each observation separately with the commands {\tt mkarf} and {\tt mkacisrmf}. 
Our default spectral analysis of unresolved sources 
uses a local background that, given the small 
extent of the sources, is directly extracted from a source-free region on the 
same chip.  This applies both to the unresolved sources and the diffuse emission. 
Below, a more careful treatment is applied to the analysis of the nucleus of the Spiderweb Galaxy, 
which is embedded in the diffuse emission.  
For most of the sources in the field, the soft and hard band fluxes are
estimated simply by converting the count rate obtained by aperture photometry 
to energy flux through the conversion factors.  We prefer aperture 
photometry to a more model-dependent photometry, relying on the modelization of the 
point spread function (PSF), 
because this last procedure can introduce systematic biasing effects that may be 
hard to control. On the other hand, the Poissonian noise associated to
aperture photometry can be handled with simple statistical methods.  
The fluxes obtained in this 
way account for the full uncertainty due to Poisson noise, allowing us to
obtain a reliable measurement for  the entire source population across three orders 
of magnitude.  The derivation of the conversion factors is detailed 
in the following subsections.

\section{X-ray sources in the Spiderweb Complex\label{xraysources}}

In this paper, we focus on the field of view within a radius
of 5 arcmin (corresponding to $\sim 2.5$ Mpc at $z=2.156$) centered on the nucleus of 
the Spiderweb, which results in a total of 78.5 arcmin$^2$.  
This region includes the solid angle covered by our observation with a 
sensitivity that is greater than about half of the highest sensitivity reached at the aimpoint, and 
with an angular resolution that is higher than 3 arcsec (FWHM).
% CHECK THE FWHM
This allows us to have a smooth sky coverage, 
thereby avoiding large corrections when evaluating the number counts.  
% We proceed with the blind search of X-ray sources as described in the next subsections.

\subsection{X-ray source detection}

X-ray source detection was performed with a hybrid procedure combining visual 
inspection and a standard detection algorithm ({\tt wavdetect}). The goal is to
maximize the completeness of the final source list.  
% Investigation of the number counts will be performed on a slightly shorter 
% list based on a well defined selection criterion, as described in detail in 
% Section \ref{numbercounts}.  
In the first step, we considered the entire FOV included in the $\sim 14$ 
arcmin radius, and visually identify 238 sources in at least one of the 
three images (soft, hard, and total band), with approximate position and size. 

We then ran {\tt wavdetect}\footnote{https://cxc.cfa.harvard.edu/ciao/threads/wavdetect/}, 
setting the scale parameter to 
{\tt scales = "2 4 8 16 32"} and the detection threshold set to 
{\tt sigthresh = 1e-05}.  This parameter should be approximately equal to the
number of pixels in the image, a choice that implies about one background fluctuation
detected as a source.  As we are searching for sources within 5 arcmin, 
we have about $1.17\times 10^6$ pixels, setting  {\tt sigthresh = 1e-05} means
approximately 12 strong background fluctuations detected as
sources\footnote{https://cxc.cfa.harvard.edu/ciao/ahelp/wavdetect.html\#plist.sigthresh}. 
Our choice is  therefore optimal for selection of unresolved sources down to low signal and  
low significance, with an acceptable level of contamination.
Considering that our final selection is based on a threshold 
in the $S/N$ of aperture photometry, our final list of
sources show a  significantly higher 
S/N than the lower limit of the {\tt wavdetect} selection, 
implying a much lower number of spurious sources with respect to the initial
selection. We expect to have a completeness of very close to unity, 
particularly in the inner 5 arcmin, where the angular resolution is a few arcsecs.
However, we note that running {\tt wavdetect} with pixel scales below and equal to $32$ 
is less efficient for detecting extended sources, or sources at
off-axis angle $\sim 7$ arcmin and larger, where the angular resolution of 
{\sl Chandra} rapidly deteriorates. 
As we focus on the sky region at off-axis angle $<5$ arcmin, this choice does not affect
our results.  In addition, visual inspection confirms that there are 
no relevant extended sources in the field within 5 arcmin.  
The extended emission surrounding the Spiderweb Galaxy 
will instead be discussed in a separate publication.

We find 249, 289, and 381 unresolved source candidates 
in the soft, hard, and total-band images, respectively, selected by {\tt wavdetect}. 
We match the {\tt wavdetect}-detected sources with the visually selected sample 
assuming a matching radius of 1 arcsec, and collect the {\tt wavdetect} 
positions for all the matched sources, with the exception of the
source candidates identified only by visual inspection.  We then 
obtain a master catalog with 398 sources. All but 11 of the visually identified sources 
are recovered by the {\tt wavdetect} algorithm, and these latter
are at more than 5 arcmin from the 
Spiderweb, and therefore have no impact on this work. We define a circular extraction 
region for each source to perform aperture photometry. 
The extraction radius $r_{\rm ext}$ is computed with a simple relation depending
only on the off-axis angle $\theta$.  We set $r_{\rm ext}=2.55\times 
(0.67791-0.0405083*\theta+0.0535066*\theta^2)$ arcsec 
\citep[see][and references therein]{2002RosatiCDFS}.  We find that 
this choice ensures that at least 95\% of the expected flux is included 
in the circular extraction region, both in the soft and hard band\footnote{For 
details see {\tt https://space.mit.edu/cxc/marx/tests/PSF.html}}.  We 
manually adjusted the position and size of the extraction regions
to avoid overlapping of nearby source candidates: in three cases we 
manually reduced $r_{\rm ext}$ by one pixel ($0.5$ arcsec) and shifted the centroid
by 1 arcsec; in three other cases we simply reduced
the $r_{\rm ext}$ by three pixels ($1.5$ arcsec).  We verified {\sl a posteriori} that, 
on average, the
photometry of the six sources (flagged in Table \ref{source_list})
changes by less than 10\%, with a negligible impact on 
our final results. The background was estimated 
locally in an annulus around each source 
(outer radius $r_{\rm ext}+3^{\prime\prime}$, inner 
radius $r_{\rm ext}+1^{\prime\prime}$) after removing nearby unresolved sources.  
This choice, thanks to the depth of the total exposure, 
ensures a fair sampling of the background, and takes into account local variation
of the background. 
% We also verify {\sl a posteriori} that photometry values
% are consistent when we use a smoothed background map.
% WE HAVEN'T TRIED A MODELED BACKGROUND

We estimated the Poisson probability of being a background fluctuation $P{i}$ for 
each candidate source $i$ following the procedure described in \citet{2020Puccetti}.
First, we computed a radius $r_{\rm max}$ (with $r_{\rm max} < r_{\rm ext}$) 
within which the $S/N$ is maximum.  We then computed  

\begin{equation}
P_{i} = e^{-B_i}\Sigma_{j=C_i,\infty} (B_i)^j/j!
,\end{equation}

\noindent
where $B_i$ is the estimated background counts within $r_{\rm max}$, and $C_i$ is the
total counts within $r_{\rm max}$. \citet{2009Puccetti} showed that assuming
$P_i<2.0\times 10^{-5}$ ensures a very high reliability, and so we can use this threshold to 
select a sample of sources with a negligible spurious component.  
On the other hand, computing the completeness as a function of $P_i$ requires 
extensive simulations to take into account the sensitivity and the PSF effects across the
field of view. 
 Therefore, for the completeness we prefer to use a criterion 
directly relying on the S/N from aperture photometry.  This approach has been shown to be
reliable in deep {\sl Chandra} surveys such as the Chandra Deep Field South (CDFS), where
we adopted a threshold of $S/N>2.0$ to compute the point-source sky coverage \citep[see][]{2001Giacconi,
2001Tozzib}. In this work we combine these two criteria to best exploit
the sensitivity of the {\sl Chandra} data whilst maintaining the very high purity
of the sample.

We split our candidate list according to the distance from the Spiderweb Galaxy, 
finding 264 candidates at more than 5 arcmin and 
134 detection candidates within 5 arcmin. For the remainder of the paper, we consider only the
candidate detections in the central FOV.
We computed $P_i$ in the soft, hard, and total (0.5-7 keV) bands for each source,  
and consider any source that satisfies $P_i<2.0\times 10^{-5}$ in any of the
three bands to be a detection.  This step selects 107 source candidates. 
We then applied our completeness criterion $S/N>2.0$, and find that 9 out of 107 sources do not reach 
the S/N threshold in any band. However, we find that
5 of them have $S/N>1.8$ in the hard band, and after a careful revision by 
visual inspection, we decided to keep them in the final source list, while discarding the
remaining 4 sources. 

In addition, we search for robust candidates that may have been missed by our 
reliability criterion by searching for sources with $P_i>2\times 10^{-5}$ in all three bands, 
but selecting only sources with $S/N>2$ in the soft or hard bands. 
We find only 
four such sources and all have a $P_i$ value very close to our threshold 
(of the order of $3-4 \times 10^{-5}$). After visual inspection, 
we decided to include them in the final sample, which, after this final step, includes
107 sources.  

Incidentally, the same selection procedure identifies a final list of 176 sources
in the outer ring, which has a field of view 6.8 times larger than
the central 5 arcmin radius circle.  This shows that the detected source density
in the outer ring is on average four times lower.  Considering that 
sources at more than 5 arcmin from the aimpoint have significantly larger 
errors on position and photometry due to the rapid degradation of the PSF, 
and that the multiwalength data coverage essentially drops at distances $>5$ arcmin, 
we strengthen the conclusion that the data quality in the field beyond 5 arcmin does not
allow us to further extend the study of the Spiderweb Complex with the current
dataset.  The analysis of the field at distances larger than 2.5 Mpc from the Spiderweb
is potentially interesting, in particular to investigate the possible extension of 
the Spiderweb Complex at radii larger than 2.5 physical Mpc, but requires additional data and must be postponed to a future work.

\subsection{Measurement of source fluxes}

The exposure-corrected count rates are estimated by rescaling the measured 
net count rate by the ratio of the exposure map values at the aimpoint to the 
emission-weighted exposure within the extraction region. The average 
monochromatic exposure map values at the aimpoint are $470.8$ cm$^2$ and 
$367.4$ cm$^2$ in the soft and hard band, respectively.  Soft- and hard-band 
fluxes are computed from the net count rates corrected for vignetting and 
assuming the typical conversion factors $C_{\rm soft}$ and $C_{\rm hard}$ in the corresponding band. 
The conversion factors clearly depend on the spectral shape, which itself is
well known to depend on average on the source flux \citep[see][]{2001Tozzib}, 
and can be approximately described with a power law with a slope 
in the range $1.2<\Gamma < 1.9$. Slopes in the range $\Gamma = 1.7-1.9$ 
are typical of unabsorbed AGN, as measured in deep surveys such as the CDFS \citep[see][and references therein]{2017Liu}.
Harder spectra are mostly due to intrinsic absorption, which depends on the 
intrinsic column density and source redshift. Clearly, this information 
is not available for the full catalog. We proceed by assuming an
average redshift of 1.5 \citep[representative of AGN in 
the CDFS; see Figure 23 in][]{2017Luo}, and a typical intrinsic absorption 
$N_H\sim 10^{22}$ cm$^{-2}$.   We find that the
conversion factor obtained with these assumptions, and allowing for at least 
a factor of three variation in the average redshift and absorption, is well inside the range that we obtain assuming $1.2<\Gamma < 1.9$.  Therefore, 
without providing a detailed description of the spectral shape of our
unresolved sources, we adopt the full range $1.2<\Gamma < 1.9$ and consider the 
corresponding variation 
in the conversion factors as a systematic uncertainty. This is a fair choice for
the estimation of the source flux and a reliable measurement of the 
number counts.  The X-ray sources identified as members of the Spiderweb 
Complex are further investigated with a detailed spectral analysis (see Section 6.4). 

After correcting for a Galactic absorption column density of 
$N_{\rm H} = 3.18\times 10^{20}$ cm$^{-2}$, 
we obtain $C_{\rm soft} = (8.77\pm 0.65) \times 10^{-12}$ cgs/
counts/s in the soft band. In the hard band, 
we derive two conversion factors of $C_{\rm hard} = (1.87\pm 0.11) \times 10^{-11}$ cgs/counts/s
and $C_{\rm hard} = (2.66\pm 0.35) \times 10^{-11}$ cgs/count/s to convert from the
2-7 keV count rate to the 2-7 keV and 2-10 keV fluxes, respectively. 
We note that we apply a factor 1.05 to account for the average flux loss
outside the extraction region,
and also note that while the net detected counts in the hard band refer
to the 2–7 keV energy range, we compute the energy fluxes
in the 2–10 keV band, for a better comparison with the literature,
despite our unresolved sources showing very little signal above 7 keV. 
The list of 107 sources with X-ray position, net count rate in both bands, and soft and
hard fluxes is shown in Table \ref{source_list}. A negative count rate may be measured when a
source is detected only in one band, in which case we report the 2$\sigma$ upper limit
for the energy flux.

\begin{table}
\caption{Conversion factors}
\label{cf}
\begin{center}
\begin{tabular}[width=0.5\textwidth]{lcc}
\hline
Count rate & Energy  & Conversion factor  \\
band (keV) &  band (keV) &   cgs/counts/s \\
\hline
0.5 - 2.0  & 0.5 - 2.0   &      $(8.77\pm 0.65) \times 10^{-12}$  \\
2.0 - 7.0        &      2.0 - 7.0        &       $(1.87\pm 0.11) \times 10^{-11}$  \\
2.0 - 7.0        & 2.0 - 10.0    &        $(2.66\pm 0.35) \times 10^{-11}$ \\
\hline
\end{tabular}
\end{center}
\tablefoot{The conversion factors are computed at the aimpoint and
corrected for Galactic absorption.  The central value and the associated error correspond to 
a power-law spectrum with a slope in the range $1.2<\Gamma<1.9$, as discussed in the text.}
\end{table}

\subsection{X-ray number counts\label{numbercounts}}

\begin{figure}
\begin{center}
\includegraphics[width=0.49\textwidth]{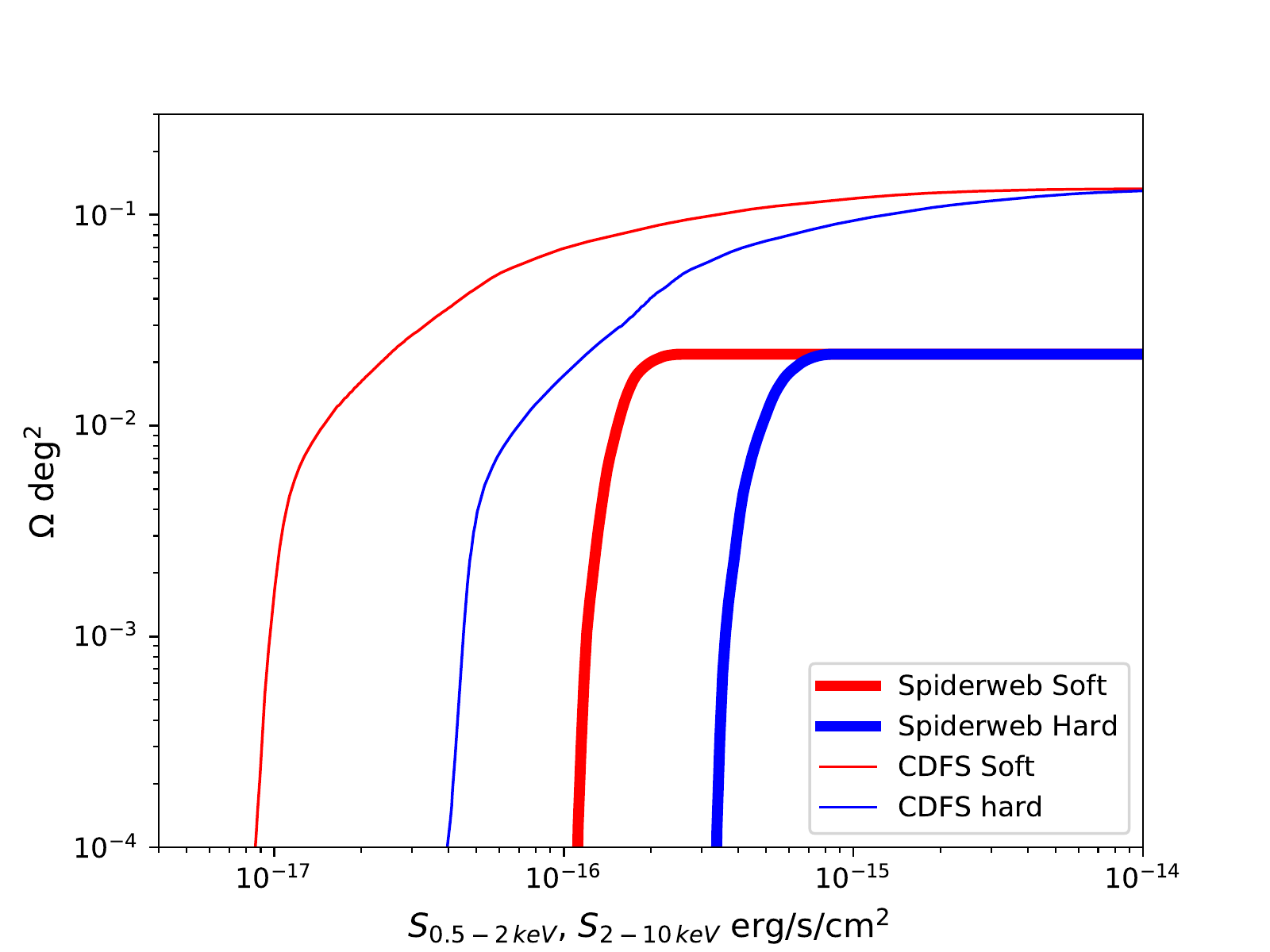}
\caption{Thick lines show the point-source sky coverage of the innermost region of 5 arcmin 
of the Spiderweb field investigated in this work, in the soft (red) and 
hard (blue) band.  Thin lines show the point-source sky coverage in the CDFS from \citet{2017Luo}.
}
\label{skycov}
\end{center}
\end{figure}

To derive the number counts, we first compute the point-source sky coverage as a function of the energy flux.
To do that, we first create images of the soft and hard background. These are obtained by removing 
all the detected sources and replacing the signal in the extraction region with a 
Poisson realization of the emission measured in the background annulus.  Using the 
same conversion factors, and the monochromatic exposure maps, we then compute the solid angle 
in which a source of a given soft or hard flux is detected according
to our completeness criterion $S/N>2.0$.
As shown in Figure \ref{skycov}, the sky coverage rapidly drops to zero at fluxes of 
$1.3 \times 10^{-16}$ erg/s/cm$^2$ and $3.9\times 10^{-16}$ erg/s/cm$^2$ 
in the soft and hard band, respectively\footnote{Formally, the flux limit is defined as
the flux at which the sky coverage drops to 1/10 of the full field of view of 78.5 arcmin$^2$.}. 
The deepest X-ray 
field to date (and until a new X-ray mission with arcsec resolution), is provided by the CDFS
\citep[][]{2001Giacconi,2002RosatiCDFS,2017Luo} which has
reached the unparalleled depth of 7 Ms.  Comparison with the sky coverage 
of the CDFS shows that the field of view analyzed in this work is five times smaller 
than the maximum solid angle covered in the CDFS.  The flux limit is about 13 and 10 
times higher than in the CDFS in the soft and hard band, respectively. We note that the
flux limits of the two fields roughly scale as the ratio of the 
total exposure times (a factor of 10), despite the sensitivity
loss due to the contamination build-up over ACIS. 
This is because we take advantage of 
the larger effective area of the back-illuminated CCD in ACIS-S with respect to the
front-illuminated CCDs in ACIS-I (used for the CDFS), partially compensating the sensitivity loss. 

The cumulative number counts (logNlogS) are computed directly as

\begin{equation}
N(>S) = \Sigma_{S_i>S}  {1\over {\Omega_i}},
\end{equation}
where $S_i$ and $\Omega_i$ are the flux and the corresponding sky coverage 
of the $i^{\rm th}$ source, respectively.
% No correction for Eddington bias has been performed. However, the impact of the 
% Eddington bias is estimated to be negligible. 
To compute the soft and hard number counts, we consider only the 
sources with $S/N>2$ in the corresponding
band.  We select 72 sources in the soft and 96 sources in the hard band, which, combined together, add up to a total of 107 unique sources within the 5 arcmin radius.

\begin{figure}
\begin{center}
\includegraphics[width=0.49\textwidth]{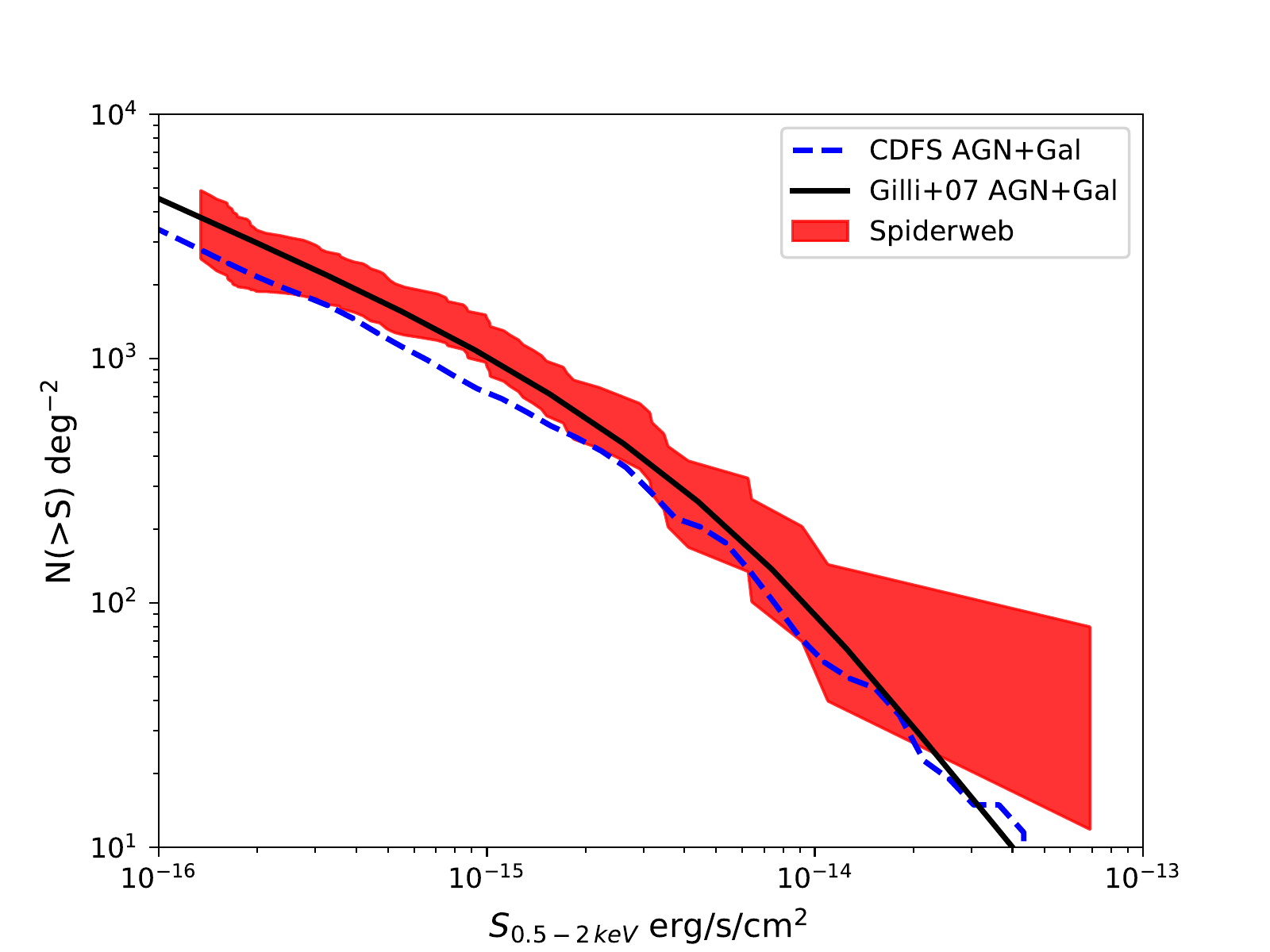}
\caption{Cumulative number counts in the soft band in the Spiderweb field
within 5 arcmin from the Spiderweb Galaxy (red shaded area).  The shaded area
shows the uncertainty due to the combination of source statistics, photometric errors,
and systematic uncertainty on the conversion factors.  The blue dashed line shows the soft counts 
in the CDFS from \citet{2017Luo}. The black solid line shows the prediction from the AGN 
model of \citet{2007Gilli} plus the contribution from normal galaxies from \citet{2017Luo}. 
}
\label{logNlogS_soft}
\end{center}
\end{figure}

\begin{figure}
\begin{center}
\includegraphics[width=0.49\textwidth]{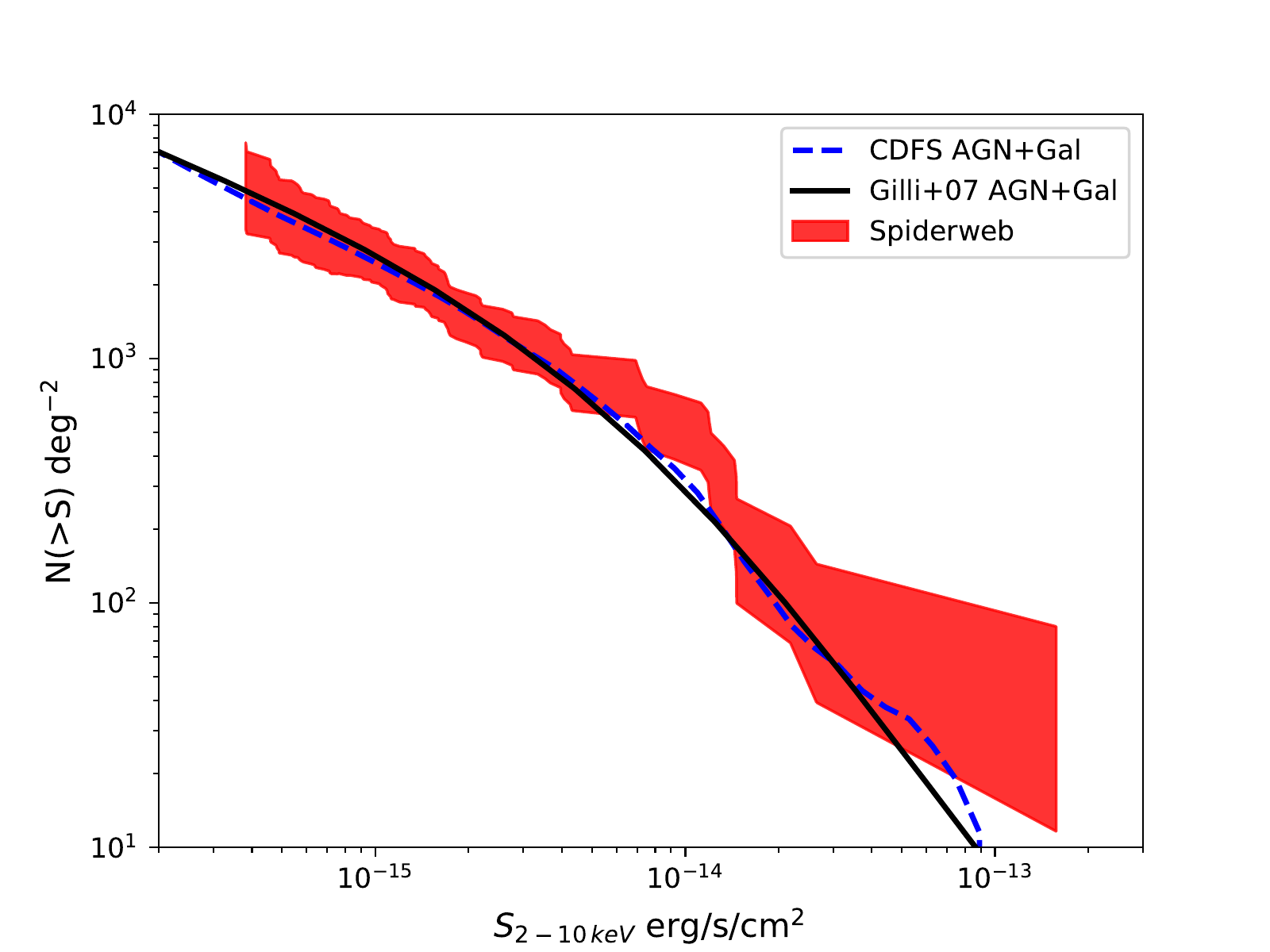}
\caption{Cumulative number counts in the hard band in the Spiderweb field
within 5 arcmin from the Spiderweb Galaxy (red shaded area).   The shaded area
shows the uncertainty due to the combination of source statistics, photometric errors,
and systematic uncertainty on the conversion factors.  The blue dashed line shows the hard counts 
in the CDFS from \citet{2017Luo}. The black solid line shows the prediction from the AGN model of
\citet{2007Gilli} plus the contribution from normal galaxies from \citet{2017Luo}. 
}
\label{logNlogS_hard}
\end{center}
\end{figure}

The soft band logNlogS is shown in Figure \ref{logNlogS_soft}.  The shaded area
shows the formal $1 \sigma$ uncertainty obtained as the sum in quadrature of the 
statistical error due to the source statistics and the error on the source fluxes.
The last error is in turn the sum in quadrature of the Poissonian photometric error and the
systematic error associated to the conversion factors.  The uncertainty is dominated
by the source statistics at large fluxes, and by the photometric errors at low
fluxes.
The same figure shows the soft band logNlogS measured in the CDFS by 
\citet{2017Luo}, and the predicted logNlogS
based on the X-ray background synthesis model developed by 
\citet{2007Gilli}\footnote{http://cxb.oas.inaf.it/,  see also \citet[][]{2005Hasinger}.}, which includes an exponential decline 
in the AGN space density at redshifts above $z = 2.7$ to cope with the 
results from wide-area surveys \citep[e.g., SDSS][]{2006Richards,2006Fan}.
As the \citet{2007Gilli} model includes only AGN, we add the steep galaxy
number counts as measured by \citet{2017Luo} in the CDFS \citep[see also][]{2020Marchesi}.  
Normal (star forming) galaxies are relevant only at fluxes well below $10^{-16}$ erg/s/cm$^2$, 
and therefore have a negligible impact on our results. 
In particular, the expected contribution of normal galaxies is equal to that of 
AGN at $10^{-17}$ erg s$^{-1}$ cm$^{-2}$ both in the 0.5-2 keV and 2-7 keV bands 
\citep[see Figure 31 in][]{2017Luo}.
The hard-band logNlogS is shown in Figure \ref{logNlogS_hard}.  In both cases, we find very good
agreement with the model expectation, while both model and counts in the Spiderweb Field appear 
to be $\sim 20$\% higher than the CDFS in the soft band at fluxes $\sim 10^{-15}$ erg/s/cm$^2$.
The measured counts in the hard band show a $\sim 30$\% excess in a small range around
fluxes $\sim 10^{-14}$ erg/s/cm$^2$ with respect to the model and CDFS. 
Overall, the logNlogS in the Spiderweb Field appear to be broadly in line with 
expectations, with the small hint of an excess on a limited flux range.  Whether or not this excess is 
related to the presence of the Spiderweb Complex is investigated 
in Section \ref{identification}.

\section{Identification of X-ray-emitting protocluster members\label{identification}}

On the basis of the  shallow {\sl Chandra} observation
\citep[Obsid 898][]{2002Pentericci}, only five X-ray sources
have been identified among the spectroscopically confirmed
or narrow-band-selected protocluster members.
%$Ly\alpha$ ID 968 and 778 plus one below threshold and only one with H$\alpha$ (ID 215).
These sources are firmly identified in the X-ray band as bright 
AGN in the 2--10 keV  luminosity range 0.8-4$\times 10^{44}$ erg s$^{-1}$. No emission has been 
detected from any of the remaining galaxies in the protocluster down to a rest-frame 
2--10 keV luminosity limit of $4 \times 10^{43}$ erg s$^{-1}$. 
% Despite the low statistics, a mild excess of sources with respect to the number counts 
% in deep X-ray surveys 
% (at least a factor of $\sim 1.6$ and possibly higher) 
% has been claimed in the soft band.  However only two of them have been confirmed 
% spectroscopically, with the other three being only consistent with $z\sim 2.156$ based 
% on their H$\alpha$ excess.  
Four out of these five sources are aligned along the radio and X-ray 
diffuse emission \citep{2005Croft}. However, this is not surprising because confirmed and 
candidate protocluster members, irrespective of their selection, appear to be aligned in the same direction 
as the radio jets and the X-ray emission.  The relatively high AGN fraction estimated around 
the Spiderweb Galaxy, despite the low statistics, suggests that the 
AGN were probably triggered by the ongoing formation of the protocluster. 
Another possible mechanism may be associated to the radio-mode feedback of the 
central radio galaxy, in a scenario similar to the observation of the
protocluster in the SDSSJ1030 field \citep{2019Gilli}.  However, the extension of the jets 
in the case of the Spiderweb Galaxy is limited to less than 100 kpc from the nucleus, 
making it a different case with respect to the SDSSJ1030 protocluster, where
the jet extension is significantly larger (about 700 kpc between the two lobes).  
% and to the simulations of very high-z, bright QSO \citep[see][]{2014Costa}. 
% in realta’ il Costa+14, cosi’ come la maggior parte delle simulazioni,
% e’ piu’ sul negative feedback.. e il Costa in particolare piu’ da SNe che da AGN.
In any case, to demonstrate a trigger effect on 
the protocluster members, it is necessary to compute the fraction of 
X-ray-emitting (at the Seyfert level or higher) members over the entire protocluster
population.  In this section, we proceed to identify all the secure X-ray-emitting protocluster
members, with the final goal to estimate the AGN fraction in the Spiderweb system.

% In addition, the alignment of the confirmed X-ray emitting members with the overdensity of the 
% Ly$\alpha$ emitters, and the radio axis of the central galaxy \citep{2005Croft}, suggests that 
% we may witness triggering of AGN and SF activity on Mpc scale 

\subsection{Matching X-ray sources to spectroscopic members}

We match the X-ray sources with the optical source list, both with spectroscopic 
and photometric redshift, or narrow-band-selected sources.  
Based on the accuracy of the X-ray centroid and of the error on the 
position of the CO-selected sources, and neglecting the errors on the 
optical and NIR positions, we assume an initial 
matching radius of 1 arcsec for optical and NIR sources, and 2 arcsec for CO sources. 
Based on these values, we can immediately compute the expected number of false matches.
Considering the entire FOV (a circle of 5 arcmin radius), and a cross-section 
of $\pi$ arcsec$^2$, we have a total fractional cross-section of 
$2.8\times 10^{-3}$ for the 252 unique sources (including
spectroscopic redshifts and narrow-band-selected members).  
If the 107 X-ray sources are randomly distributed in this field of view, the 
expected number of random matches to protocluster members (confirmed or candidate)
is 0.30.  Therefore we expect significantly less than one spurious counterpart.  

We obtain a list of 14 X-ray-emitting protocluster members (including the Spiderweb), 
13 of which matched with members with spectroscopic redshift (including the Spiderweb), 
and 1 (XID 73) with an ERO from \citet{2004aKurk}. Incidentally, 
we note than none of the {"flies"} are X-ray emitting.  This is 
not surprising, because given the measured mass range $10^8<M_*<10^{10} M_\odot$ and 
SFR range $0.5-26\,  M_\odot$ yr$^{-1}$ \citep{2009Hatch}, the corresponding 
cumulative X-ray emission from star formation only is estimated to be more than 
an order of magnitude lower than our current flux limit.
The selected sources were visually checked on the X-ray, HST, and Subaru 
images to identify potentially spurious counterparts.  We find that
all the counterparts are unique within 1 arcsec except XID 90, which
has a secondary counterpart that 
has also been selected as an ERO candidate at $z\sim 2.16$ \citet{2004aKurk}.  Therefore, 
XID 90 may be considered, at worst, to be another X-ray member candidate without spectroscopic confirmation, 
similarly to XID 73. 

\begin{figure}
\begin{center}
\includegraphics[width=0.49\textwidth]{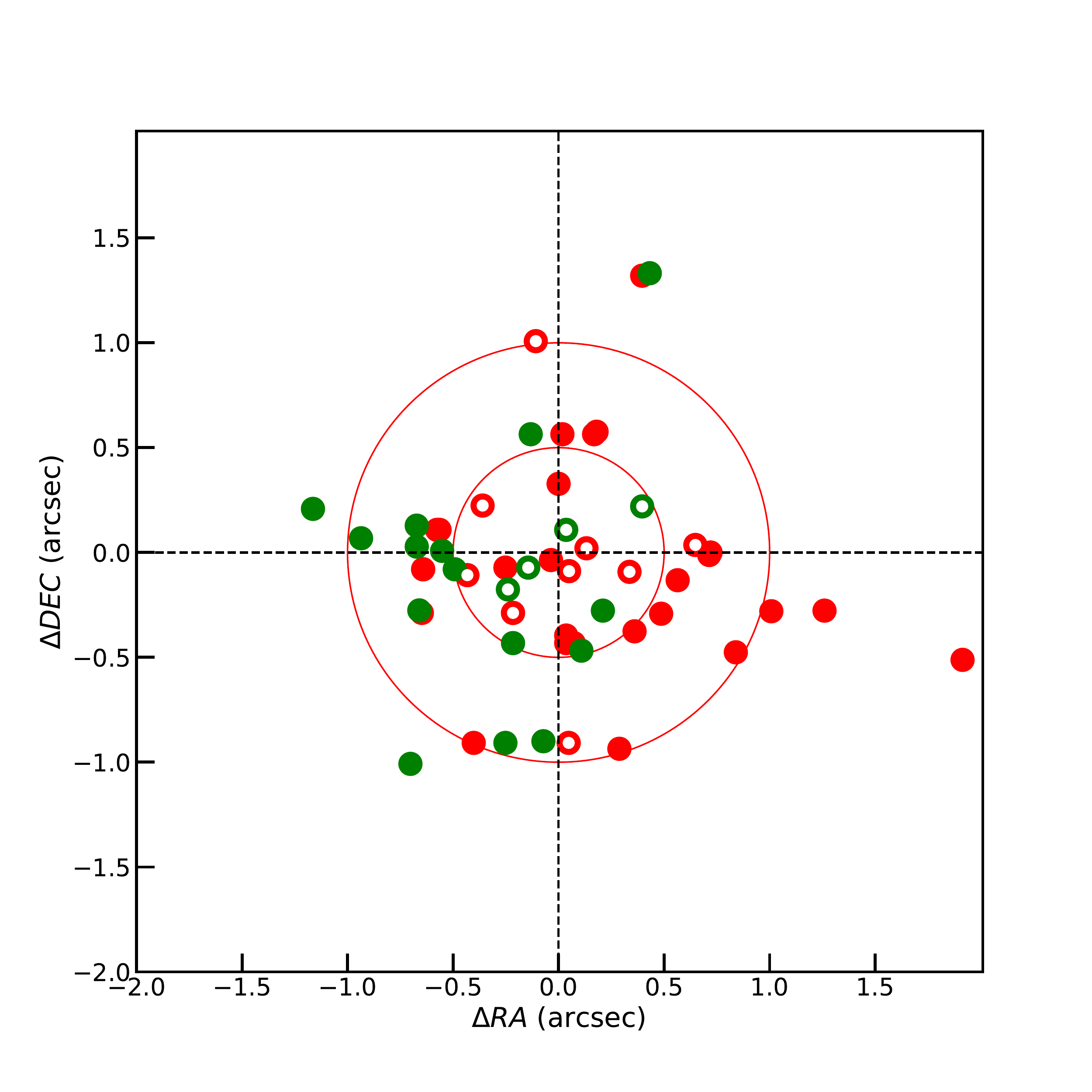}
\caption{Scatter plot of the offsets for the 13 X-ray sources 
identified as members of the Spiderweb Complex, excluding the Spiderweb Galaxy.  
Red points corresponds to spectroscopic counterparts, while green points to 
color-selected counterparts.   
The offsets of the closest counterpart for each source are shown
with a white dot inside the red or green points, while other points corresponds 
to other counterparts listed in independent catalog from Table \ref{catalogs}. 
The large red circles correspond to 0.5 and 1 arcsec offsets.}
\label{scatter}
\end{center}
\end{figure}

The scatter plot of the offsets between all the X-ray counterparts and all 
the optical matches from the catalogs we used in this work is shown 
in Figure \ref{scatter}.  The closest counterparts for each single source are marked
with a white dot inside the red (spectroscopic) or green (photometric) point, showing that all the 
counterparts are well within 1 arcsec from the 
X-ray centroid, with most of them within 0.5 arcsec.  We inspected the 
two sources (XID 12 and XID 80) that have an offset of $\sim 1$ arcsec, 
and find that their X-ray centroid is not coincident
with the X-ray peak, but is slightly displaced because of an apparent extension of the
X-ray emission.  The signal is too weak to classify these X-ray sources as extended
\footnote{The extension may be associated to radio jets, an occurrence that 
will be tested when the wide-field, high-resolution radio images are available.}, 
but when considering the X-ray peak instead of the X-ray centroid, the offsets drop
below $0.5$ arcsec.  Therefore, we find that, overall, we can safely assume that the offsets
between the X-ray sources and the spectroscopic counterparts can be considered equal 
to or less than 0.5 arcsec, which brings the expected number of spurious matches below 0.1.  
We carefully inspected all the counterparts in the HST and Subaru 
images, finding no alternative optical counterpart candidates except those
already mentioned. In the rest of the paper, we consider a sample of 12 X-ray 
spectroscopically confirmed members (excluding the 
Spiderweb) and an additional X-ray color-selected member candidate. 
Considering the 112 and 96 spectroscopically confirmed sources in the range $2.0<z<2.3$ 
and $2.11<z<2.20$, this corresponds to a fraction of $11.0\pm 0.5$\% and $13.0\pm 0.5$\%, 
respectively.  However, this fraction does not take into account the 
dependence on the stellar mass, which is a key   quantity that must be considered
when computing the AGN fraction.  This aspect is discussed in Section \ref{fraction}.

\begin{figure*}
\begin{center}
\includegraphics[width=0.98\textwidth]{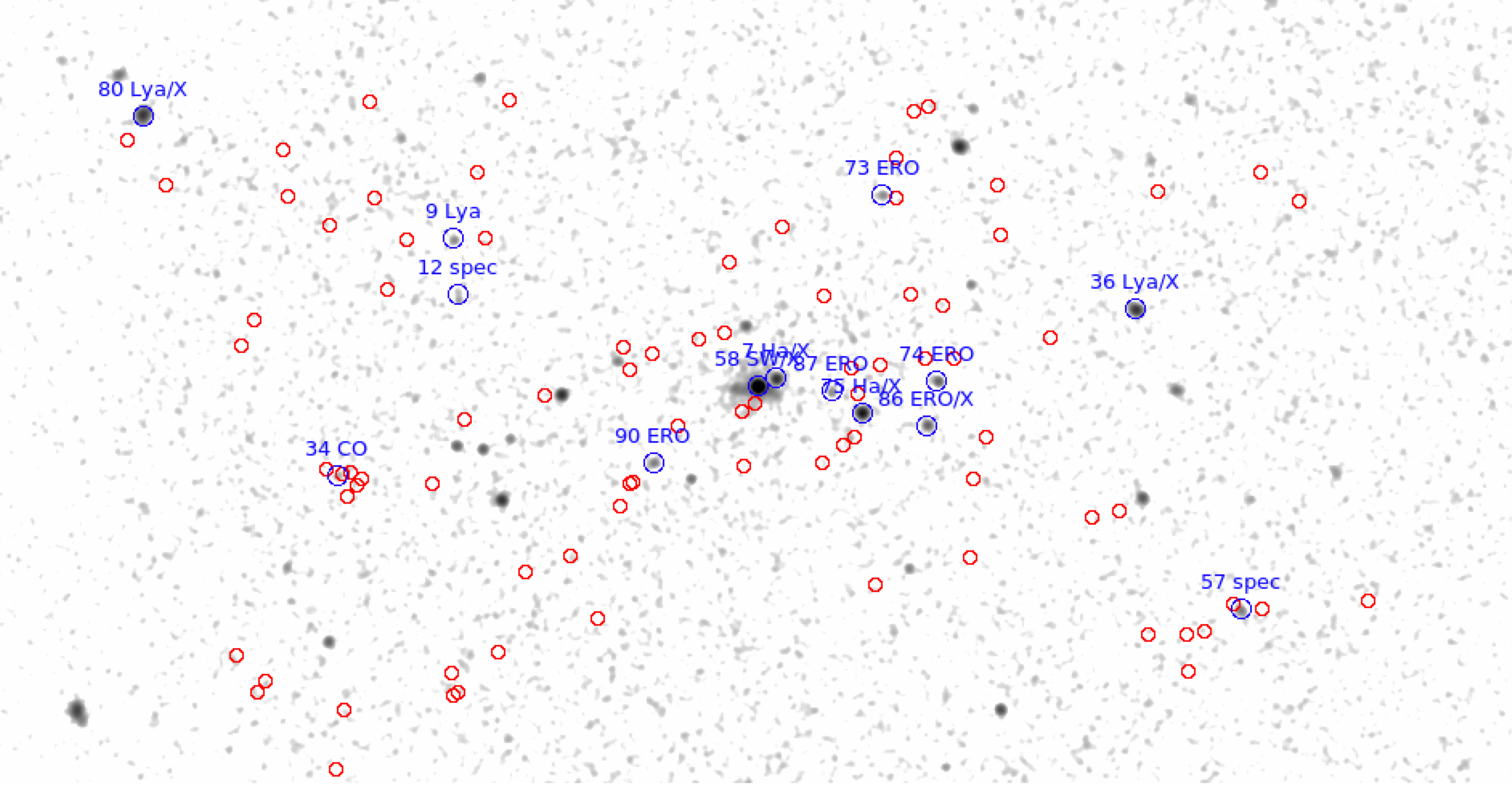}
\caption{Hard (0.5-7.0 keV) band image of the center of the field (a box of $460\times 240$ arcsec)  
with the X-ray-emitting protocluster members marked with blue circles, 
and all the other spectroscopically confirmed protocluster members ($2.0<z<2.3$)
marked with red circles.  The image has been smoothed with a 1 arcsec Gaussian filter for clarity. 
X-ray members are labeled with the source ID, the original selection: Ha, Lya, ERO, and CO for $H_\alpha$, 
$Ly_\alpha$, red color, and CO-line selection, respectively, and "spec" for spectroscopic selection. 
Finally, the sources labeled with "X" were previously identified as X-ray sources in the literature, as described in Section 2.2.
% We also included one source without spectroscopic confirmation
% among the X-ray emitting members.
}
\label{zspec_Xray}
\end{center}
\end{figure*}

In Figure \ref{zspec_Xray} we show the X-ray image of the field in the 0.5-7 keV band
with the X-ray-emitting protocluster members marked with blue circles, 
and all the spectroscopically confirmed protocluster members marked with red circles.
For better visualization, here we show only a box of $460\times 240$ arcsec 
entirely comprised within 5 arcmin from the Spiderweb Galaxy (the FOV we consider), 
which includes all but two (which are nonX-ray-emitting) of the spectroscopic protocluster members.
The filamentary distribution appears to be more evident in the X-ray sources than 
in the spectroscopically confirmed members, but given the low statistics, 
the difference is not significant.

\begin{figure}
\begin{center}
\includegraphics[width=0.49\textwidth]{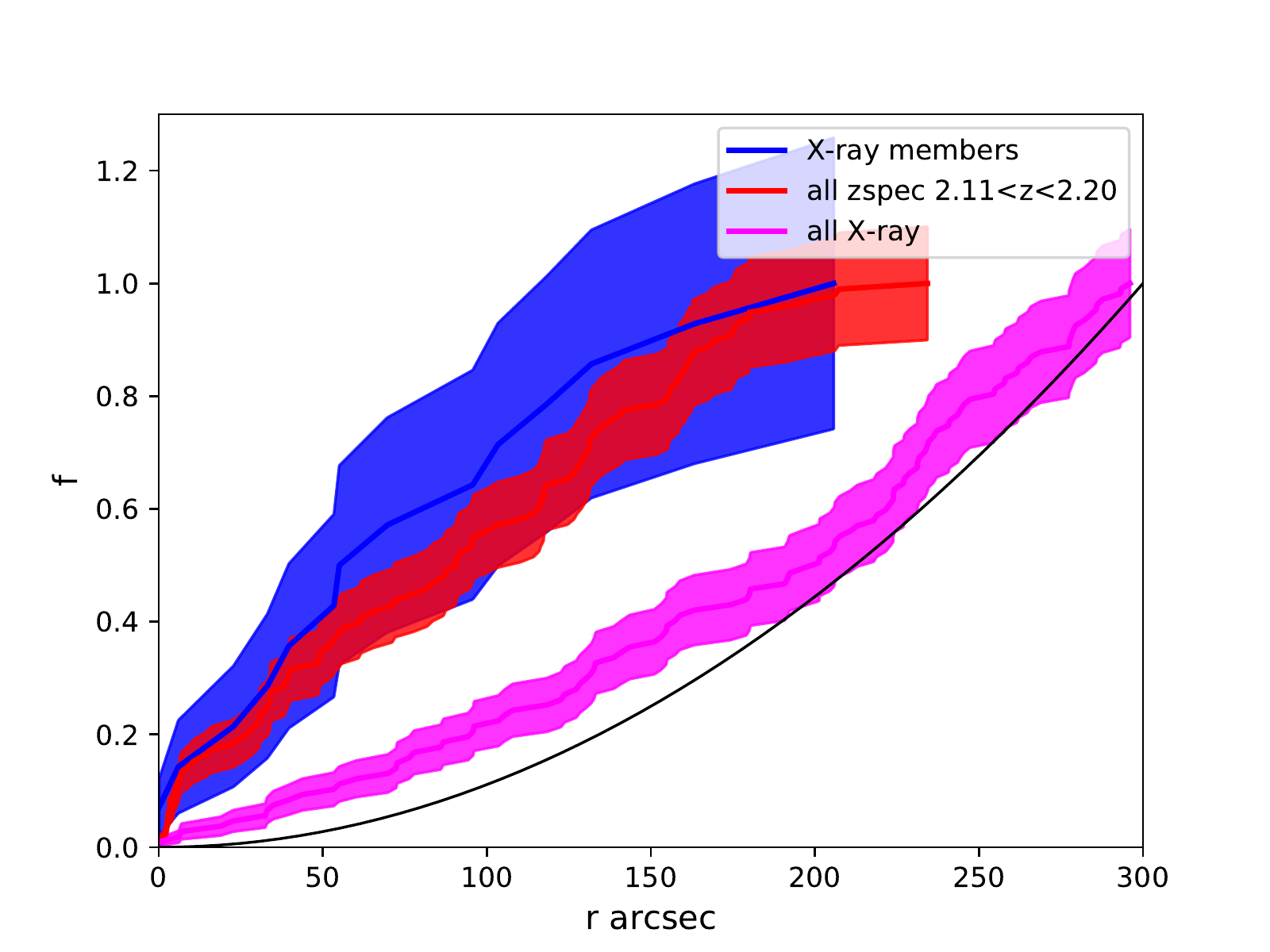}
\caption{Fraction of sources within a given radius from the Spiderweb Galaxy.  
X-ray cluster members (shown in blue) have the same distribution as all the 
spectroscopically confirmed protocluster members in the $2.11<z<2.20$ range
(shown in red).  On the other hand, 
the distribution of all the X-ray sources (dominated by the field) is clearly less
clustered.  The thin black line shows a perfectly uniform source distribution.
The shaded area corresponds to an uncertainty of $1\sigma$.
}
\label{zspec_dist}
\end{center}
\end{figure}

In Figure \ref{zspec_dist} we plot the azimuthally averaged fraction of sources of a 
given class as a function of the distance from the Spiderweb Galaxy,
finding that the protocluster members 
are clearly more clustered than the bulk of the X-ray sources.
This is not surprising
because the clustered distribution of the spectroscopic members was already noted 
in all the works listed in Table \ref{catalogs}.
On the other hand, when compared to the spectroscopically confirmed protocluster members, 
the X-ray-emitting members show only a slightly higher concentration, which is not
statistically significant.  
This suggests that, based on current data, there is no evidence of enhanced X-ray activity
closer to the Spiderweb Galaxy.   We checked 
whether or not the 15 sources considered to be part of the Spiderweb Complex but
not in the protocluster (therefore belonging to the ranges $2.00<z<2.11$ and $2.20<z<2.30$)
show a different behavior, but we find their distribution similar to that of 
the protocluster members. We note that our conclusion on the   
X-ray activity being independent on the distance from the protocluster center relies
on the assumption of spectroscopic completeness, an aspect that nevertheless
cannot be properly quantified given the composite nature of our spectroscopic sample.

\begin{figure}
\begin{center}
\includegraphics[width=0.49\textwidth]{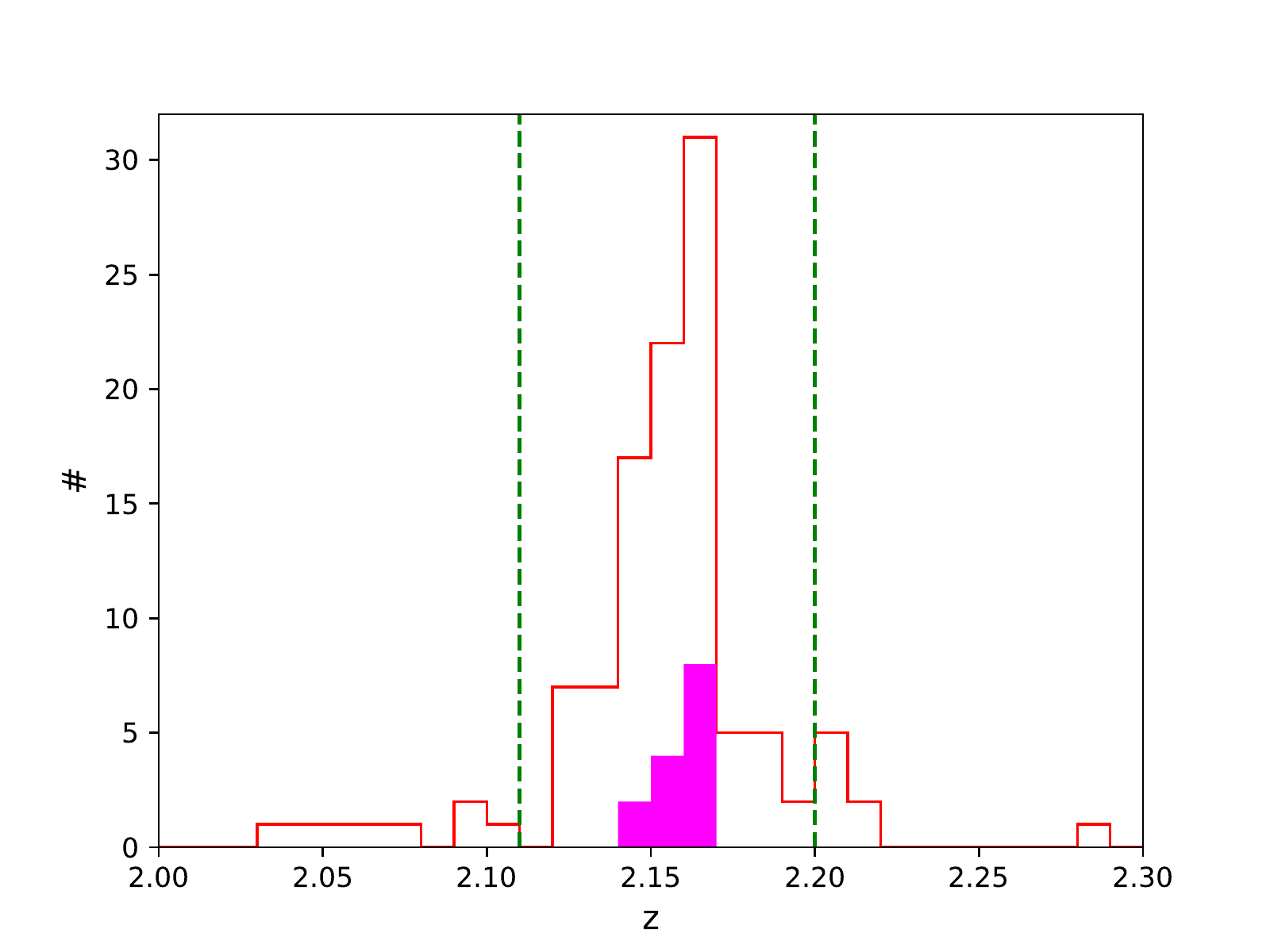}
\caption{Redshift distribution of the X-ray sources identified as 
protocluster members, including the one without spectroscopic confirmation 
(magenta filled histogram) compared to the 
redshift distribution of all the sources in the field with spectroscopic redshift in 
the range $2.0<z<2.3$. The dashed, vertical lines correspond to the protocluster redshift range.
}
\label{zdistX}
\end{center}
\end{figure}

Finally, in Figure \ref{zdistX} we plot the redshift distribution of the X-ray sources identified as 
protocluster members including the ERO (magenta filled histogram) compared to the 
redshift distribution of all the sources in the field with spectroscopic redshift in 
the range $2.0<z<2.3$.  The two distributions are statistically equivalent, 
showing again that the X-ray-emitting members are drawn uniformly in redshift among the 
spectroscopically confirmed members.

\subsection{ A note on the X-ray sources not associated to the protocluster}

In addition to the X-ray sources identified as cluster members in the 
innermost 5 arcmin (13 spectroscopically confirmed, including the 
Spiderweb Galaxy, plus 1 color-selected member), 
we find counterparts for 9 sources with redshift $z<2$, which are therefore in the 
foreground (6 of them are identified in the spectroscopic survey of 
X-ray sources in the field by \citet{2005Croft}, and 3 are found in NED by chance).
In the innermost 5 arcmin, we have 84 X-ray sources 
that have no redshift information.  Clearly, the most important question is 
whether or not we should expect more X-ray-emitting protocluster members among this 
set of sources.  To give a first-cut estimate, we report here some preliminary results
on the photo-z measurements of $K_{\rm S}$ band-detected sources in the field (Pannella et al. 
in preparation) obtained with the method described in Section 3. 
Driven by the comparison of spectroscopic and photometric redshifts
in the field, we conservatively
consider all the sources with $1.7< z_{\rm phot}<2.6$ as potentially associated to the
protocluster.  We have 20 additional X-ray-emitting sources in this sample. If we 
assume a flat distribution of the true redshift in this range, we expect approximately two additional X-ray sources belonging to the protocluster ($2.11<z<2.20$). 

Also, there are 58 X-ray sources within the central 5 arcmin
that are matched with $K_{\rm S}$ sources with $z_{\rm phot}<1.7$
or $z_{\rm phot}>2.6$. As it is very unlikely for these sources to have redshift in the 
protocluster range, we consider them as foreground and background sources. These leave 
out 15 X-ray sources without any redshift information.  By visual inspection, most of 
these sources do not have a possible counterpart in the HST and Subaru images, and therefore
some of them may be high-z sources.  Therefore, a reasonable assumption is to 
consider that all the 15 sources without any redshift information are outside the 
$2.11<z<2.20$ range.  We also note that the fraction of X-ray sources
without optical counterparts ($\sim 10\%$) is larger than that found in deep 
multiwavelength surveys like the CDFS (a few \%).  This effect is not significant and 
depends on the depth and solid angle of the  available optical and NIR coverage. This aspect 
will be further investigated in forthcoming papers (Pannella et al. in preparation).

As a final step, we selected all the $K_{\rm S}$ band detected sources complementary to 
the spectroscopy- and narrow-band-selected candidates we considered in this work and
find 380 sources within 5 arcmin with $1.7< z_{\rm phot}<2.6$. Following the previous
argument, we may expect an additional approximately 38 protocluster members without X-ray emission.  
If we focus on sources with ${\rm log}(M_*/{\rm M}_\odot)>10.5$, we may expect about 
75\% of these sources above this threshold (see Section 8), therefore an additional 28 
protocluster members.  
In summary, preliminary results on the SED of galaxies detected in the Spiderweb field
show that further spectroscopic follow-up campaigns may increase the 
number of protocluster members by a $\sim$ 20\%--30\%. Nevertheless, this simple order-of-magnitude 
estimate of the X-ray protocluster members that could have been missed is 
useful to bracket the AGN fraction and AGN enhancement factor
in the Spiderweb protocluster, as discussed in Section 
\ref{fraction} and Section \ref{enhancement}. 

% Rifinire la stima di photo-z aspettati nel protocluster.  Per il
% contributo dello spiderweb complex nel zphot range 1.7<zp<2.6,
% all'ordine zero: prendendo il catalogo di Muzzin e tagliando a
% logM>10.8 Salpeter (il nostro mass cut, tirato...) e 1.7<zp<2.6,
% vengono 0.9-1 sorgente per arcmin^2. Per quello che dicevamo l'altra
% volta si puo' considerare (ordine zero) statisticamente che ogni
% eventuale eccesso in questo mass e redshift range e' legato alla
% spiderweb (o struttura attorno collegata). Nel catalogo che ti ho
% mandato, tagliando in chi2 (<30), odds (>0.5), starflag (<2) e Nbands
% (>4), ci sono 132 oggetti (tra cui quasi 20 spec members). L'area del
% campo e' sui ~100 sq. arcmin quindi ci si aspetterebbe ordine ~90-100
% field contaminants nella selezione zphot. Quindi l'eccesso netto e' di
% una trentina di oggetti di cui 20 spec members, quindi tra i zphot
% selected probabilmente ne abbiamo ancora una decina associati a
% spiderweb complex. 

\subsection{Number counts in the $2.11<z<2.20$ redshift range}

To quantify the overdensity of X-ray sources with respect to the field, we computed the
X-ray number counts separately in a narrow redshift bin centered on the 
Spiderweb Galaxy. We decided to adopt the range $2.11<z<2.20$, and 
compute the logNlogS inside and outside this redshift range, 
both in the data and in the model of \citet{2007Gilli}. 

\begin{figure*}
\begin{center}
\includegraphics[width=0.49\textwidth]{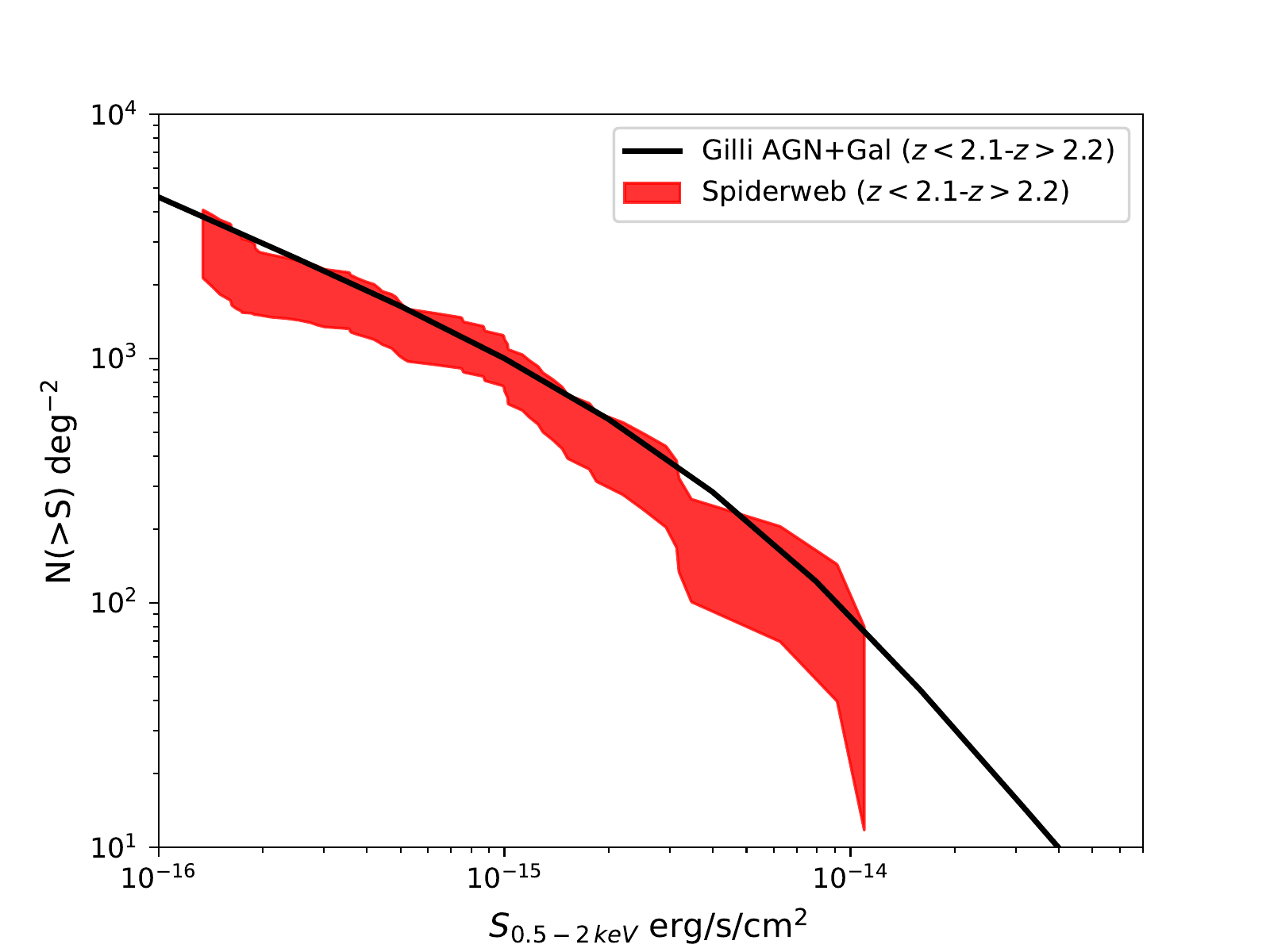}
\includegraphics[width=0.49\textwidth]{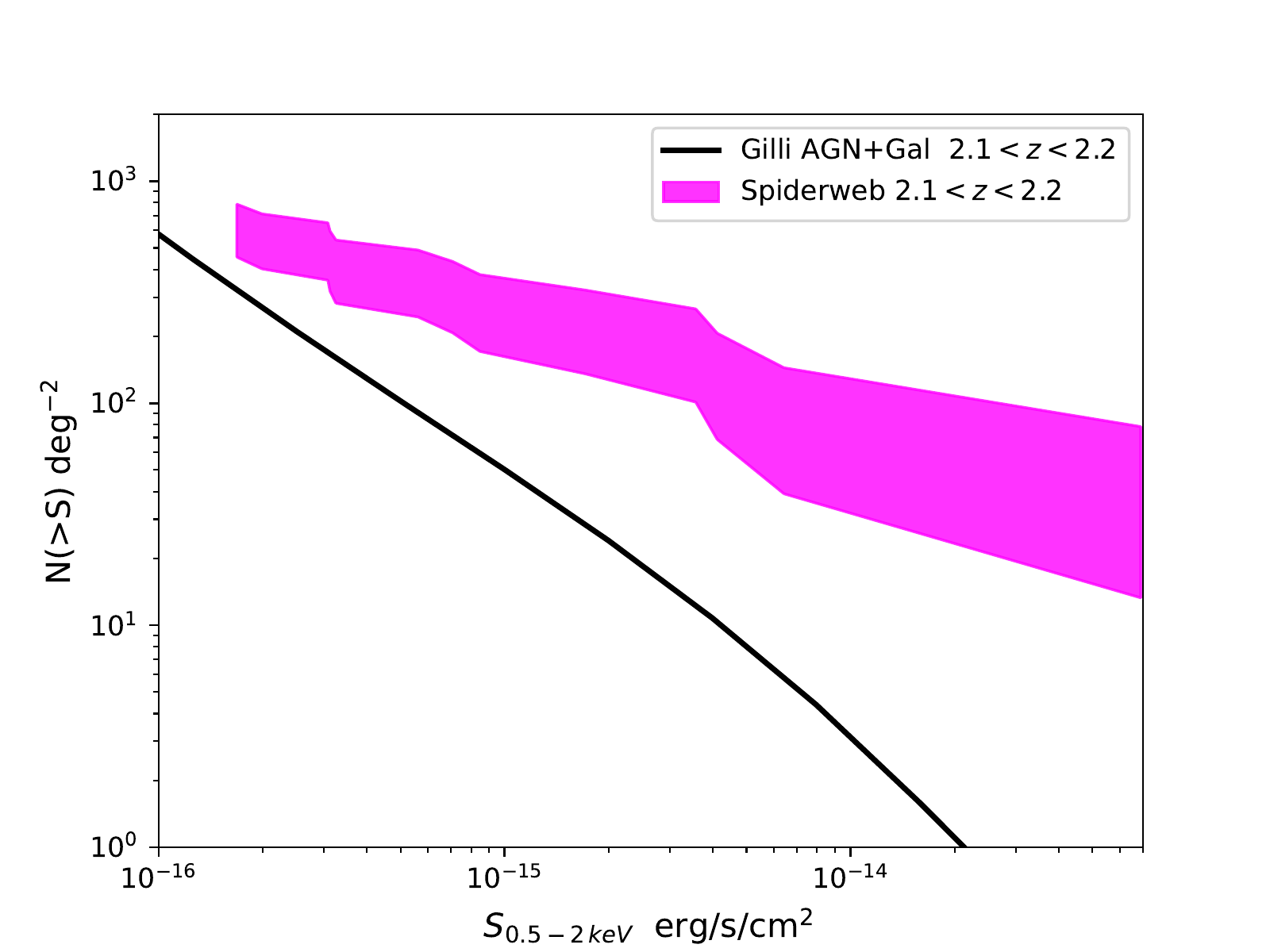}
\caption{Soft-band cumulative number counts in different redshift ranges. Left panel: Soft-band logNlogS in the Spiderweb field 
(within a radius of 5 arcmin) after excluding the X-ray sources identified 
in the $2.11<z<2.20$ range compared to the 
expectations from the model of \citet{2007Gilli} in the same redshift range.  The shaded area corresponds
to an uncertainty of $1\sigma$.  Right panel: Soft-band logNlogS in 
the Spiderweb field only for the redshift interval $2.11<z<2.20$ compared to the 
expectations from the model of \citet{2007Gilli} in the same redshift range.
The shaded area corresponds to an uncertainty of $1\sigma$.  
We note the different scales between the two panels.
}
\label{logNlogS_soft_zbin}
\end{center}
\end{figure*}

\begin{figure*}
\begin{center}
\includegraphics[width=0.49\textwidth]{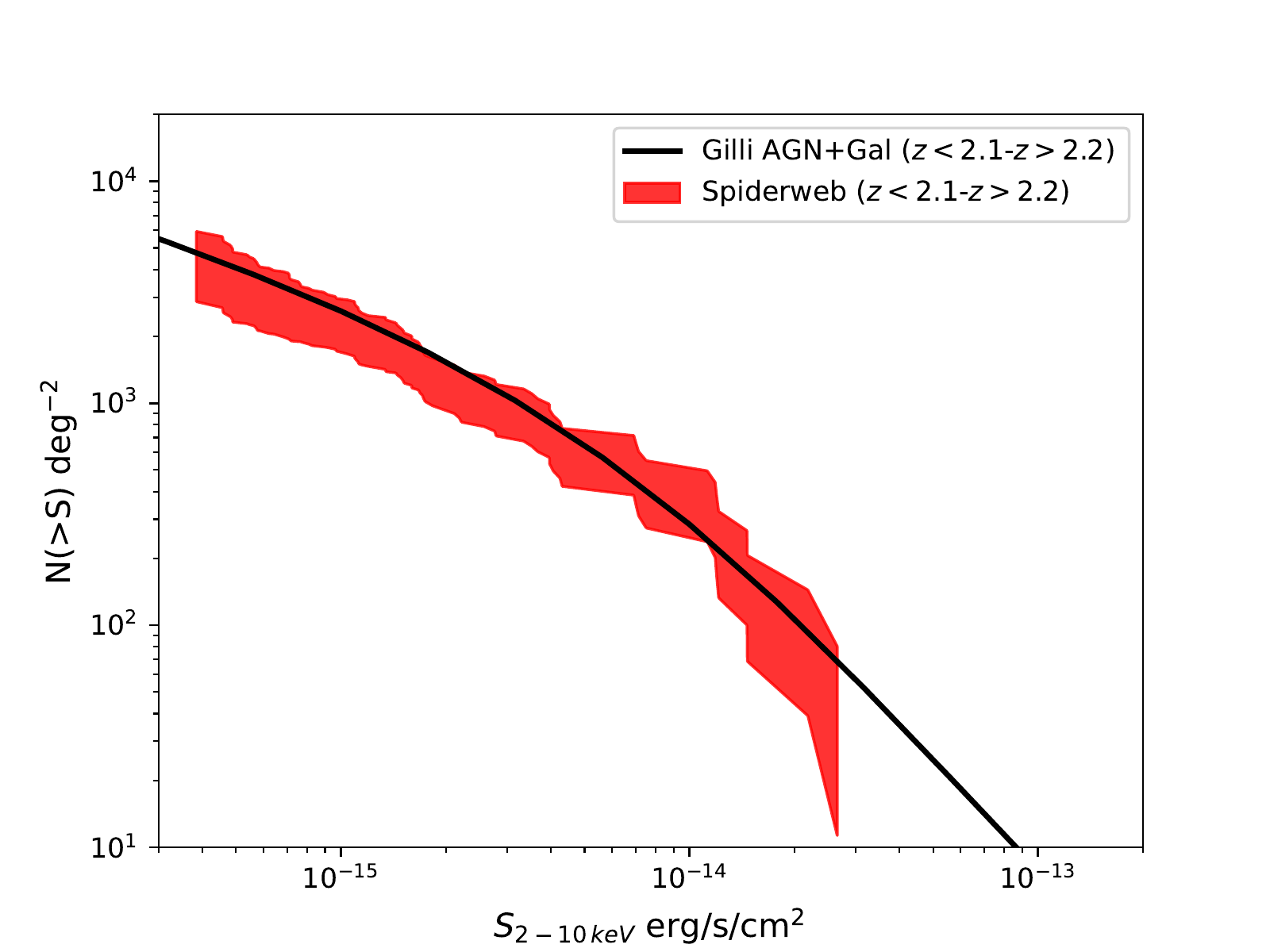}
\includegraphics[width=0.49\textwidth]{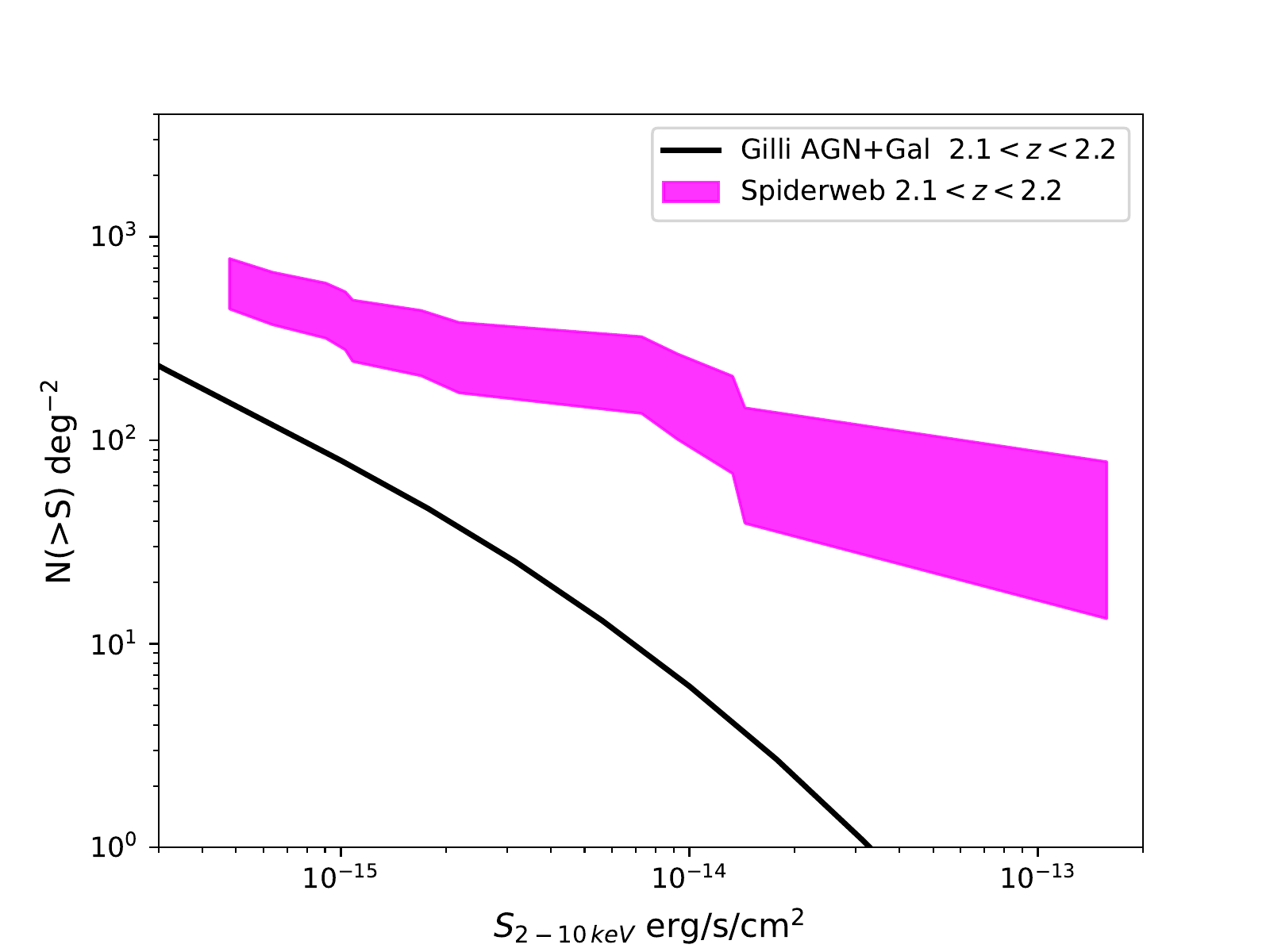}
\caption{Hard-band cumulative number counts in different redshift ranges. 
Left panel: Hard-band logNlogS in the Spiderweb field (within a radius of 5 arcmin) 
after excluding the X-ray sources identified in the $2.11<z<2.20$ range, compared to the 
expectations from the model of \citet{2007Gilli} in the same redshift range.   The shaded area corresponds
to an uncertainty of $1\sigma$.   Right panel: 
Hard-band logNlogS in the Spiderweb field only for the redshift interval $2.11<z<2.20$, compared to the 
expectations from the model of \citet{2007Gilli} in the same redshift range.
The shaded area corresponds to an uncertainty of $1\sigma$.  
Note the different scales between the two panels.
}
\label{logNlogS_hard_zbin}
\end{center}
\end{figure*}

The left panel of Figure \ref{logNlogS_soft_zbin} shows the logNlogS 
in the soft band after excluding the X-ray sources identified in the $2.11<z<2.20$ range
compared to the model expectations.  We find substantial agreement, with the counts in the 
Spiderweb field being slightly below the model, in line with the CDFS data. The right panel 
of  Figure \ref{logNlogS_soft_zbin} shows the soft-band logNlogS only for 
the redshift interval $2.11<z<2.20$ compared with the expected number counts 
from the model (black solid line). A much higher normalization is observed, and a 
different slope, showing that the excess is at relatively bright fluxes.  Clearly, the excess 
above fluxes $10^{-14}$ at soft erg/s/cm$^2$ is merely due to the presence of the 
Spidwerweb Galaxy. However, at fluxes $\sim 10^{-15}$ erg/s/cm$^2$ the excess 
can be robustly estimated and is found to be more than a factor of $\sim 5$, 
and more than an order of magnitude at $\sim 10^{-14}$ erg/s/cm$^2$.  
We find the same behavior in the hard band, as shown in Figure  \ref{logNlogS_hard_zbin}. 
Also in this case, the excess is more than a factor of $\sim 5$ at fluxes 
$\sim  10^{-15}$ erg/s/cm$^2$, and up to $\sim 10$ at $\sim  10^{-14}$ erg/s/cm$^2$. 
Finally, we note that the strikingly different slopes in the range $2.11<z<2.20$ 
is due to the presence of relatively bright ($\sim 10^{-15}$ erg s$^{-1}$ cm$^{-2}$) sources, 
which are therefore not affected by any means by our completeness correction.
Therefore, we are able to confirm an excess in the number of X-ray sources per 
square degree at the same level 
estimated from spectroscopy-, narrow-band-, or color-selected sources. 
% Clearly this
% excess need to be normalized by the number of galaxies in the same redshift range, 
% for which we have only a lower limit given by the current list of 
% spectroscopically confirmed members.  
The different shape of the 
number counts with respect to the model 
may be a sign that the X-ray population in this redshift range is not
uniformly extracted from the field population, but there may be some 
environmental effect at play. 
% The limiting factors here are the low statistics and the lack of a complete census of the
% galaxy population in the Spiderweb Complex.  However, we will attempt to constrain 
% the AGN fraction after completing the study of the X-ray properties of the 
% spectroscopically confirmed members and color-selected member candidates.

\subsection{Properties of X-ray-emitting protocluster members}

\begin{figure}
\begin{center}
\includegraphics[width=0.49\textwidth]{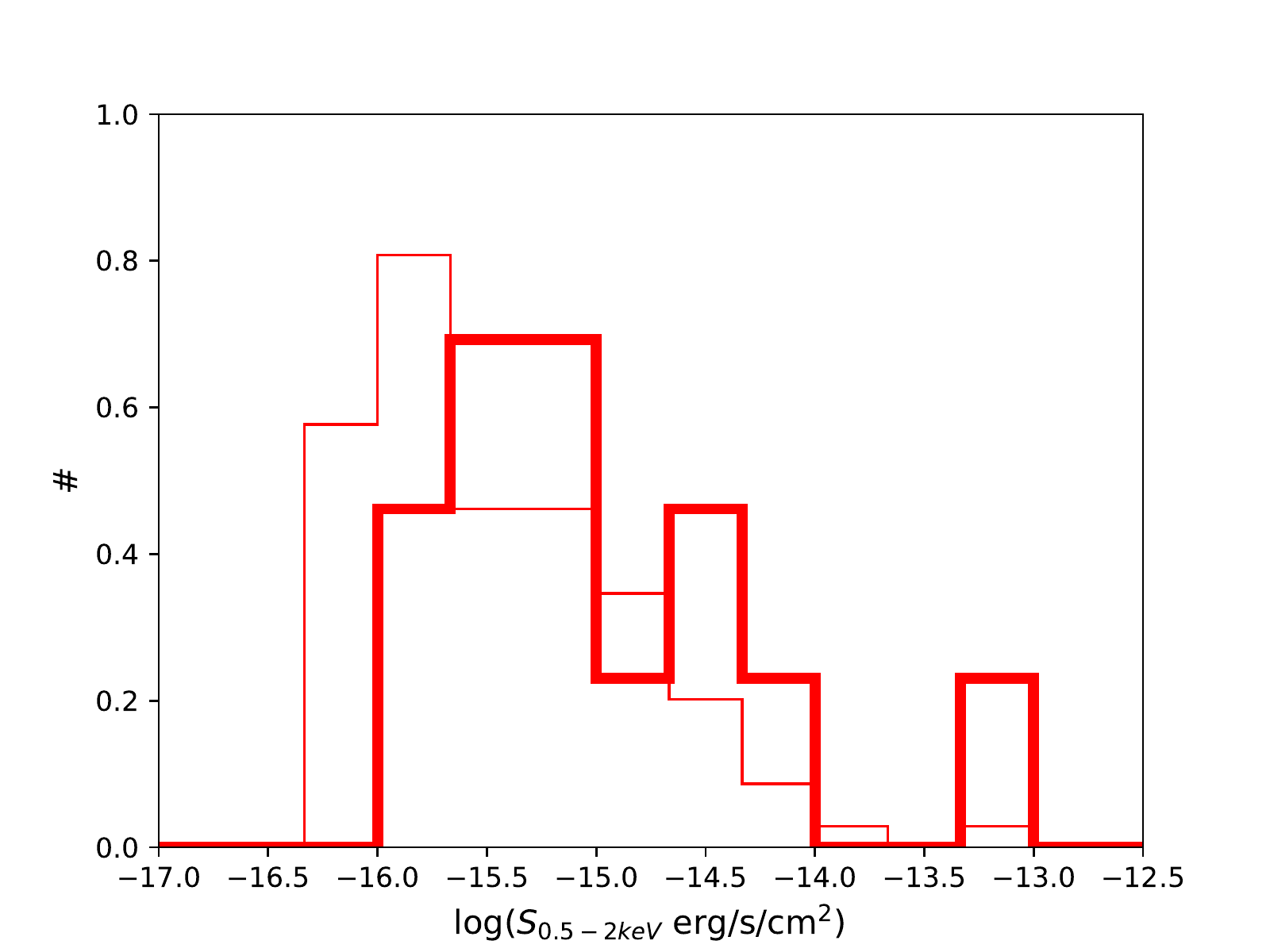}
\includegraphics[width=0.49\textwidth]{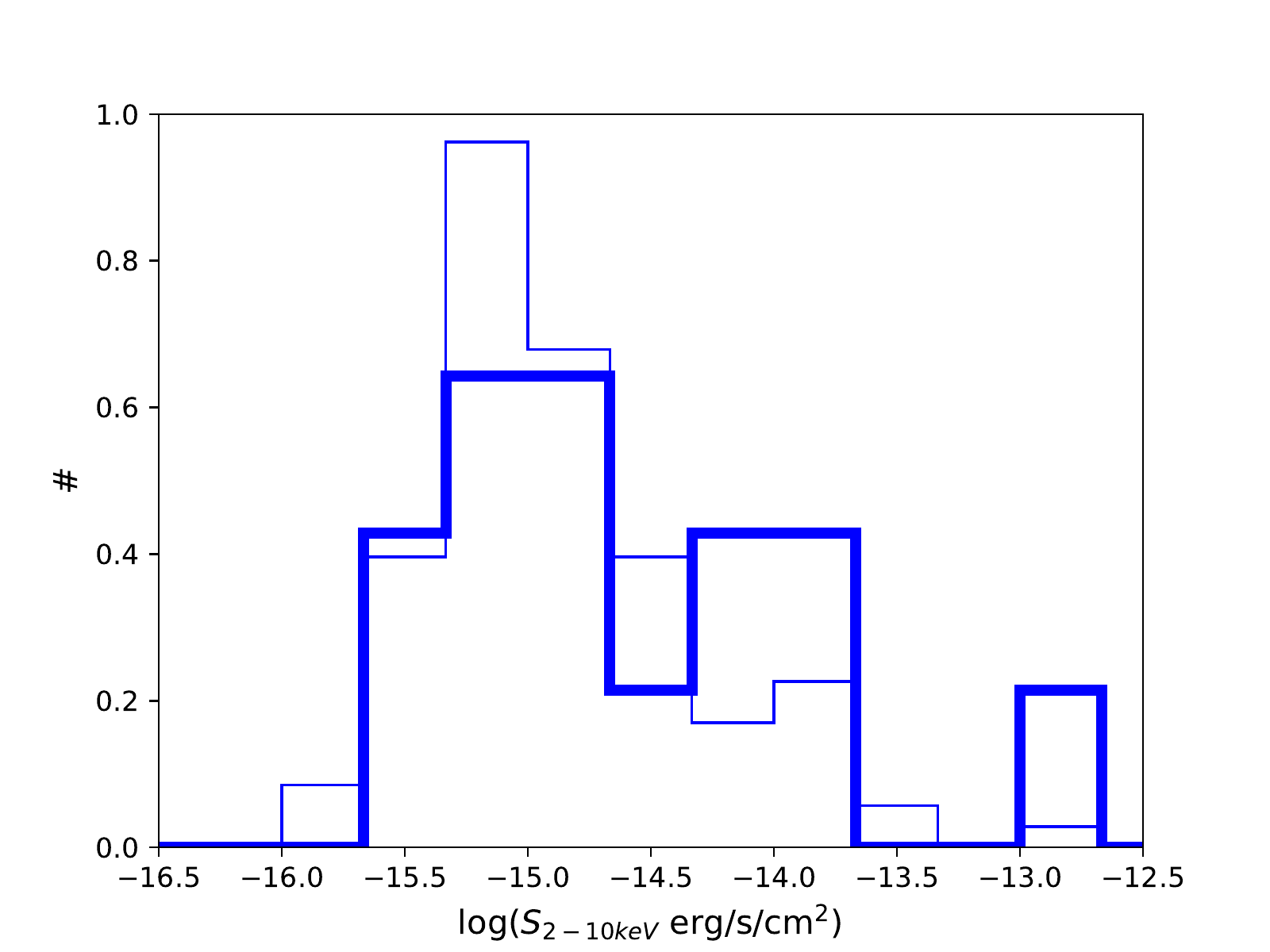}
\caption{Flux distributions.  Top panel: Differential normalized distribution of the soft fluxes in the 
Spiderweb Galaxy field.  The total sample of X-ray sources is shown with a thin line, 
while the subsample of the 14 protocluster members is shown with a thick line. 
Bottom panel: Same as in the top panel but in the hard band.  }
\label{flux_histo}
\end{center}
\end{figure}

We then investigated how different the population of X-ray protocluster members is with respect to the
bulk of the field X-ray sources in the Spiderweb Galaxy field.  As a first check,
we plotted the differential distribution of soft and hard-band fluxes; see Figure \ref{flux_histo}.
We note that in the soft and hard bands, the distribution of X-ray fluxes of the protocluster 
members is slightly shifted towards higher values with respect to the entire X-ray sample. 
% At the same time, X-ray cluster members are detected down to the 
% lowest fluxes reached in our survey.  

\begin{table*}
\caption{X-ray properties of the 14 protocluster members}
\label{src_Xrayfit}
\begin{center}
\begin{tabular}{lccccc}
\hline
ID &    $N_H$             & $\Gamma$  & EW  & $L_X (0.5-2 keV)$ & $L_X (2-10 keV)$ \\
   & $10^{22}$ cm$^{-2}$ &         &  keV  & $10^{44}$ erg/s   &  $10^{44}$  erg/s   \\
\hline
7  &  $40.1_{-8.4}^{+9.7}$      &  $ 2.00_{-0.28}^{+0.29} $     & $ 0 $                         & $3.47\pm 0.33 $ & $4.01\pm 0.27 $       \\ 
36 &  $1.13_{-1.05}^{+1.16}$    &  $ 1.86_{-0.14}^{+0.15} $     & $ 0.22_{-0.17}^{+0.20} $       & $1.27\pm 0.08 $       & $1.86\pm 0.15 $       \\ 
57 &  $<1$                      &  $ 1.89_{-0.28}^{+0.30} $     & $ 0 $                         & $0.21\pm 0.04 $         & $0.28\pm 0.07 $       \\ 
58 &  $4.26_{-0.32}^{+0.33}$    &  $ 1.90_{-0.031}^{+0.032} $   & $ 0.016_{-0.016}^{+0.024} $       & $31.92\pm 0.44$       & $42.85\pm 0.68 $      \\ 
74 &  $<2.7$                    &  $ 1.59_{-0.20}^{+0.21} $     & $ 0.47_{-0.31}^{+0.37} $       & $0.17\pm 0.02 $       & $0.39\pm 0.07 $       \\ 
75 &  $0.33_{-0.32}^{+0.65}$    &  $ 1.83_{-0.09}^{+0.10} $     & $ 0.07_{-0.07}^{+0.15} $       & $2.05\pm 0.09 $       & $3.11\pm 0.17 $       \\ 
80 &  $18.0_{-3.9}^{+4.6}$      &  $ 1.96_{-0.20}^{+0.23} $     & $ 0.34_{-0.12}^{+0.13} $       & $3.26\pm 0.21 $       & $4.15\pm 0.25 $       \\ 
86 &  $<2$ ($>150$)$^*$         &  $ 1.10\pm 0.20 $             & $ 0.51_{-0.32}^{+0.38}$         & $0.07\pm 0.01 $ $^\dagger$    & $0.34\pm 0.05 $ $^\dagger$    \\ 
\hline
\hline
9  &  $<2.2$                    &  $ 1.80 $                     & $ - $                         & $0.086\pm 0.020 $       & $0.135\pm 0.051 $     \\ 
12 &  $19.4_{-19.4}^{+19.8}$    &  $ 1.80 $                     & $ - $                         & $0.12\pm 0.04 $         & $0.18\pm 0.10 $       \\ 
34 &  0 ($>150$)$^*$    &  $ -0.76_{-0.84}^{+0.70} $    & $ - $                         & $<0.01$  $^\dagger$             & $0.026\pm 0.009$  $^\dagger$          \\ 
73 &  $13.3_{-6.1}^{+7.6}$      &  $ 1.80 $                     & $ - $                         & $0.15\pm 0.05 $         & $0.23\pm 0.05 $       \\ 
87 &  0                         &  $ 2.2\pm 0.4$                & $ - $                         & $0.18\pm 0.04 $         & $0.15\pm 0.09 $ \\ 
90 &  $14.8_{-5.8}^{+7.5}$ &    $ 1.80 $                        & $ - $                         & $0.18\pm 0.04 $         & $0.28\pm 0.07  $ \\ 
\hline
\end{tabular}
\end{center}
\tablefoot{X-ray properties from the  spectral fit for the 14 protocluster members
identified in the X-ray, including the Spiderweb Galaxy (XID 58). The two sources 
flagged with the symbol $^*$  are
considered Compton-thick candidates, and therefore have an estimated intrinsic 
absorption $>1.5\times 10^{24}$ cm$^{-2}$.  The corresponding luminosity values 
flagged with the symbol $^\dagger$  correspond to the observed
luminosity and not to the intrinsic, unabsorbed values, as for the other sources.}
\end{table*}

We then proceeded with a standard X-ray spectral analysis of the sources.  We adopted a simple
model consisting in an intrinsically absorbed power law, using the model components {\tt zwabs}
and {\tt powerlaw} within Xspec. Galactic absorption is described with the model {\tt tbabs}, 
and its value was fixed to $3.18\times 10^{20}$ cm$^{-2}$ \citep{2016HI4PI}.  
We also added a Gaussian line with 
intrinsic width below the spectral resolution, at an energy corresponding to the redshift 
neutral iron K$_\alpha$ line at 6.4 keV rest-frame, which was allowed to vary in energy by 5\% around this
interval (corresponding to a redshift uncertainty of $\Delta z\sim 0.1$).  We formally measured
the equivalent width in the observed band of this Gaussian component without 
making an attempt to quantify the confidence level of a line detection 
given the low signal regime. We leave the two shape parameters
$N_H$ and $\Gamma$ both free to vary when a source has more than 40 counts 
in at least one observed band (0.5-2 keV or  2-7 keV), while we fix 
$\Gamma = 1.8$ in the other cases.  The best-fit values 
are shown in Table \ref{src_Xrayfit}.  In addition, we performed 
a simplified spectral analysis for the remaining six sources, where the intrinsic spectral slope
is frozen to $\Gamma = 1.8$, and the intrinsic absorption is estimated. 
In Appendix B, we show the spectra of the 8 sources
suitable for spectral analysis, along with the best-fit model, and the spectra 
of the fainter 6 sources, despite the coarse binning.  In this section, we discuss 
the X-ray properties of the 13 protocluster members, while the 
Spiderweb Galaxy, apart from a few basic spectral parameters, will be discussed 
in a companion paper (Tozzi et al. in preparation).

First, we note that we are able to measure the intrinsic slope with good accuracy
in eight sources, finding an average $\langle \Gamma \rangle = 
1.75$.  Source XID 86 has un unusual, hard slope (with an hardness ratio $HR\sim 0$) 
and a measured equivalent width for the 
neutral Fe $K_\alpha$ line of $ \sim 0.5 $ keV, close to the value expected for 
reflection-dominated, Compton-thick sources 
\citep[see, e.g.,][]{2002Norman,2011Gilli,2020Iwasawa}. 
We consider it a Compton-thick
candidate with $N_H>1.5\times 10^{24}$ cm$^{-2}$.
Therefore, the slope of its spectrum is not representative of the 
spectral shape of the intrinsic emission. If we exclude this source, we measure
an average intrinsic slope of $\langle \Gamma \rangle = 1.86\pm 0.05$, in line with expectations 
\citep[see, e.g.,][]{2006Tozzi,2020Iwasawa}.  
Another source with an unusually flat spectrum is XID 34. This source 
is the faintest among the protocluster X-ray members (17 net counts in the 0.5-7 keV band) 
and cannot be meaningfully fitted with a frozen $\Gamma = 1.8$. We tentatively 
considered it another  Compton-thick candidate.

\begin{figure}
\begin{center}
\includegraphics[width=0.49\textwidth]{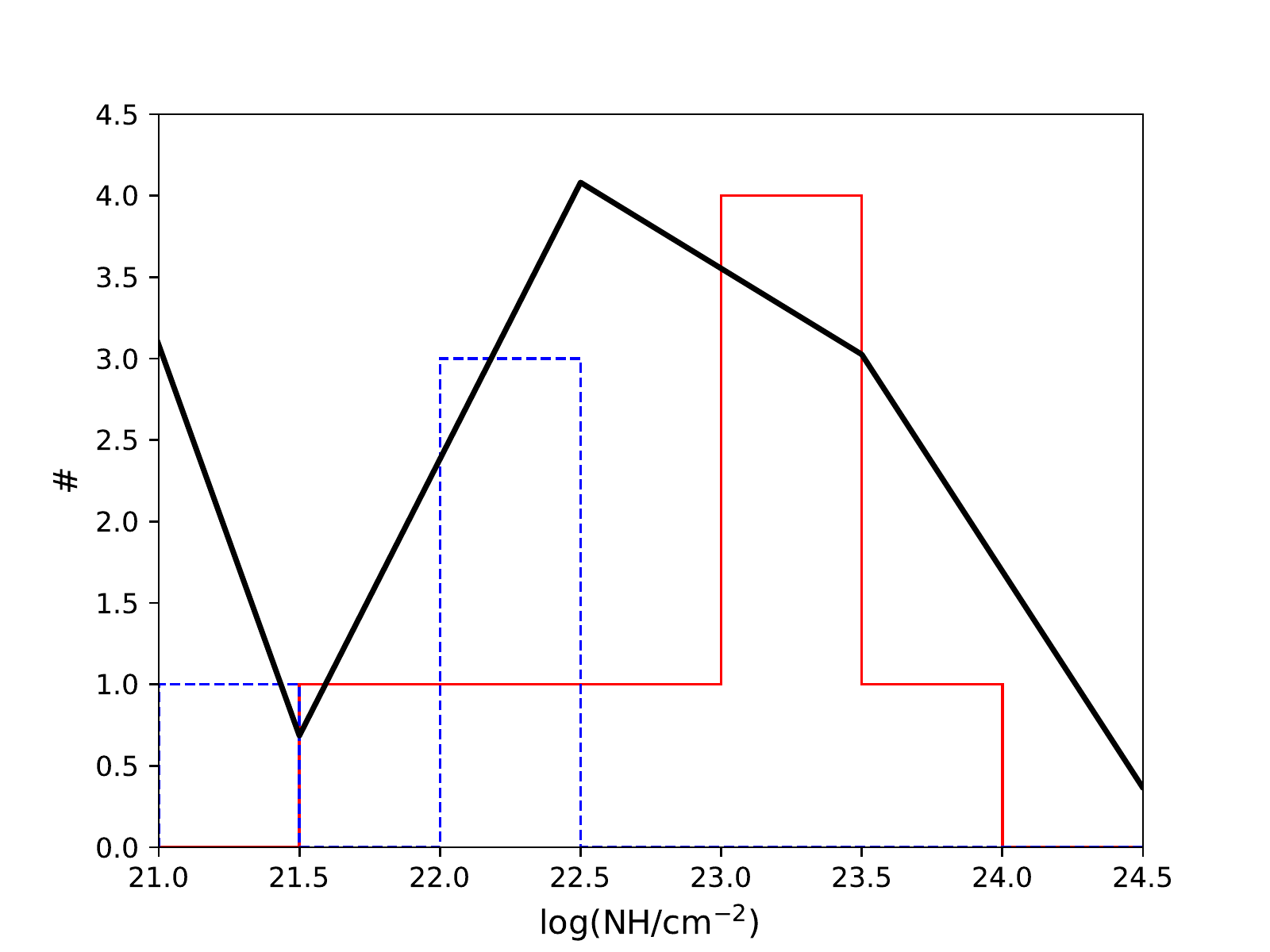}
\caption{Intrinsic absorption in AGN protocluster members.
The red histogram shows the distribution of the intrinsic absorption of the
eight protocluster members (including the Spiderweb) for which a robust measurement of $N_H$ 
has been possible. The blue dashed histogram shows the upper limits on $N_H$ obtained
for six sources.  The black solid line shows the distribution of $N_H$ expected
in the model of \citet{2007Gilli} at the flux limit of our observations.  
The two Compton-thick candidates with a reflection-dominated spectrum are not included.
% The two sources that require a very flat $\Gamma$ ($\leq 1$) are
% considered Compton-thick candidates with a reflection dominated spectrum, and 
% shown at $N_H> 10^{24}$ cm$^{-2}$.
}
\label{NH_histo}
\end{center}
\end{figure}

For all the remaining sources, we are able to directly measure the intrinsic absorption or
an upper limit for it.
In Figure \ref{NH_histo} we show the distribution of the intrinsic absorption for the 
14 protocluster members (including the Spiderweb Galaxy).  We also include upper
limits (shown with the dashed histogram), while we exclude the two Compton-thick candidates. 
The five sources (not including the Spiderweb) for which we robustly measure $N_H$ unsurprisingly
have $N_H>10^{23}$ cm$^{-2}$, which is simply a consequence of the high-redshift
and the low-signal regime. We also show $1\sigma$ upper limits for six sources, finding
values of $N_{\rm H}$ of less than a few times 10$^{22}$ cm$^{-2}$. 
The limited statistics and the low signal prevent us from making any general conclusions 
as to the distribution of $N_{\rm H}$ among the protocluster members.  A visual
comparison with the distribution of $N_{\rm H}$ expected in the model by
\citet{2007Gilli} in the redshift bin $2.0<z<2.2$ (black line) normalized to the
number of sources in our small sample suggests that the 
number of strongly absorbed sources (7 sources with $N_{\rm H}>10^{23}$ cm$^{-2}$) is 
broadly consistent with that expected in the field at the same redshift. 
However, we note that the Compton-thin source with the highest intrinsic absorption 
that we are able to measure (XID 7, $N_{\rm H}=4.0\times 10^{23}$ cm$^{-2}$) 
is still more than four times less absorbed than the most absorbed sources detected
in protoclusters \citep[see][]{2019Gilli,2020Vito}. 

\begin{figure}
\begin{center}
\includegraphics[width=0.49\textwidth]{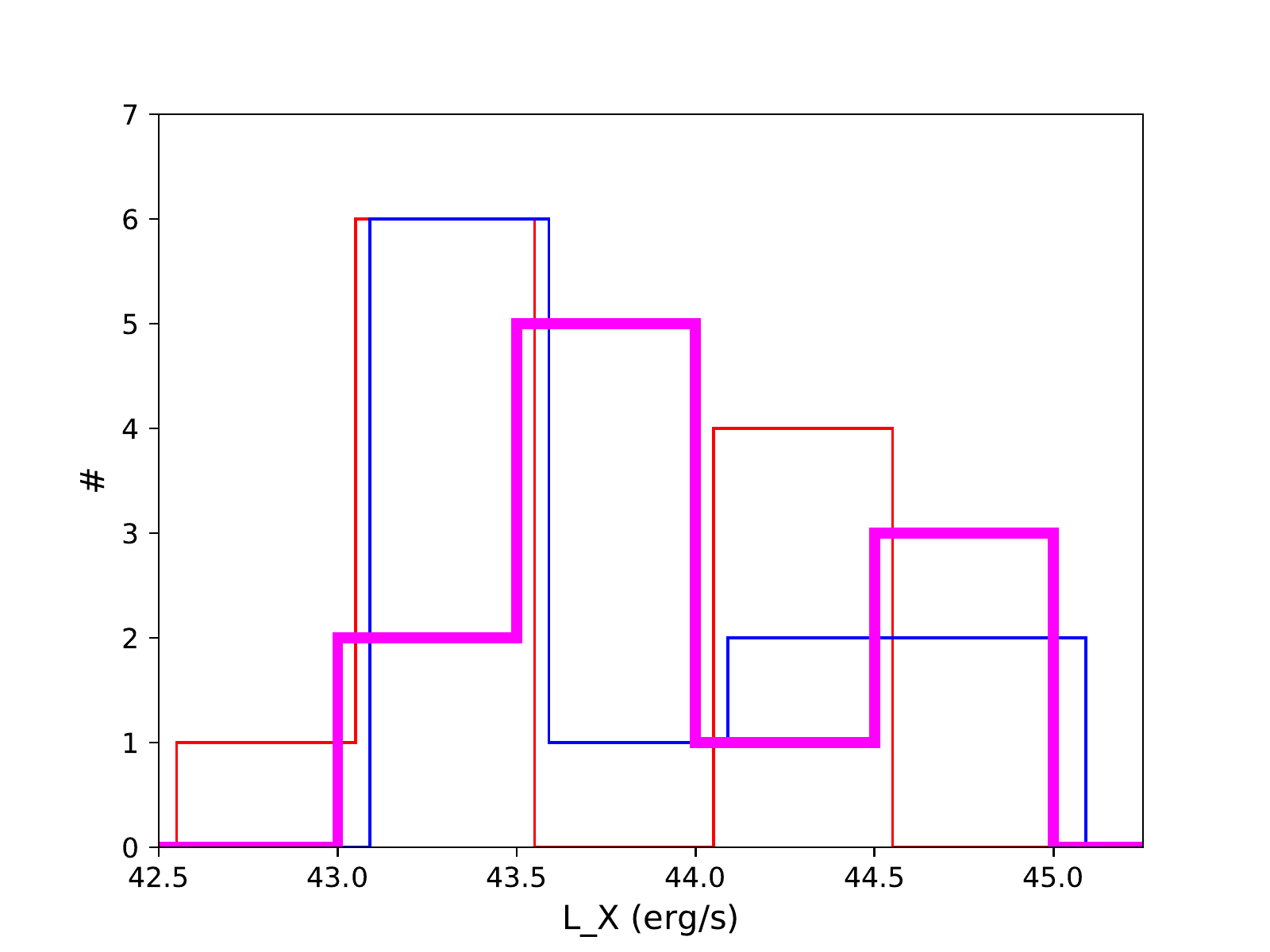}
\caption{Distribution of the X-ray luminosity of the 
11 protocluster members (excluding the Spiderweb and the two 
Compton-thick candidates) in the soft (thin red line), 
hard (thin blue line), and total band (thick magenta line).  For clarity, the histograms
for the soft and hard luminosities are shifted by 0.05 and 0.1 dex, respectively.
}
\label{LX_histo}
\end{center}
\end{figure}

\begin{figure}
\begin{center}
\includegraphics[width=0.49\textwidth]{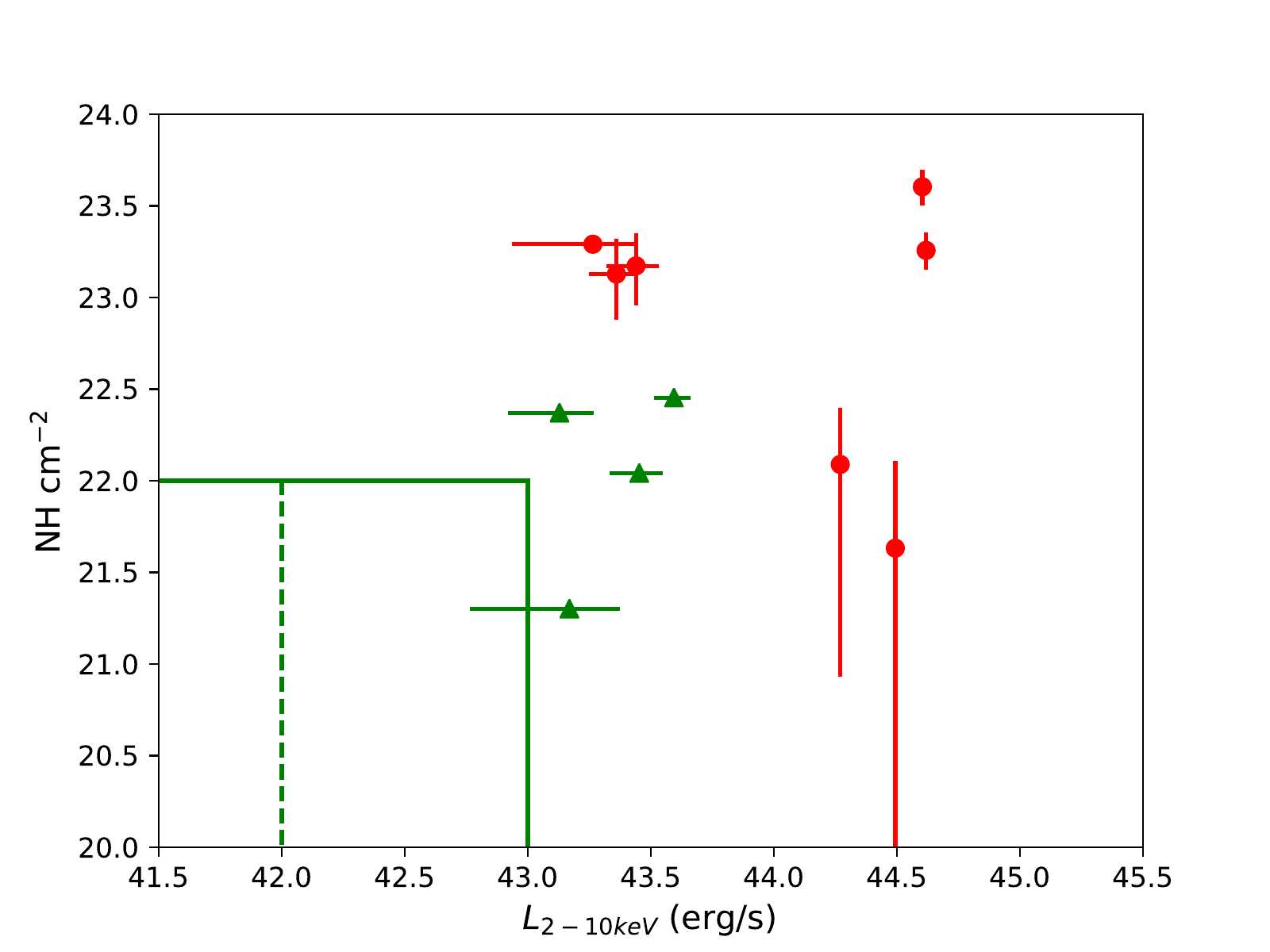}
\caption{Scatter plot of $N_H$ and $L_{2-10 keV}$ for 11 sources 
(excluding the Spiderweb and the two Compton-thick candidates).  Red circles correspond
to measured $N_H$ values, while green triangles to $1\sigma$ upper limits.  The green box shows the
parameter space still consistent with star formation at the level of $1000$ M$_\odot$/yr or 
$100$ M$_\odot$/yr (vertical dashed line). }
\label{nhlx}
\end{center}
\end{figure}

Having the intrinsic absorption, and an estimate of the 
intrinsic power law, we can compute the unabsorbed luminosity 
in the rest frame 0.5-2 and 2-10 keV bands.  
The unabsorbed luminosities  
of all sources but the two Compton-thick candidates and the Spiderweb Galaxy 
are shown in Figure \ref{LX_histo}.  All sources have a rest-frame, hard-band 
luminosity of greater than $10^{43}$ erg/s. 
The luminosity range puts all these sources in the Seyfert regime, far from normal
star forming galaxies. In principle, we may expect obscured star formation at a high
level, up to 1000 $M_\odot$/year.
Using the average $L_{2-10 keV} - SFR$ relation in the redshift range 2.0-2.5 as
measured from deep fields by \citet{2016Lehmer}, we obtain 
$L_{2-10 keV} \sim 10^{42}$ and $\sim 10^{43}$ erg/s for a SFR of 100 and 1000 $M_\odot$/yr, 
respectively.  On the other hand, sources with high intrinsic absorption are 
unambiguously dominated by nuclear emission with absorption 
due to an optically thick torus, while absorption at the level
of $N_H\sim 10^{22}$ cm$^{-2}$ and below can be associated to the 
interstellar medium of the host, if sufficiently dense (but see
recent results in Norman et al. 2021, where column densities of 
$N_H\sim 10^{23}$ cm$^{-2}$ can be reached at $z\sim 2$).
In Figure \ref{nhlx} we plot the intrinsic $N_H$ as a function of  
$L_{2-10 keV}$ and show with green lines the 
corner where $N_H$ and $L_{2-10 keV}$ are consistent with being powered by star formation 
at the level of 100 and 1000 $M_\odot$/yr.  We conclude that all the 
X-ray sources detected in the protocluster 
are consistent with being in the Seyfert and QSO regime. 

We made an attempt to detect variations in the average properties of 
the X-ray protocluster members as a function of the distance from the Spiderweb Galaxy.  
We find that half of the spectroscopically confirmed protocluster members are distributed
within a radius of 90 arcsec (corresponding to $\sim 760$ kpc). Therefore, 
we consider the properties of the X-ray-emitting protocluster members below and 
above this radius.  As already shown in Figure \ref{zspec_dist}, the distribution of 
the X-ray sources is very similar to that of all the spectroscopically confirmed members.
The fraction within and beyond 90 arcsec is $14 \pm 5$\% and  $13 \pm 5$\%, respectively.
If we consider only protocluster members with $ log(M_*)>10.5$ (see Section 8), we find 
$33 \pm 13$\% and  $29 \pm 12$\% within and beyond 90 arcsec, respectively. 
The X-ray properties do not change either.  The distribution of intrinsic absorption in the 
two subsamples is almost exactly symmetric (with one C-thick candidate per side) apart
from source ID=7, which shows the highest measured $N_H$ and is the closest
to the Spiderweb.  Also, the average soft (hard) band luminosity is $0.87\,  (1.21) \times 10^{44}$ erg/s and
$0.82 \, (1.1)\times 10^{44}$ erg/s, within and beyond 90 arcsec, respectively. 
Our preliminary investigation shows (possibly) marginally higher X-ray fraction, 
intrinsic absorption, and lower luminosities towards the protocluster center. 
However, this trend, if present, can only be reliably measured with a much larger sample, 
implying population studies of protoclusters with comparable X-ray depth.

Finally, apart from the X-ray-detected sources, we note that there are no protocluster members that have been identified
as AGN from the optical and NIR spectrum.  A detailed 
discussion of the single sources can be found in Appendix C.

\subsection{Average X-ray properties of X-ray silent, spectroscopic, and color-selected protocluster members}

\begin{figure}
\begin{center}
\includegraphics[width=0.49\textwidth]{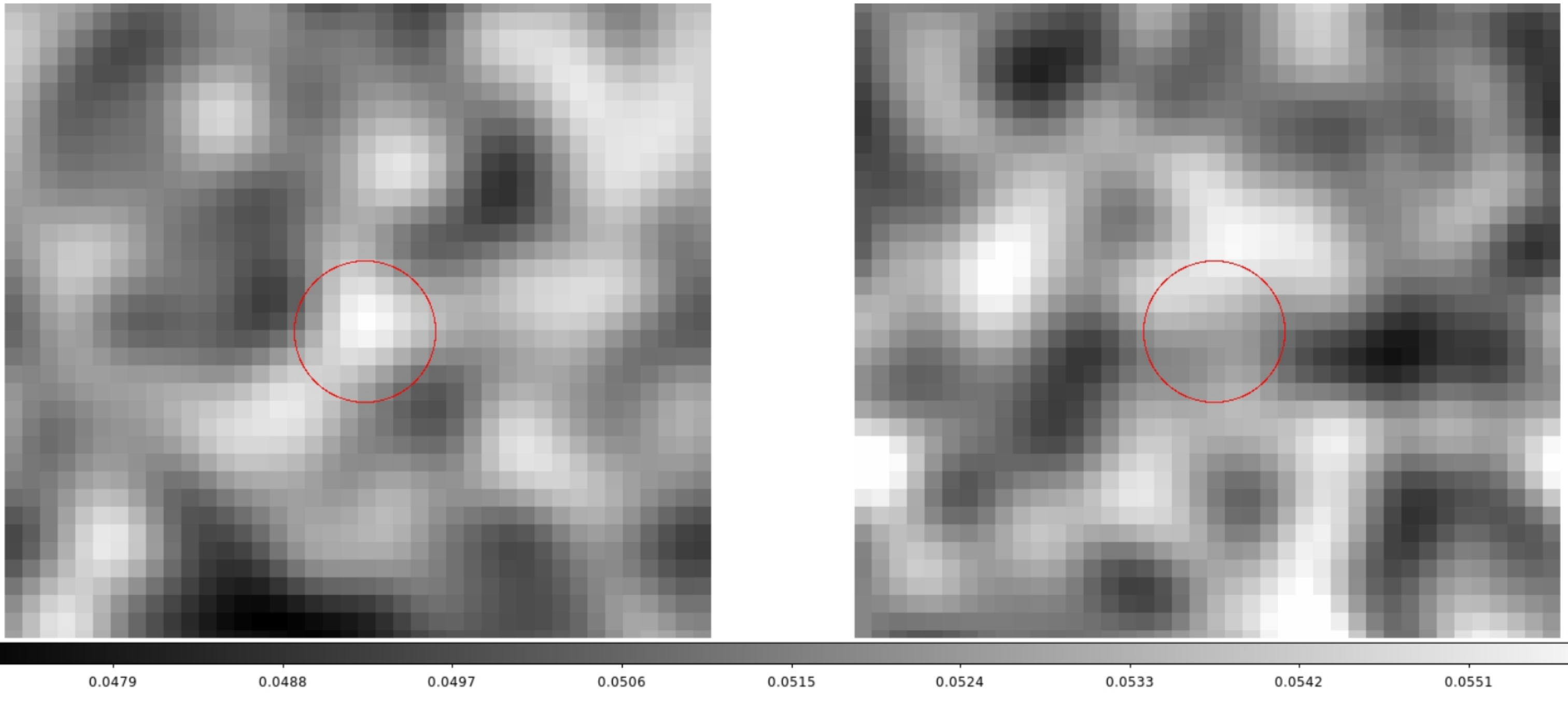}
\caption{Stacked X-ray images.  Left panel: Stacked image of the 82 spectroscopically confirmed protocluster
members in the soft band.  The size of the cutout is $20\times 20$ arcsec, 
while the red circle shows the 2 arcsec radius
region used to perform aperture photometry on the stacked image.
Right panel: Same as in the left panel but in the hard band.
}
\label{stacked_specz}
\end{center}
\end{figure}

We investigated the presence of X-ray emission below our detection
threshold in the other 
spectroscopically identified protocluster members.  We extracted a list of 82 sources 
with spectroscopic redshift in the range $2.0<z<2.3$ 
after removing the sources already detected in the X-ray band, 
and all the sources embedded in the diffuse emission close to the 
Spiderweb Galaxy within a radius of 12 arcsec (the so called flies). 
We then used the background 
images already used to compute the sky coverage, where all the 107 identified 
X-ray sources have already been removed.
These soft and hard-band ``background-only'' images were obtained simply by cutting out
the X-ray-detected sources and replacing the signal at their position with random 
noise consistent with the signal in the annular region surrounding it. These images therefore 
contain only the sky X-ray emission from sources below our detection threshold.
We then simply stack the images at the position of the 82 optical sources previously
selected.
The resulting images, shown in Figure \ref{stacked_specz}, provide an estimate of the 
cumulative emission from spectroscopically confirmed, nonX-ray-detected protocluster members 
in the soft and hard band.   If we perform a simple aperture
photometry in a circle of 2 arcsec (corresponding to a reasonable photometric
aperture enclosing 95\% of the flux in our field), we do  not obtain significant
detection in either band, formally measuring  $39 \pm 30 $ and $49 \pm 40 $ in the 
soft and hard band, respectively. However, in the total (0.5 - 7 keV) band
we have $89 \pm 50$ net counts, which formally has a statistical significance of 
$\sim 2\, \sigma$.  If we consider this as a marginal detection, after correction for average vignetting 
($\sim 5$\%), we obtain a soft-band flux of  
$ \sim 6.2 \times 10^{-18}$ erg/s/cm$^{-2}$ and a hard-band flux 
of  $ \sim 1.6 \times 10^{-17}$ erg/s/cm$^{-2}$ per source. Assuming a typical 
spectrum with $\Gamma = 1.8$ and $N_{\rm H}\sim 10^{22}$ cm$^{-2}$, this gives 
an intrinsic, rest-frame luminosity range of 
$2.3-3.0\times 10^{41}$ erg/s and $3.6-4.8 \times 10^{41}$ erg/s 
(after correcting for intrinsic and Galactic absorption) 
in the 0.5-2.0 and 2-10 keV bands, respectively.  
The average, total band X-ray luminosity of these sources is
therefore well below $\sim 10^{42}$ erg/s, 
which in turn corresponds to a maximum SFR of significantly less than $ 100$ M$_\odot$/yr per source.
This shows us that, with current {\sl Chandra} capabilities, 
we can barely detect the average X-ray signal from spectroscopically confirmed
members at the level of approximately one net count in our deep 715 ks 
exposure.  At the same time, this would imply a widespread star formation 
activity at the level of  a few tens of M$_\odot$/yr on average
for all the protocluster members, while optical and NIR spectroscopy do not 
confirm this.  Therefore, if such a weak X-ray signal were confirmed,
it may be ascribed to the combination of low-level nuclear activity and moderate star formation. 
Clearly, to explore the low-X-ray-luminosity regime of 
the bulk of the protocluster members, we should increase the current
exposure by at least a factor of five.
This would allow us to explore the star formation regime 
corresponding to $20-50$ M$_\odot$/yr in each single source, implying that
all the strong-starburst galaxies may be detected.
% On the other hand, 
% at present we can conclude that, if due to star formation, the stacked
% X-ray signal of the spectroscopically confirmed, X-ray silent members is well below
% the level of $100 M_\odot$/year.  In fact, an average level of $100 M_\odot$/year
% would have provided a $2\sigma$ detection in each band, and a robust $>3 \sigma$ detection 
% in the total band.  The same argument of course applies for low-level AGN activity.  
We conclude that X-ray silent spectroscopic members do not contribute 
to the AGN luminosity function of the protocluster, but they may be
hosting starburst events or low-level AGN activity. This is currently beyond the capability of 
{\sl Chandra}  (considering regular and large programs) and it definitely 
constitutes an interesting luminosity range to be explored by future high-angular-resolution X-ray missions, such as Lynx \citep{2018LynxTeam} 
and AXIS \citep{2019Mushotzky,2020Marchesi}.

\begin{figure}
\begin{center}
\includegraphics[width=0.49\textwidth]{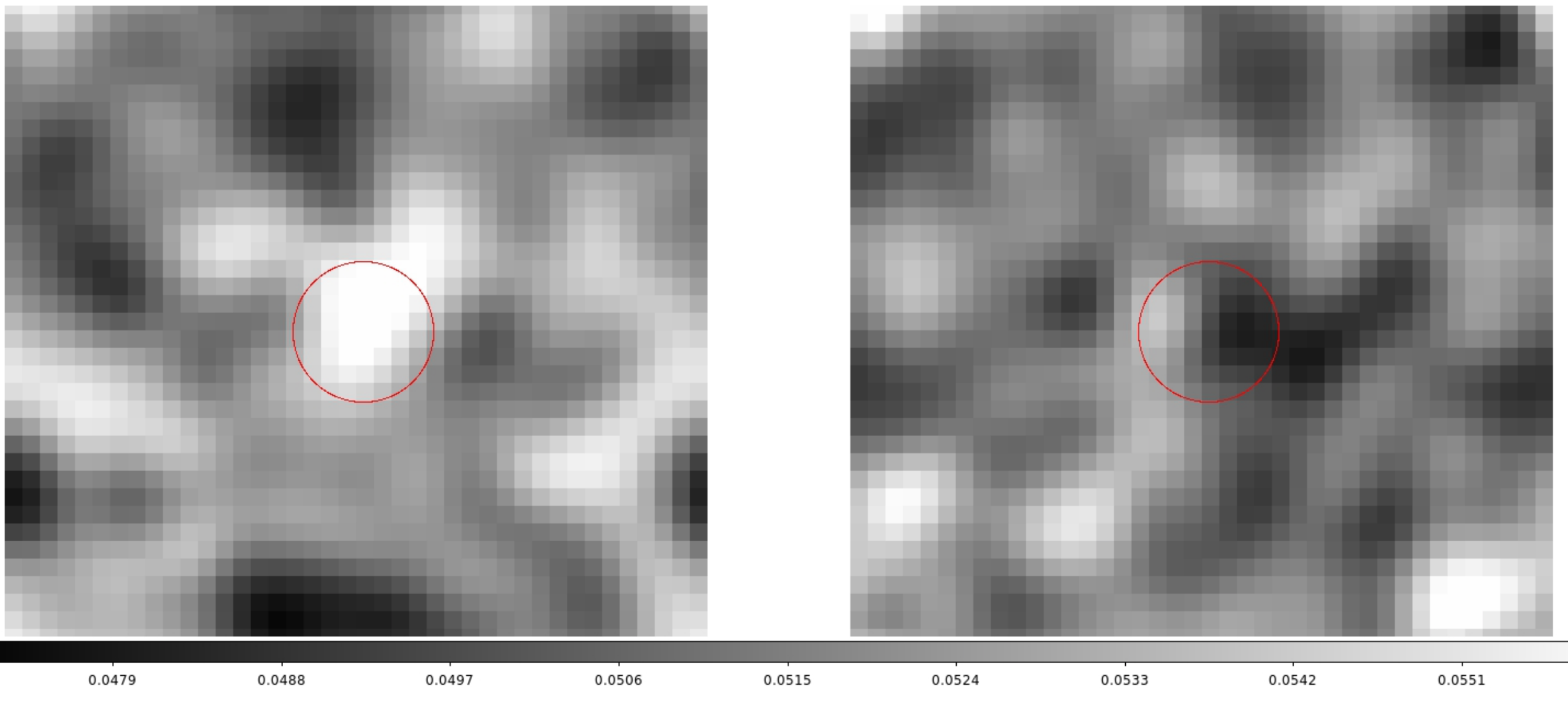}
\caption{Stacked X-ray images.  Left panel: Stacked image of the 81 color- or narrow-band-selected
protocluster member candidates in the soft band.   The red circle shows the 2 arcsec radius
region used to perform aperture photometry on the stacked image.
Right panel: Same as in the left panel but in the hard band.
}
\label{stacked_colorsel}
\end{center}
\end{figure}

We then applied the same procedure to the 81 color- or narrow-band-selected sources that 
have no X-ray detection, and find similar results, possibly with a more
convincing detection in the soft band, as shown in Figure 
\ref{stacked_colorsel}.   The usual aperture photometry within a circle of 2 
arcsec provides $60\pm 30$ net counts in the soft band, and therefore a formal
detection at $2\sigma$.  The soft band photometry 
corresponds to a soft flux of $(9.4\pm 5) \times 10^{-18}$ erg/s/cm$^{-2}$, 
and, assuming the usual mildly absorbed power law and  $z=2.156$, to a 
luminosity of $(3.6 \pm 1.8) \times 10^{41}$ and   $(5.6\pm 2.8) \times 10^{41}$ erg/s per source
in the soft and hard band, respectively. 
This luminosity range is typical of starburst galaxies
with SFR of the order of $30-50$ M$_\odot$/yr.
In the hard band, we obtain only an upper limit (formally the photometry gives
$13\pm 39$ net counts), which translates to $2\sigma$ upper limits
of $4.0 \times 10^{41}$ and $6.0 \times 10^{41}$ erg/s per source
in the soft and hard band, respectively.  This tells us that the average source is softer than the adopted, AGN-like model. 
Despite the small signal obtained in the soft band, we further
split the 81 color- or narrow-band-selected sources without X-ray detection
into four subsamples according to their selection criteria: $H_\alpha$ emitter, 
$Ly_\alpha$ emitter, ERO \citep{2004aKurk}, and photo-z \citep{2014Dannerbauer,2018Shimakawa,2019Tadaki}.
We find 1 $\sigma$ positive 
photometry in the soft band for $Ly_\alpha$  (13)- and ERO (29)-selected sources, and 1.5 $\sigma$ positive 
photometry for photo-z (30). In the hard band we have both positive and negative photometry. 
Therefore, the $2 \sigma$ signal in the soft band is 
confirmed, and is contributed mostly by the 30 protocluster member candidates selected
by photo-z. However, we have no control
on the contamination of this list of protocluster member candidates.  
This aspect, coupled to the large statistical uncertainty associated to 
stacked emission measurements, 
hampers any conclusion on the X-ray emission from 
color-selected protocluster member candidates, except that we do not
expect them to host significant nuclear activity. 

Our general conclusion is that there is no significant X-ray emission that we have missed
so far among all the confirmed or candidate protocluster members that have not
been X-ray detected. However, the obtained marginal detections 
confirm that potentially we are able to detect 
star formation in the protocluster galaxies at the level of $\sim 30-50$ M$_\odot$/yr 
by stacking their X-ray emission.  Additional work must be done before we can draw 
any strong conclusion, in particular it is necessary to increase the number of spectroscopically 
confirmed members by more than a factor of two.  From this perspective, 
future observations of this target with JWST and further spectroscopic follow-up may have strong synergy with the available X-ray data and 
reveal further relevant properties of the galaxy population in the Spiderweb 
protocluster.

\section{X-ray luminosity function of protocluster members\label{function}}

To compute a cumulative luminosity function in the protocluster region, 
we need a comoving volume representative of our sample.  
If we consider the comoving volume corresponding to a circle of 5 arcmin 
centered on the Spiderweb Galaxy, and the 
difference in the comoving radial distance between $z=2.11$ and $z=2.2$, 
we find a value of $2.584\times 10^{4}$ Mpc$^3$.  
The comoving mass corresponding to this volume in the assumed cosmology
is $M_{\rm com}=1.05\times 10^{15} M_\odot$, 
which is in the range of predicted virial mass at $z=0$ for the
Spiderweb protocluster.  However, this is rather a lower limit to the mass
included within a radius of 5 arcmin, because the central part of the protocluster
is definitely overdense with respect to the ambient cosmic density.
% assuming CRITICAL DENSITY RHO = 2.778 10^11 H^2 MSUN MPC^{-3}
% ASSUMING H=0.7, AND OMEGAM=0.3, RHO=0,408366 \times 10^11 MSUN MPC^{-3}

Second, we need an estimate of the intrinsic luminosity of the
Compton-thick candidates.  As mentioned above, the luminosity values 
listed in Table \ref{src_Xrayfit} for XID 86 and XID 34 simply refer the observed 
luminosity that,  in the case of a reflection-dominated spectrum, is only a fraction of the 
intrinsic power. We conservatively assume that the observed hard (2-10 keV) luminosity 
is $\sim 10$\% of the intrinsic one \citep[corresponding to the upper envelope of the
values observed for local Compton-thick AGN; see][]{2019Marchesi,2021Zhao}, 
and that the soft-band luminosity is 
1.6 times the hard-band luminosity, roughly corresponding to a power-law spectrum with $\Gamma = 1.8$. 
This very crude approximation allows us to obtain a first-order estimate of the
intrinsic luminosity of the two Compton-thick candidates, and therefore to have
the soft- and hard-band luminosities for all the 13 protocluster members.

Finally, we need to include the sky coverage correction in the 
luminosity function.  We collected the weight factor corresponding to the 
soft- and hard-band fluxes of each source,  which is similar to our method used to 
compute the number counts. We then considered the smallest weight for 
each source, computed according to Equation 2.  
Given the relatively high luminosities of the protocluster members, 
we only have a small ($\sim 1.15$) correction for the two faintest sources (XID 12 and XID 34). 
For the sake of clarity, we note that their S/N in the corresponding detection bands 
is $\sim 3$, which is far enough from the detection limit and therefore corresponds
to a very mild sky coverage correction.

\begin{figure*}
\begin{center}
\includegraphics[width=0.49\textwidth]{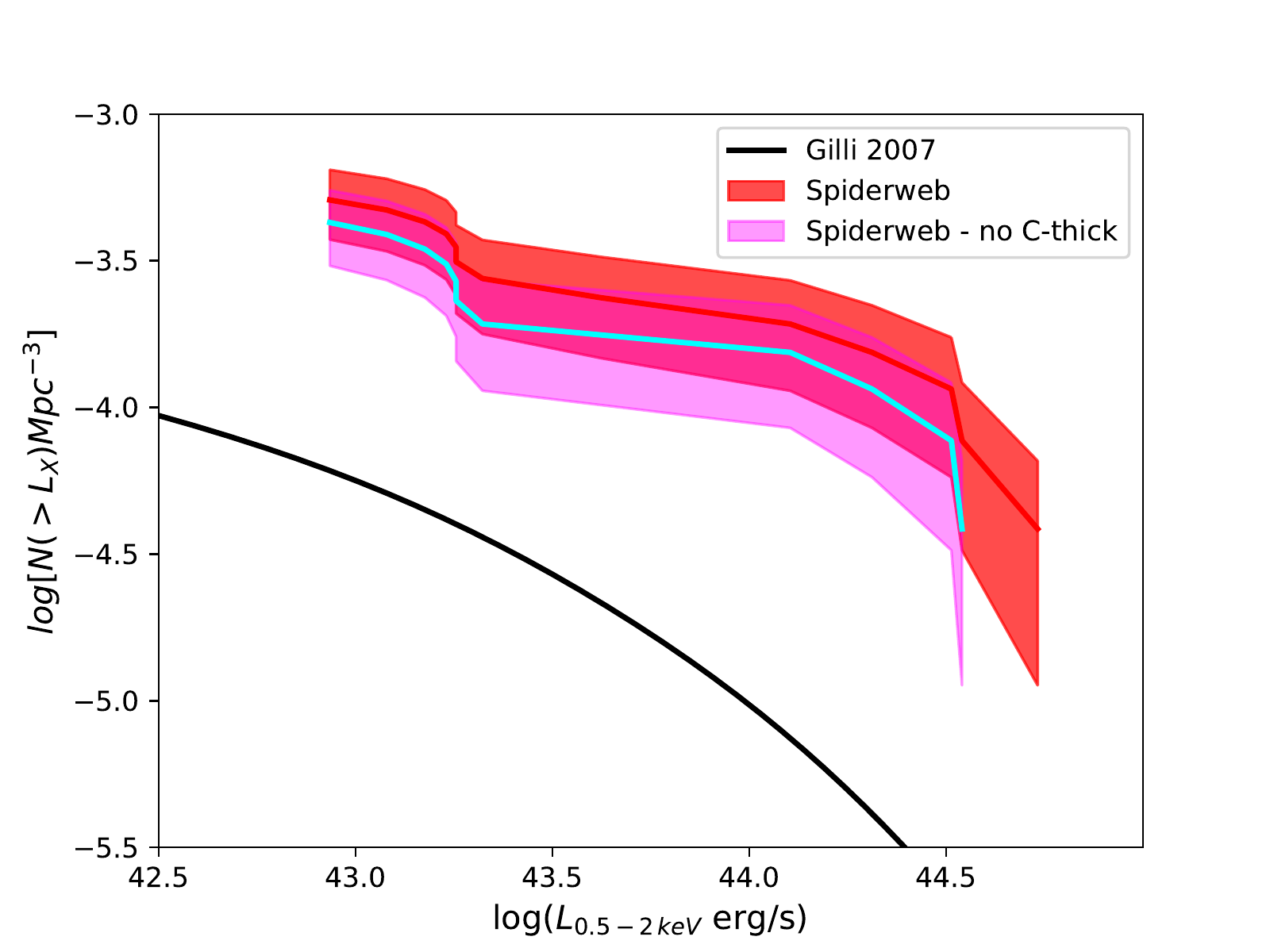}
\includegraphics[width=0.49\textwidth]{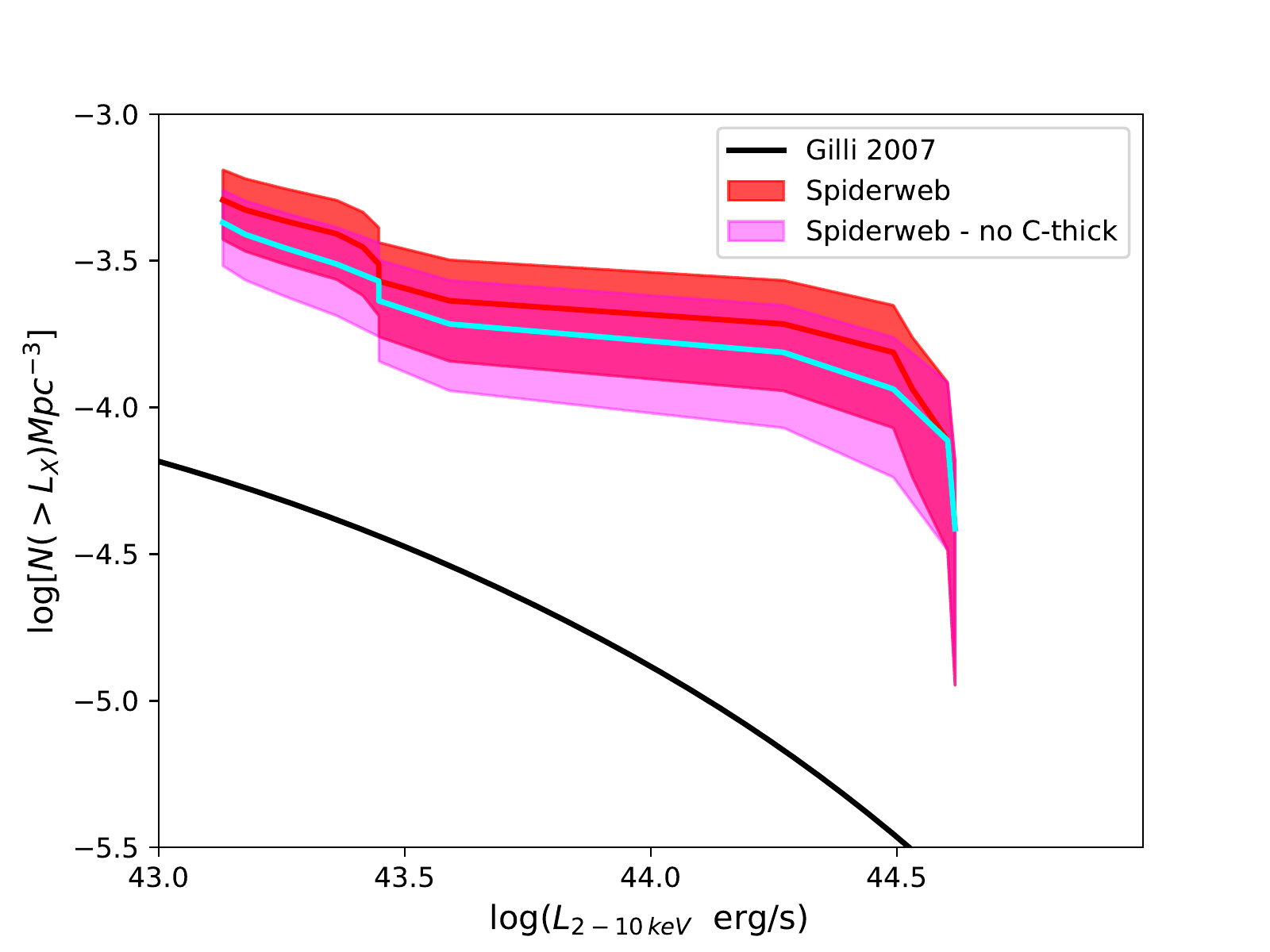}
\caption{ AGN luminosity functions.  Left panel: Cumulative rest-frame, soft-band 
luminosity function of the AGN in the Spiderweb protocluster normalized to the 
comoving volume defined by $2.11<z<2.20$ shown by the red, solid line with
the red-shadowed area, and a circle of 5 arcmin centered 
on the Spiderweb Galaxy. The cyan solid line with the pink-shadowed area 
is the same without the luminosity correction for the 
two C-thick candidates.  The black solid line is the XLF in the field from the model of 
\citet{2007Gilli}.  Shaded areas correspond to 
$1\, \sigma$ uncertainty.  The Spiderweb Galaxy is not included in the luminosity function.  
Right panel: Same as left panel but in the rest-frame, 2-10 keV hard band.
}
\label{NGTLX}
\end{center}
\end{figure*}

\begin{figure}
\begin{center}
\includegraphics[width=0.49\textwidth]{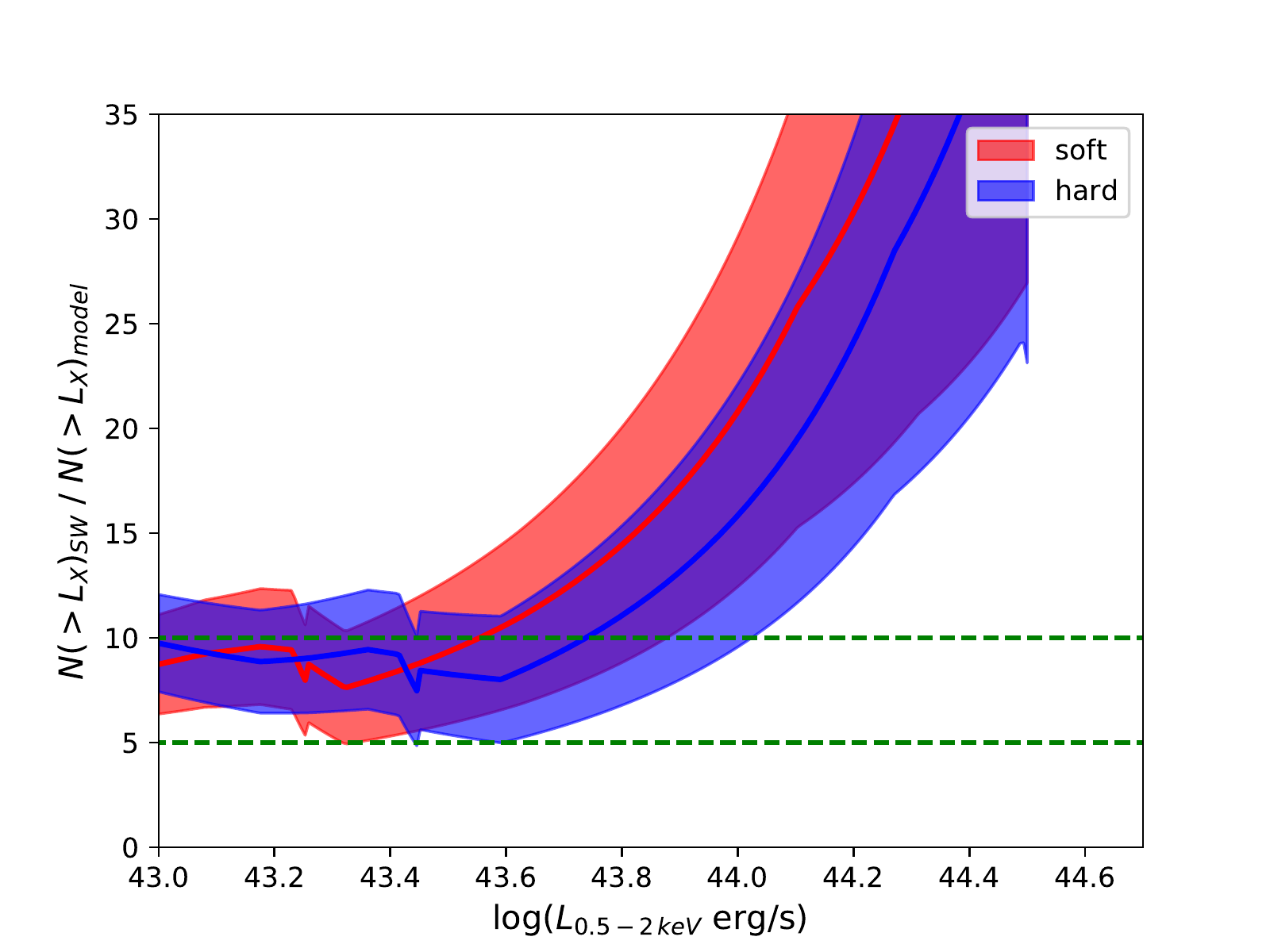}
\caption{Ratio of the soft and hard band X-ray AGN luminosity functions in the Spiderweb protocluster
to the field luminosity function from the model by \citet{2007Gilli}.
}
\label{NGTLX_ratio}
\end{center}
\end{figure}

In Figure \ref{NGTLX} we show the cumulative rest-frame luminosity in the soft and hard band 
after excluding the Spiderweb Galaxy.  In both cases we show this luminosity with and without the
two Compton-thick candidates, given the large uncertainties in their estimated 
luminosities.  We also plot the cumulative luminosity 
function predicted at $z\sim 2.2$ in the model of \citet{2007Gilli}.  As expected, 
the excess measured in the Spiderweb protocluster volume is consistent 
with the excess previously measured in the number counts.  The excess is always 
about an order of magnitude
at the lowest end and increases with luminosity, reaching a
factor of 20 and beyond at $L_X\sim 10^{44}$ erg/s in both bands. 
This is also clear in Figure \ref{NGTLX_ratio}, where we show the ratio of the X-ray 
AGN luminosity function in 
the Spiderweb protocluster to the field luminosity function in the field from the 
model by \citet{2007Gilli}.  
To compute the ratio we include the Compton-thick candidates. 
Removing them will result in a lower excess, decreased by a factor of 
$\sim 1.5$, as shown in Figure \ref{NGTLX}.
As discussed above, this excess is due to the
combination of the average galaxy overdensity in the protocluster
region and the enhancement factor. In Section 9, we provide an 
estimate of the enhancement factor alone.

\section{AGN fraction in the Spiderweb protocluster\label{fraction}}

The measurement of the AGN fraction as a function of the environment is an important
observable that can provide insight into the SMBH feeding mechanisms.  Several studies in the
X-ray band have been performed for relatively low redshift in massive systems.  Typically, 
the AGN fraction among cluster members increases with distance from the center, and 
reaches values comparable with the field (of the order of $\sim 1$\%) when considering
the full X-ray luminosity range ($10^{41}$-$10^{44}$ erg/s), 
or significantly lower than the field when focusing on luminosities $>10^{43}$ erg/s
\citep[see][and discussion therein]{2020Puccetti}.
Overall, the X-ray nuclear emission in cluster galaxies (with the noticeable exception of 
the central galaxy) is expected to be triggered by interactions \citep[see][]{2015Ehlert}, 
or by ram pressure in the outskirts \citep{2017Poggianti}, while several other mechanism
are contributing in the field.  The situation is expected to be significantly
different in protoclusters, where the lower density of galaxies, the different
velocity distribution, and the much larger availability of cold gas, may 
contribute to produce an AGN fraction that is significantly larger than in the field.  
% Despite the AGN fraction is 
% significantly higher also in the field at the typical 
% protocluster redshift ($z\sim 2$), 
% GENERAL TEXT, TOO GENERAL FOR THIS SECTION
% We recall that protoclusters here are defined to be the progenitors of today massive 
% clusters.  They differ from more sparse large-scale structures that may 
% eventually evolve into smaller mass halos ($M<10^{14}M_\odot$). 
% However, in simulations, it is not efficient identifying
% protoclusters just looking at core overdensity, since many ``protocluster
% candidates'' are observed to evolve in groups or small clusters. 
% In the introduction we already commented on the difficulty of defining a protocluster
% in a consistent way.  Clearly, a robust classification is key to investigate the 
% evolution of star formation (SF) and AGN in protocluster.  The risk is to consider a mixed bag of large-scale
% structures, with different dynamical status, and to obtain a highly noisy measurement of the
% activity in the member galaxies.  
Here we focus on the AGN fraction, and try to compare
the results we obtained in the Spiderweb Complex with the literature, 
focusing on the best-studied protoclusters.  

The first aspect to notice is that 
the presence of activity (star formation and/or nuclear) in galaxies has some 
dependence on the environment and on the cosmic epoch at the same time.  
The two dependencies cannot 
be easily decoupled, because the main characteristics of dense environments 
change with epoch as well.  Locally, it is well established that star formation 
and nuclear activity ---at least at the bright end--- are suppressed in 
dense environments such as massive, virialized halos. At the same time, the AGN fraction 
seems to increase in cluster progenitors at redshift $z\geq 2$ 
\citep{2013Martini,2016Alberts,2021Kalita}, 
while several studies have found intense star formation in at 
least some cluster cores at $z\geq 1.5$ \citep{2010Tran,2013Brodwin,2014Santos,2015Santos,2018Coogan}, 
suggesting an inversion of the SF-density anticorrelation observed 
in local clusters. An important aspect is the
large diversity among the cluster population in the range $1.5<z<2.0$, with some of them
hosting passive evolved galaxies, and some a rapidly evolving active population. 

This is consistent with an increase (possibly a co-evolution) of SF and AGN in massive
halos in the range $1.5<z<2.0$ when approaching their formation epoch.  
At present, we do not have a comprehensive understanding of how
the connection between AGN activity and large-scale galaxy environment
proceeds with redshift.  This connection may be particularly effective 
when entering the regime of protoclusters. 
The deep {\sl Chandra} observation of the Spiderweb Complex presented in this work 
gives us the possibility to measure the X-ray AGN fraction in a protocluster environment, 
and the X-ray AGN enhancement factor with respect to the field at the same redshift.  
% We note that we include only X-ray detected AGN, and exclude the (few)
% cases when AGN are identified in other wavebands (thanks to high radio power or 
% optical high ionization emission lines).
Clearly, the member galaxy selection is key in this case, because protocluster
members are usually picked up from a selection of Ly$\alpha$ emitters, or
from color--color-selected subsamples, potentially introducing a bias associated to galaxy type. Here, a convenient parameter is the stellar mass. 

We recall that the fraction of X-ray AGN among Ly$\alpha$ emitters is usually low, both
in the field and in protoclusters \citep[see, e.g.,][]{2003Malhotra,2016Zheng}.  However, 
several works have claimed that there is enhanced nuclear activity in massive protoclusters. 
Apart from early claims in the Spiderweb based on two spectroscopically 
confirmed X-ray-detected AGN and three X-ray member candidates \citep[][]{2002Pentericci}, 
a high AGN fraction at the level of $15-20$\% has been found in 
2QZ \citep[][]{2013Lehmer} at $z\sim 3$, DRC at $z\sim 4$ \citep[][]{2020Vito}, 
% Recently, similar behaviour has been 
% observed at $z\sim 4$ by \citet{2020Vito}, where two X-ray emitting members have been 
% found to be the most gas-rich in the structure.
and at lower redshift in Cl0218.3-0510 at $z=1.62$ \citep{2017Krishnan}.
Recently, AGN identified through optical spectroscopy of X-ray data in the 
{\sl Planck}-selected protocluster at $z\sim 2.16$ have been shown to
amount to $(20\pm 10)$\% of all the identified protocluster members 
\citep{2021Polletta}.
In all cases, the limited statistics make the uncertainty on the 
AGN fraction significantly larger than $30\%-50$\%.
On the other hand, other protoclusters in the range $2<z<3$ appear to have
a significantly lower fraction, implying an enhancement factor consistent
with unity \citep{2009Lehmer,2010Digby-North,2019Macuga}.
The fraction of X-ray-emitting AGN in protoclusters 
measured so far  in the literature is shown in Table \ref{AGNfraction}.  

% VERSION WITH ENHANCEMENT FACTOR
% \begin{table*}
% \caption{The AGN fraction, and AGN enhancement with respect to the field, 
% measured for a sample of protoclusters from the literature, and compared to the Spiderweb
% Protocluster. The AGN enhancement factor quoted in this Table is derived assuming
% $^*$ The fraction in \citet{2020Vito} refers to the fraction of X-ray sources among sub-mm galaxies.}
% \label{AGNfraction}
% \begin{center}
% \begin{tabular}[width=0.5\textwidth]{lcccc}
% \hline
% Protocluster & z & X-ray AGN fraction & AGN enhancement factor  & Ref\\
% \hline
% Cl0218.3-0510 & 1.62 & $17^{+6}_{-5} $ & $ 2.1\pm 0.7$ & \citet{2017Krishnan} \\
% Spiderweb & 2.156 & $ 14.5\pm 2.0$\% & $6.4^{+7.4}_{-3.6}$   & This work\\
% 2QZ       & 2.23  & $17^{+16}_{-9}$\% &       $3.5^{+3.8}_{-2.2}$ & \citet{2013Lehmer} \\
% HS1700  & 2.3 &  $2.9^{+2.9}_{-1.6}$\% &      $\sim 1$ & \citet{2010Digby-North} \\
% USS 1558-003    & 2.53 & $2.0^{+2.6}_{-1.3}$\% &      $\sim 1$ & \citet{2019Macuga} \\
% SSA22       & 3.09  & $5.1^{+6.8}_{-3.3}$\% &         $6^{+10.3}_{-3.6}$ & \citet{2009Lehmer} \\
% DRC & 4.002 & $15^{+20}_{-10}$\% & $\sim 1$ & \citet[][]{2020Vito}\\
% \hline
% \end{tabular}
% \end{center}
% \end{table*}

% VERSION WITHOUT ENHANCEMENT FACTOR
\begin{table*}
\caption{X-ray AGN fraction in a sample of protoclusters}
% from the literature, and compared to the Spiderweb Protocluster. 
\label{AGNfraction}
\begin{center}
\begin{tabular}[width=0.5\textwidth]{lccc}
\hline
Protocluster & z & X-ray AGN fraction &  Ref\\
\hline
Cl0218.3-0510 & 1.62 & $17^{+6}_{-5} $\% &  \citet{2017Krishnan} \\
Spiderweb & 2.156 & $ 25.5\pm 4.5$\%   & This work\\
PHz G237.01 & 2.16  &  $20\pm 10$\%  & \citet{2021Polletta}\\
2QZ       & 2.23  & $17^{+16}_{-9}$\%  & \citet{2013Lehmer} \\
HS1700  & 2.3 &  $2.9^{+2.9}_{-1.6}$\% & \citet{2010Digby-North} \\
USS 1558-003    & 2.53 & $2.0^{+2.6}_{-1.3}$\% & \citet{2019Macuga} \\
SSA22       & 3.09  & $5.1^{+6.8}_{-3.3}$\% &  \citet{2009Lehmer} \\
DRC & 4.002 & $15^{+20}_{-10}$\% &  \citet[][]{2020Vito}\\
\hline
\end{tabular}
\end{center}
\end{table*}

\begin{figure}
\begin{center}
\includegraphics[width=0.49\textwidth]{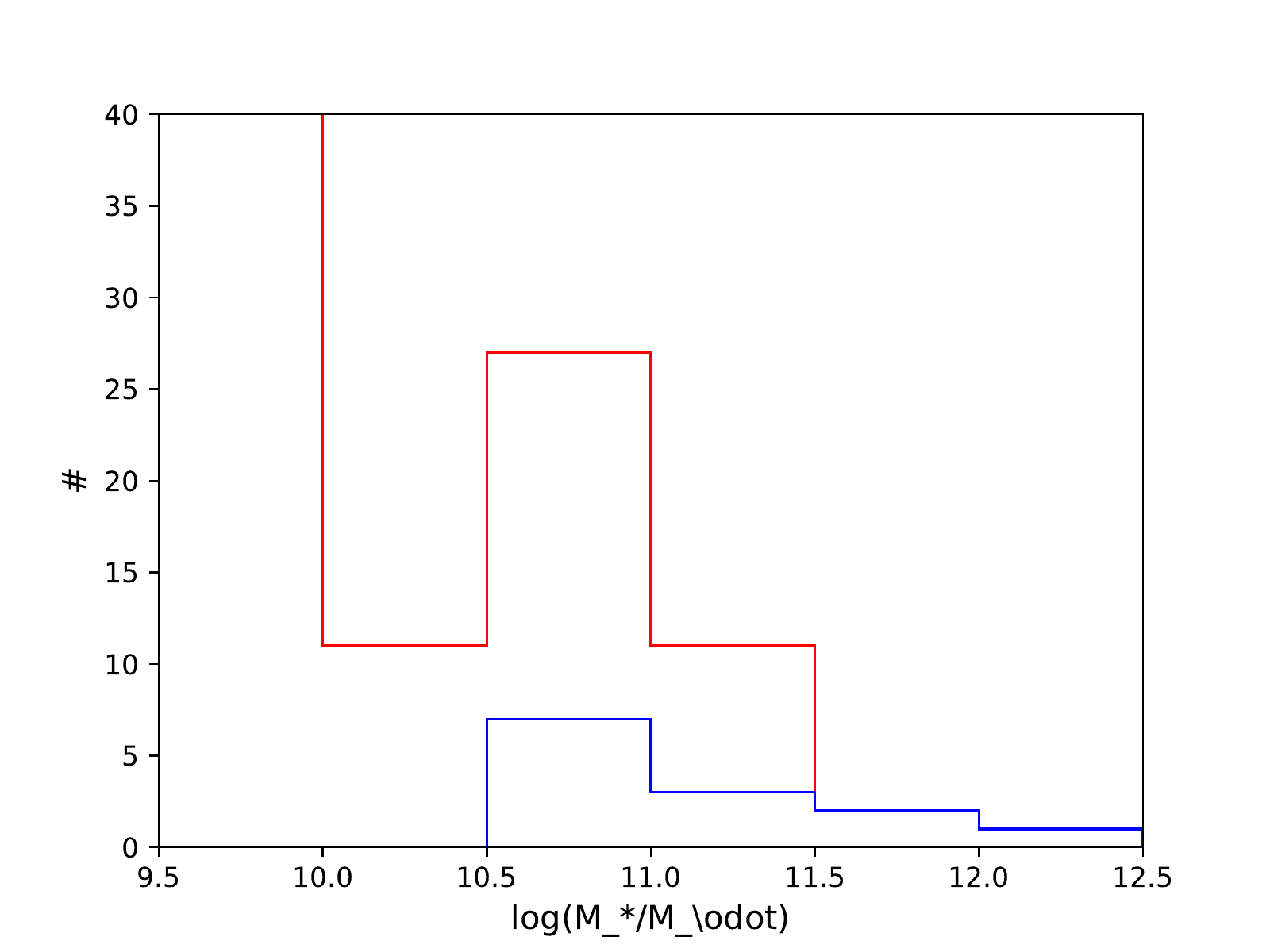}
\caption{Distribution of $M_*$ among the protocluster members in the 
redshift range $2.11<z<2.20$ (red histogram)
and among the X-ray-emitting members (blue histogram).  The first bin ($9.5<{\rm log}(M_*)<10.0$)
is an upper limit where we include all the sources that do not have a $K_{\rm S}$-band counterpart in our image. }
\label{mstar}
\end{center}
\end{figure}
 
To measure the AGN fraction in the Spiderweb protocluster, we need to control the selection 
in stellar mass. As detailed in Section 3, we estimated stellar masses $M_*$
for about 80\% of the sample.  Among the 96 protocluster members ($2.11<z<2.20$), we have a
robust measurement of $M_*$ for 62 of them, while 34 are not detected.  
We assume a nominal upper limit of ${\rm log}(M_*/{\rm M}_\odot) = 10$ 
(which is irrelevant in this work).  We note that most of the sources without a $K_{\rm S}$ counterpart 
are from the COALAS survey.  In fact, from Table 3 of \citet{2021Jin}, we notice that
half of them are not detected in the $K_{\rm S}$ band, while 90\% of them are detected in the $I$ band, 
and therefore these sources should correspond to relatively low-mass galaxies. 
% We list them in the mass bin $9.5<{\rm log}(M_*)<10$ for completeness.

The distribution of $M_*$ in the protocluster is shown in Figure \ref{mstar}.  
We note that we include all the sources that do not have a 
$K_{\rm S}$ band counterpart in our image in the first bin $9.5<{\rm log}(M_*)<10.0$. This is clearly 
an oversimplification but has no effect on our results.
Ideally, we may compute 
an AGN fraction as a function of $M_*$, but given the low statistics, we simply consider 
all the sources with ${\rm log}(M_*/{\rm M}_\odot)>10.5$. 
For completeness, we briefly discuss preliminary results on the distribution of protocluster
sources in the $L_X-M_*$ plane in Section \ref{LXMstar}.
Finally, we reiterate that, apart from  the X-ray-detected sources, there are no protocluster members that have been identified
as AGN from the optical and NIR spectrum.
Therefore, excluding the Spiderweb Galaxy, we have 13 X-ray AGN (11 excluding the source
with a photometric counterpart and the one with a double counterpart) over a total of 
44 sources for an X-ray AGN fraction with $L_{2-10 keV}>10^{43}$ erg/s of $29.5$\% ($25$\%).
To bracket this value, we consider a maximum of $\sim 28$ additional members, 
two with X-ray emission, as discussed in Section 6.2.  Their inclusion would bring the 
AGN fraction to $21$\%.  Considering this as a lower limit, we can estimate the
AGN fraction in the Spiderweb protocluster as $25.5\pm 4.5$\%, where the error bar also accounts for
the maximum systematic uncertainty.
We stress that the uncertainty we quote here, based on the 
photometric study of Pannella et al. (in preparation), also includes
the effect of the different area coverage between X-ray data
and spectroscopic or color/emission line selection. A detailed treatment of the effects
of the protocluster member selection within 5 arcmin would imply a significant
additional effort and would only mildly reduce the uncertainty we have estimated.

\begin{figure}
\begin{center}
\includegraphics[width=0.49\textwidth]{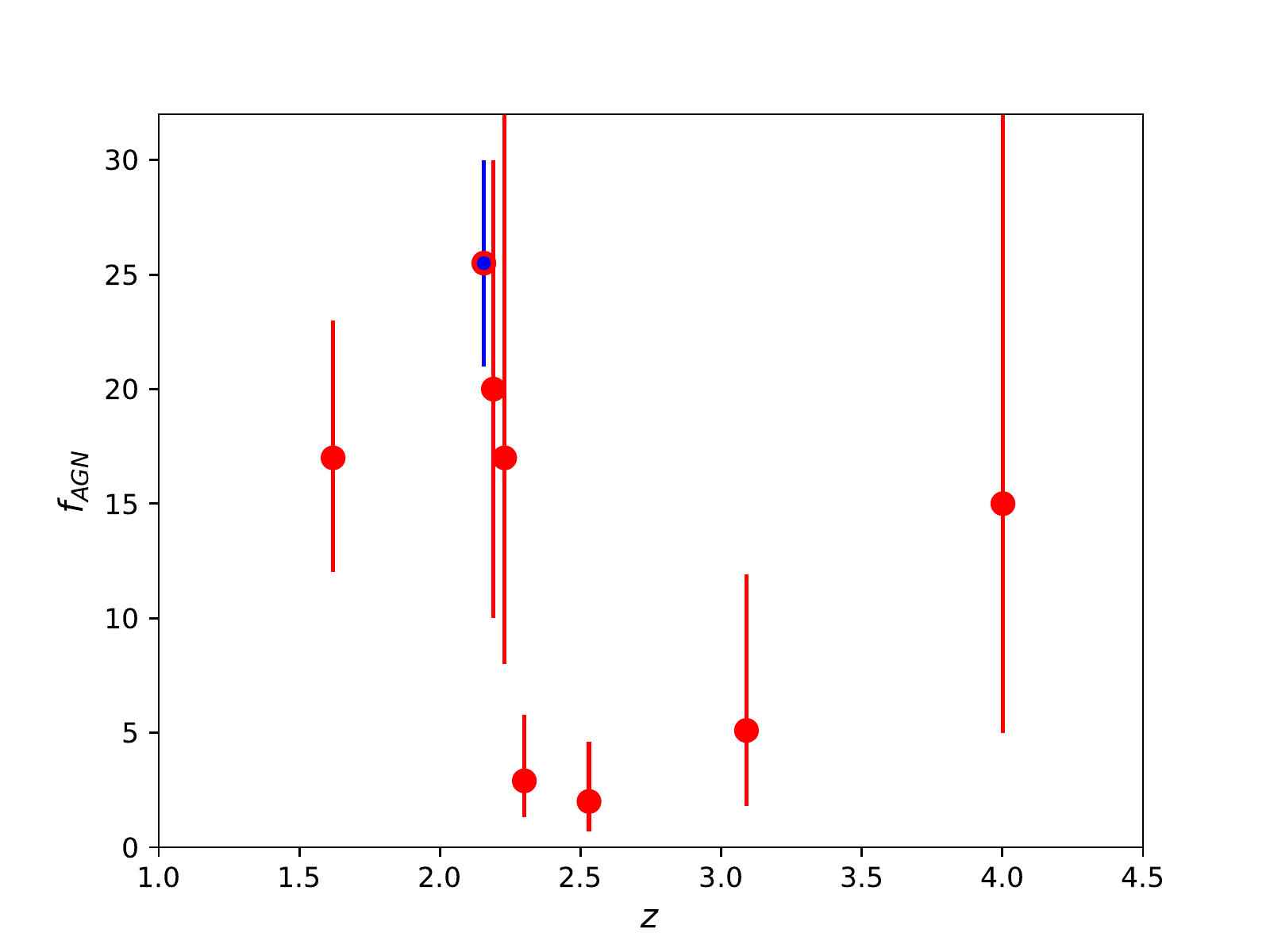}
\caption{Fraction of X-ray AGN over the total number of identified protocluster members
from Table \ref{AGNfraction}.  The Spiderweb Protocluster
is marked with a blue dot.  The plot is meant to show measurements of the AGN fraction 
available in the literature, and necessarily includes measurements for different populations and 
different stellar mass ranges.}
\label{AGN_fraction}
\end{center}
\end{figure}

As shown in Table \ref{AGNfraction}, this value is somewhat higher but 
consistent with the measurement from \citet{2017Krishnan} at $z=1.62$ 
and to the highest values found at $2<z<4$ by \citet[][2QZ]{2013Lehmer}, 
and by \citet[][DRC]{2020Vito}\footnote{We note that \citet[][]{2020Vito} considers only 
X-ray selection among submm-detected galaxies, revising for example the value 
found in SSA22 to $50^{+39}_{-24}$\%.}.  
Three different works instead find values in the range of 2\%--5\%, within an
uncertainty of a factor of 2.  We plot these values in Figure \ref{AGN_fraction}, 
where the measurement in this work is marked with a blue dot.  
We present this plot simply to compare the different values obtained in the 
literature so far, with no attempt to explore a possible behavior with 
redshift. The main problem with the measurement of the evolution of the AGN fraction in protoclusters
is that these values were obtained from data
with different X-ray depth and different selection for the protocluster membership; 
the first aspect is not an issue, because all the medium-deep {\sl Chandra} data used 
in these works reach luminosity levels of 
$\sim 10^{43}$ erg/s in both bands. On the other hand, 
we lack an homogeneous and complete selection of the protocluster members in the
optical.  As of today, the only robust constraint on the evolution of the X-ray AGN fraction
is a clear increase from the low local values at the level of $1$\% to $>10$\% 
in a few protoclusters at $z>2$. Similar but slightly larger uncertainties 
are found for the enhancement factor, which is discussed in the following section. 

% In other words, the selection of
% protocluster regions should be assessed on a quantitative measurement that
% is not available for most of the protoclusters studied so far. 

\section{AGN enhancement factor in the Spiderweb protocluster\label{enhancement}}

To compare the AGN fraction in the Spiderweb protocluster with that in the field, we refer to the few works in the literature
that provide an estimate of this latter, key quantity. \citet{2009Lehmer} estimate an average 
fraction of $1.9^{+2.6}_{-1.3}$\% and $0.7^{+1.6}_{-0.6}$\% from $z \sim 2 - 3.4$ Lyman 
Break Galaxies and $z = 3.1$ Ly$\alpha$ emitters in the CDF + SSA22 field, respectively, 
for ${\rm log}(M_*/M_\odot)>10.5$.  These values were 
obtained for $\log(L_{8-32 keV})>43.5$, which, for a power law with $\Gamma = 1.8$, corresponds
to $\log(L_{0.5-10 keV})>43.66$. Only seven sources satisfy this threshold, or nine if we
consider the Compton-thick candidates and assume that the hard-band intrinsic luminosity 
is ten times the observed one. Therefore, the fraction of 
AGN with $\log(L_{0.5-10 keV})>43.66$ in our spectroscopic sample is  $18.0\pm 2.0$\%.
In this case, we find the enhancement factor for $\log(L_{0.5-10 keV})>43.66$ in the Spiderweb 
protocluster to be $f_{\rm enh}=9.5^{+16}_{-4.0}$ or $f_{\rm enh}=26^{+100}_{-16}$, depending on the assumed field
value.

Another estimate of the field X-ray AGN fraction is provided by \citet{2013Lehmer}, 
who use an independent sample of $z = 2.216 - 2.248$ 
HAEs  from the High Redshift Emission Line Survey (HiZELS)
in the wide-area $\sim 1.6 $ deg$^2$ COSMOS survey field \citep{2013Sobral}.
\citet{2013Lehmer} found 10 {\sl Chandra} sources matched to the 210 HiZELS HAEs 
within the C-COSMOS footprint, giving an AGN fraction
in the low-density environment of  $3.0^{+4.0}_{-2.0}$\%.  
The COSMOS flux limit of $1.3 \times 10^{-15}$ erg/s/cm$^2$
in the 0.5–10 keV \citep[see][]{2016Marchesi}  corresponds to $L_{0.5-10 keV}\sim 4\times 
10^{43}$ erg/s.  This limit is equivalent to that previously used for a comparison
with LBG and HAE sources in the CDF + SSA22 field.  In this case, the enhancement factor
is found to be $6.0^{+9.0}_{-3.0}$.

Previous estimates of the enhancement factor range from $ 2.1\pm 0.7$ \citep{2017Krishnan} 
to $3.5^{+3.8}_{-2.2}$ \citep{2013Lehmer} and $6^{+10.3}_{-3.6}$ \citep{2009Lehmer}, in 
broad agreement with our results.  However, as mentioned above,  there are works
that measure substantially lower
X-ray AGN fractions in protoclusters, concluding that there is no enhancement
with respect to the field \citep[][]{2010Digby-North,2019Macuga}.  This difference may simply
reflect the diversity among protoclusters and their different dynamical stage, as expected 
in such a rapidly evolving phase.

% Another estimate can be derived from \citet{2018Aird}, where the fraction of all the galaxies
% hosting an AGN and with $3\times 10^{10}< M_*< 3\times 10^{11}M_\odot$ 
% is about $20\pm 5$ \% at $z\sim 2$ (see their Figure 6, left panel). This 
% fraction is comparable to what we find in the Spiderweb
% Protocluster for $L_{2-10 keV}>2\times 10^{43}$ erg/s.  However, 
% the results found by \citet{2018Aird} is valid for
% $\lambda_{sBHAR}>0.01$, which corresponds to $L_{2-10 keV}\sim  10^{42}$ erg/s, 
% a factor of 20 below our luminosity lower bound.  It is not clear
% how to correct the measurement in \citet{2018Aird} for this. 
% BE CAREFUL, THIS IS TRUE FOR 10^10 MASS INTERVAL - TO BE CHECKED. 
% ALSO Haggard PROVIDE SIMILAR INFO, BUT AT LOWER REDSHIFT

In all these cases, the uncertainties on the AGN fraction in the field clearly 
dominate the error budget.  Nevertheless, we find that we always obtain an 
enhancement factor in the range 5-20 for X-ray AGN 
with $L_{0.5-10 keV}> {\rm few}\times 10^{43}$ erg/s and ${\rm log}(M_*/{\rm M}_\odot)>10.5$. 
On the other hand, it is well known that the enhancement factor drops below 1 in all the 
virialized halos at relatively low redshift ($z<1$), except in the cluster
outskirts \citep{2019Koulouridis}.  Therefore, we confirm the existence 
of a strong evolution of the X-ray AGN enhancement factor in the central regions
of clusters and protoclusters with redshift. 
However, we do expect to also see a strong dependence on the environment, and therefore
a large diversity among the observed protoclusters, as probably suggested 
by the different conclusions on the enhancement factor at $z>2$
presented in the literature.  Only by disentangling the
complex pattern due to the combination of environmental effects and cosmic
evolution can we hope to reach a comprehensive picture of the galaxy population 
evolution in clusters.  

\section{Future perspectives\label{perspectives}}

There are several promising approaches to extending this work in order to gain further insight
into the evolution of nuclear activity in protoclusters and connect it to the 
dearth of activity in virialized, mature clusters.
In addition to the difficult characterization of the protocluster candidates already discussed 
in Sect. 1, the first obvious key aspect is the selection of 
the protocluster members, which largely differ among the few cases presented in the literature.  
Ideally, the protocluster membership should have a well-defined completeness
as a function of stellar mass and distance from the center.  Stellar
mass should be used to compute AGN fraction above a given mass limit (typically, $M_* >
{\rm few}\times 10^{10}M_\odot$).  
%  The original  source of uncertainty is the size of the 
% protocluster.  Ideally, protoclusters should be classified on the basis of the estimated
% virial mass that they would achieve at $z\sim 0$.   Again, a deep census in stellar mass of the
% protocluster members is key to measure an overdensity profile, that could be used
% to classify the size and the dynamical phase of each protocluster.
The information on the stellar mass of the active galaxies may also provide
clues as to the physical mechanisms that trigger the activity. 
\citet[][]{2017Krishnan} explore the distribution of the stellar mass of the 
X-ray-emitting AGN in the field and in protoclusters, and find no differences.  
These authors suggest that triggering mechanisms are similar in both environments and that the
higher X-ray AGN fraction in protoclusters is simply due to the same mechanism occurring 
more frequently because of the denser environment.  
However,  \citet{2021Monson} recently speculated that the enhanced AGN fraction in the protocluster
SSA22 at $z=3.1$ is due to the larger masses (a factor of 2) of the LBG with respect to the 
field rather than an enhanced merger activity. 
This issue is also relevant for the process quenching the rapidly star forming galaxies that
are observed as quiescent galaxies in the protocluster, forming
the mentioned nascent red sequence.  Similarly, a census of the nuclear activity 
in the radio band is needed to complete the picture offered by the X-ray data and 
provide robust constraints on the feedback processes in the member galaxies.

Focusing on the Spiderweb protocluster, we foresee the following next steps.  First, 
an extension of the protocluster membership based on NIR images and eventual 
spectroscopic follow-up, to achieve a well-defined and complete stellar-mass function. 
Then, a detailed investigation of the archival high-resolution HST data, that, coupled 
to the submm data, may provide clues as to the role of interactions and/or the presence of 
cold gas in triggering nuclear activity.  Finally, we plan to exploit new radio 
data in the L band with JVLA and GMRT to characterize the protocluster population 
in this band. 

Such efforts would make the Spiderweb an ideal science case for the 
study of protoclusters.  In this respect, the role of X-ray data with high-angular
resolution will be key for the extension of this approach to a large sample.  Currently, 
{\sl Chandra} is the only facility that can provide data of the required quality. 
The future of this science case heavily depends on the future of high-resolution
X-ray astronomy, which is currently embodied in the Lynx \citep{2018LynxTeam} and AXIS \citep{2019Mushotzky}
mission concepts.

\section{Conclusions\label{conclusions}}

We present the deep (715 ks) X-ray observation of the 
radio galaxy J1140-2629 (the Spiderweb Galaxy) and the
surrounding large-scale structure with {\sl Chandra}.  The deep X-ray data allow us to 
map the nuclear activity in a dense environment down to the low-luminosity end of the Seyfert galaxy population, and to test whether 
% a phase of vigorous star 
% formation and baryon assembly coexists with a black-hole accretion phase, 
% and whether 
dense large-scale structure environments enhance 
X-ray nuclear activity, a phenomenon that has been investigated only recently at 
such a high redshift and only with shallower X-ray data.  In addition, 
the sensitivity and exquisite angular resolution of {\sl Chandra} allow us to 
characterize the extended emission around the Spiderweb Galaxy and possibly identify 
thermal X-ray emission from the ICM, a key piece of information 
with which to constrain both the dynamical state of the protocluster and the feedback 
mechanism at work in the Spiderweb complex, and  ultimately, to provide an 
overview of the baryon cycling in such 
an active environment at $z\sim 2$.  
% and between the assembly of dark matter halos and the formation of 
% cD galaxies, at a cosmic epoch corresponding to the peak of activity of star 
% formation and of accretion onto SMBH, to measure the X-ray luminosity function 
% in the protocluster region,
In this first paper, we focus on the nuclear 
activity of the galaxies in the Spiderweb protocluster, 
exploiting the combination of our deep X-ray data in the field 
and the large body of multi-wavelength  and spectroscopic data available
from the literature. The main results and the data products used in this work 
are made available on the project webpage\footnote{\tt http://www.arcetri.inaf.it/spiderweb/}.
Our results can be summarized as follows:

\begin{itemize}

\item We identify 107 X-ray sources within a radius of 5 arcmin 
centered on the Spiderweb Galaxy, down to flux limits of $1.3\times 10^{-16}$
erg/s/cm$^2$ and $3.9\times 10^{-16}$ erg/s/cm$^2$ in the soft and hard band, respectively.
We compute the cumulative number counts which are found to be
consistent with results from X-ray deep fields in the literature.

\item We identify 13 X-ray sources with spectroscopically confirmed counterparts
belonging to the Spiderweb Complex
(including the Spiderweb Galaxy itself) and 1 with a color-selected member candidate.
As these sources are found in the redshift range $2.11<z<2.20$, we classify them 
as protocluster members (as opposed to member of the Spiderweb Complex, which 
includes a larger redshift window $2.0<z<2.3$).  The X-ray-emitting protocluster 
members are distributed over an approximately 
rectangular region with $\sim 3.2$ and $1.3$ Mpc by side.

\item The X-ray spectral analysis finds five strongly absorbed AGN 
($N_H>10^{23}$ cm$^{-2}$), two sources with flat spectra, 
which we classify as Compton-thick candidates with reflection-dominated spectrum, 
and seven sources with moderate or low absorption.  
After correcting for Galactic and intrinsic absorption, and with an educated
guess for the intrinsic luminosity of the two Compton-thick candidates, the
rest-frame 0.5-10 keV luminosities are found to be above $ 10^{43}$ erg/s for 
all the sources. By combining intrinsic absorption and intrinsic 
luminosities, we conclude that all the X-ray-emitting protocluster members are 
powered by nuclear activity.  We do not find hints of a change of
the X-ray properties of the protocluster members with the distance from the 
Spiderweb Galaxy.

\item We stack the images of all the X-ray silent, spectroscopically confirmed members, 
finding a marginal detection corresponding to average luminosities below $\sim 
10^{41}$ erg/s, showing that there is no significant nuclear activity 
at about one order of magnitude below our detection threshold.  A slightly stronger detection is 
obtained for the stacked emission of the color or narrow-band selected members, 
consistent with being powered by star formation at the level of 
$30-50$~M$_\odot$/year, but the unknown contamination 
fraction prevents us from deriving robust conclusions on their nature.

\item The X-ray luminosity function of AGN in the volume associated to the 
Spiderweb protocluster is more than an order of magnitude larger than 
in the field at the same redshift in the range $10^{43}<L_X<10^{44}$ erg/s in 
both bands, and 
significantly flatter, implying an increasing excess at the bright end.

\item For $L_{2-10 keV}>10^{43}$ erg/s, we measure an X-ray AGN fraction 
of $25.5\pm 4.5$\% among galaxies with 
${\rm log}(M_*/M_\odot)>10.5$, which is higher but still consistent 
with the values found at least in four protoclusters at similar redshift. 

\item The measured AGN fraction in the Spiderweb protocluster at 
$L_{0.5-10 keV}> 4\times 10^{43}$ erg/s
corresponds to an enhancement factor with respect to the field 
that ranges from 5 to 20 (with a preferred value $f_{enh}=6.0^{+9.0}_{-3.0}$
with respect to the COSMOS field at comparable redshifts and stellar mass range), suggesting a 
significant environmental effect triggering nuclear activity in the Spiderweb protocluster.

\end{itemize}

Unfortunately, the decreasing sensitivity of {\sl Chandra} due to the 
contamination layer on the ACIS detector makes future studies of this kind 
very hard, and we can realistically foresee that only a handful of new 
targets will be studied with comparable detail in the future. 
Nevertheless, the confirmation of an enhanced X-ray AGN activity in protoclusters is an 
important ingredient with which to plan studies of high-z large-scale structure with 
future X-ray facilities.  By exploiting the available 
multiwavelength dataset on the Spiderweb field, we plan to further 
explore the properties of the X-ray protocluster members to investigate 
the main physical mechanism responsible for triggering the X-ray emission.

\begin{acknowledgements}
This work was carried out during the ongoing COVID-19 pandemic.
The authors would like to acknowledge the health workers
all over the world for their role in fighting in the frontline of this
crisis.  We thank the anonymous referee for detailed comments and
positive criticism that helped improving the quality of the paper. 
We thank Stefano Marchesi for useful discussions. 
P.T. and R.G. acknowledge financial contribution from the agreement 
ASI-INAF n.2017-14-H.0. M.N. acknowledges INAF-1.05.01.86.20.
S.B. acknowledges partial financial support from the Indark INFN Grant.
A.S. is supported by the FARE-MIUR grant
’ClustersXEuclid’ R165SBKTMA and INFN InDark grant. AS, MP, VS, LDM
are supported by the ERC-StG ‘ClustersXCosmo’ grant agreement 716762.
HD acknowledges financial support from the AEI-MCINN under grant 
% (La evolución de los cíumulos de galaxias desde el amanecer hasta el mediodía cósmico) with reference 
with reference PID2019-105776GB-I00/DOI:10.13039/501100011033, and from the ACIISI, 
Consejería de Economía, Conocimiento y Empleo del Gobierno de Canarias 
and the European Regional Development Fund (ERDF) under grant with reference PROID2020010107.
We thank Hans Moritz Günther for help with the use of the MARX software. We thank
Malgorzata Sobolewska for assistance during the Chandra observations.
This work is based in part on data collected at Subaru Telescope and obtained from the SMOKA, which is operated by the Astronomy Data Center, National Astronomical Observatory of Japan, \citep{2002Baba}.
This research has made use of the services of the ESO Science Archive Facility.

\end{acknowledgements}

\bibliographystyle{aa}
\bibliography{main}

\clearpage
\onecolumn

\begin{appendix}

\section{Source list}

In Table \ref{source_list} we list the properties of the 107 X-ray sources 
detected in the FOV included in a circle of 5 arcmin centered on the Spiderweb 
Galaxy.  Soft (0.5-2 keV) and hard (2-7 keV) net counts are obtained with aperture 
photometry.  Values have been rounded to the first decimal place except in a few cases. 
Negative values are sometimes obtained due to background fluctuations.  
Soft and hard (2-10 keV) band energy fluxes
are obtained multiplying the net counts, corrected for vignetting, by the 
corresponding conversion factors listed in Table \ref{cf}. 
The errors are obtained combining independently the statistical photometric error with the 
systematic uncertainty on the spectral shape (hence, on the conversion factors). 
In case the $S/N$ in a given band does not reach our threshold value $S/N>2$, 
we quote the $2\, \sigma$ upper limit to the energy flux. The original {\tt ASCII} file of
Table 1 can be retrieved from the webpage\footnote{\tt http://www.arcetri.inaf.it/spiderweb/}.

\vskip 1cm

% \begin{table}
% \centering
\begin{longtable}{lcccccc}
\caption{Catalog of the X-ray detected sources within 5 arcmin 
from the Spiderweb Galaxy. }
\\ \hline \\
XID & RA & DEC & Cts/s (soft) & Cts/s (hard) & $F_S$  & $F_H$    \\
   &    &     &              &              & $10^{-16}$ erg/s/cm$^2$ & $10^{-16}$  erg/s/cm$^2$   \\
   &    &     &              &              &   &   \\
\hline
   &    &     &              &              &   &   \\
   &    &     &              &              &   &   \\
1        &      175.18593        &      -26.54934        & $26.2 \pm 6.6$ & $164.0 \pm 13.8$ & $ 4.3 \pm 1.1 $ & $ 71.4 \pm 11.2$ \\
2        &      175.27884        &      -26.50376        & $174.8 \pm 13.9$ & $246.6 \pm 17.5$ & $ 34.7 \pm 3.8 $ & $ 121.2 \pm 18.1$ \\
3        &      175.24638        &      -26.50113        & $6.9 \pm 4.0$ & $19.4 \pm 6.1$ & $ <2.0 $ & $ 7.6 \pm 2.6$ \\
4        &      175.22588        &      -26.49542        & $220.5 \pm 15.2$ & $174.6 \pm 13.7$ & $ 29.3 \pm 3.0 $ & $ 69.2 \pm 10.6$ \\
5        &      175.20786        &      -26.49363        & $29.0 \pm 5.9$ & $28.1 \pm 6.0$ & $ 3.7 \pm 0.8 $ & $ 11.0 \pm 2.8$ \\
6        &      175.22015        &      -26.48645        & $241.1 \pm 15.8$ & $189.1 \pm 14.2$ & $ 31.9 \pm 3.2 $ & $ 75.1 \pm 11.4$ \\
7        &      175.19974        &      -26.48512        & $132.6 \pm 12.6$ & $235.3 \pm 16.0$ & $ 17.1 \pm 2.1 $ & $ 92.4 \pm 13.7$ \\
8        &      175.28694        &      -26.47671        & $19.0 \pm 5.0$ & $12.9 \pm 5.2$ & $ 4.4 \pm 1.2 $ & $ 7.6 \pm 3.2$ \\
9        &      175.23038        &      -26.47326        & $23.5 \pm 5.4$ & $12.2 \pm 4.7$ & $ 3.1 \pm 0.7 $ & $ 4.8 \pm 1.9$ \\
10       &      175.25893        &      -26.51378        & $18.5 \pm 5.2$ & $-1.7 \pm 4.3$ & $ 2.9 \pm 0.8 $ & $ <3.0$ \\
11       &      175.29327        &      -26.48855        & $7.3 \pm 4.7$ & $18.6 \pm 7.1$ & $ <3.8 $ & $ 10.4 \pm 4.2$ \\
12       &      175.22990        &      -26.47804        & $15.2 \pm 4.9$ & $9.8 \pm 5.2$ & $ 2.0 \pm 0.7 $ & $ 3.8 \pm 2.1$ \\
13$^{r3}$        &      175.28102        &      -26.44954        & $8.3 \pm 3.5$ & $-4.6 \pm 2.2$ & $ 1.9 \pm 0.8 $ & $ <1.0$ \\
14       &      175.13669        &      -26.53742        & $22.0 \pm 6.3$ & $6.7 \pm 6.1$ & $ 4.9 \pm 1.5 $ & $ <10.0$ \\
15       &      175.12944        &      -26.49295        & $16.6 \pm 5.6$ & $3.2 \pm 5.5$ & $ 2.5 \pm 0.9 $ & $ <6.0$ \\
16       &      175.23132        &      -26.54720        & $22.4 \pm 5.9$ & $38.1 \pm 8.2$ & $ 3.9 \pm 1.1 $ & $ 16.7 \pm 4.2$ \\
17$^{r1}$        &      175.18042        &      -26.53844        & $3.7 \pm 3.1$ & $11.5 \pm 4.8$ & $ <1.5 $ & $ 4.9 \pm 2.1$ \\
18       &      175.21500        &      -26.52062        & $5.3 \pm 3.7$ & $14.5 \pm 5.3$ & $ <1.7 $ & $ 5.7 \pm 2.2$ \\
19       &      175.17738        &      -26.51848        & $-2.4 \pm 2.1$ & $13.9 \pm 5.4$ & $ <0.3 $ & $ 5.6 \pm 2.3$ \\
20       &      175.11499        &      -26.48742        & $-3.7 \pm 3.8$ & $24.8 \pm 7.7$ & $ <0.8 $ & $ 15.2 \pm 5.1$ \\
21       &      175.21033        &      -26.55919        & $44.8 \pm 7.7$ & $56.6 \pm 9.1$ & $ 10.2 \pm 1.9 $ & $ 33.4 \pm 7.0$ \\
22       &      175.20374        &      -26.55444        & $141.0 \pm 12.5$ & $259.1 \pm 16.9$ & $ 25.2 \pm 2.9 $ & $ 118.5 \pm 17.4$ \\
23       &      175.18608        &      -26.54500        & $378.1 \pm 19.9$ & $263.3 \pm 17.1$ & $ 62.6 \pm 5.7 $ & $ 119.6 \pm 17.6$ \\
24       &      175.25270        &      -26.53977        & $83.4 \pm 9.8$ & $86.2 \pm 10.7$ & $ 15.2 \pm 2.1 $ & $ 39.6 \pm 7.2$ \\
25       &      175.26349        &      -26.53928        & $5.0 \pm 4.0$ & $25.6 \pm 7.6$ & $ <3.0 $ & $ 15.1 \pm 4.9$ \\
26$^{r1}$        &      175.16380        &      -26.53818        & $9.9 \pm 4.6$ & $1.8 \pm 4.5$ & $ 1.6 \pm 0.8 $ & $ <4.5$ \\
27       &      175.24773        &      -26.53220        & $619.8 \pm 25.4$ & $574.2 \pm 24.6$ & $ 109.5 \pm 9.3 $ & $ 265.3 \pm 36.7$ \\
28       &      175.24940        &      -26.53080        & $29.9 \pm 7.2$ & $32.2 \pm 7.5$ & $ 5.3 \pm 1.3 $ & $ 14.8 \pm 3.9$ \\
29       &      175.24535        &      -26.53001        & $28.6 \pm 6.6$ & $14.9 \pm 6.1$ & $ 5.1 \pm 1.2 $ & $ 7.1 \pm 3.0$ \\
30       &      175.27881        &      -26.52859        & $339.7 \pm 18.7$ & $337.3 \pm 19.1$ & $ 91.4 \pm 8.4 $ & $ 218.5 \pm 31.3$ \\
31       &      175.18436        &      -26.52488        & $107.7 \pm 11.0$ & $277.6 \pm 17.0$ & $ 14.7 \pm 1.8 $ & $ 112.4 \pm 16.3$ \\
32       &      175.17839        &      -26.51327        & $74.8 \pm 9.0$ & $98.4 \pm 10.5$ & $ 10.0 \pm 1.4 $ & $ 39.7 \pm 6.7$ \\
33       &      175.24586        &      -26.50403        & $2.7 \pm 2.7$ & $7.7 \pm 4.2$ & $ <1.0$ & $ 3.0 \pm 1.5$ \\
34       &      175.24149        &      -26.49342        & $1.1 \pm 2.9$ & $16.2 \pm 5.5$ & $ <1.0 $ & $ 6.4 \pm 2.3$ \\
35       &      175.19250        &      -26.48165        & $10.3 \pm 3.9$ & $5.6 \pm 3.8$ & $ 1.34 \pm 0.52 $ & $ <5.0$ \\
36       &      175.16551        &      -26.47922        & $260.8 \pm 16.6$ & $175.2 \pm 14.0$ & $ 35.7 \pm 3.5 $ & $ 72.9 \pm 11.2$ \\
37       &      175.18231        &      -26.46532        & $164.2 \pm 13.3$ & $359.2 \pm 19.5$ & $ 22.0 \pm 2.4 $ & $ 146.5 \pm 20.9$ \\
% \endfirsthead
% \newpage
% \caption{continued}
% \\ \hline \\
% XID & RA & DEC & Cts/s (soft) & Cts/s (hard) & $F_S$  & $F_H$    \\
%    &    &     &              &              & $10^{-16}$ erg/s/cm$^2$ & $10^{-16}$  erg/s/cm$^2$   \\
%    &    &     &              &              &   &   \\
% \hline
%    &    &     &              &              &   &   \\
%    &    &     &              &              &   &   \\
% 
38       &      175.26686        &      -26.46100        & $-1.5 \pm 2.1$ & $8.1 \pm 4.3$ & $ <0.5 $ & $ 3.5 \pm 1.7$ \\
39       &      175.26224        &      -26.45923        & $77.5 \pm 9.4$ & $19.0 \pm 6.3$ & $ 12.5 \pm 1.8 $ & $ 8.3 \pm 2.9$ \\
40       &      175.28162        &      -26.45306        & $16.0 \pm 5.7$ & $62.8 \pm 9.4$ & $ 3.6 \pm 1.3 $ & $ 36.7 \pm 7.3$ \\
41       &      175.27260        &      -26.44206        & $13.2 \pm 5.1$ & $18.3 \pm 6.8$ & $ 3.0 \pm 1.2 $ & $ 11.2 \pm 4.4$ \\
42       &      175.16692        &      -26.43220        & $90.8 \pm 10.8$ & $101.6 \pm 12.3$ & $ 13.8 \pm 1.9 $ & $ 42.5 \pm 7.6$ \\
43       &      175.20650        &      -26.42832        & $119.2 \pm 12.0$ & $28.0 \pm 7.7$ & $ 17.6 \pm 2.2 $ & $ 11.5 \pm 3.5$ \\
44       &      175.14221        &      -26.54348        & $1.6 \pm 4.7$ & $24.5 \pm 7.7$ & $ <2.4 $ & $ 14.4 \pm 4.9$ \\
45       &      175.15338        &      -26.53828        & $8.9 \pm 5.3$ & $14.9 \pm 6.7$ & $ <3.2 $ & $ 6.4 \pm 3.0$ \\
46       &      175.13422        &      -26.51784        & $7.7 \pm 5.0$ & $12.6 \pm 6.7$ & $ <2.8 $ & $ 5.5 \pm 2.7$ \\
47       &      175.26782        &      -26.45793        & $3.6 \pm 4.4$ & $11.6 \pm 6.4$ & $ <2.0 $ & $ 4.9 \pm 2.5$ \\
48       &      175.15282        &      -26.45285        & $12.4 \pm 5.1$ & $22.5 \pm 6.8$ & $ 1.86 \pm 0.77 $ & $ 9.7 \pm 3.2$ \\
49       &      175.16095        &      -26.44319        & $7.5 \pm 4.4$ & $11.0 \pm 5.5$ & $ <2.3 $ & $ 4.6 \pm 2.4$ \\
50       &      175.14491        &      -26.44095        & $2.2 \pm 5.3$ & $51.5 \pm 9.7$ & $ <2.0 $ & $ 22.2 \pm 5.1$ \\
51       &      175.23083        &      -26.43491        & $11.7 \pm 4.9$ & $8.7 \pm 5.9$ & $ 1.68 \pm 0.70 $ & $ 3.5 \pm 2.3$ \\
52       &      175.20361        &      -26.43283        & $6.3 \pm 4.3$ & $68.0 \pm 9.3$ & $ <2.0 $ & $ 27.6 \pm 5.3$ \\
53       &      175.19453        &      -26.42296        & $1.2 \pm 4.0$ & $44.7 \pm 8.6$ & $ <1.4 $ & $ 18.3 \pm 4.3$ \\
54       &      175.13251        &      -26.46288        & $3.3 \pm 4.8$ & $17.4 \pm 7.3$ & $ 1.9 $ & $ 7.2 \pm 3.2$ \\
55       &      175.18709        &      -26.50135        & $1.4 \pm 2.7$ & $38.6 \pm 6.8$ & $ <0.8 $ & $ 16.0 \pm 3.5$ \\
56       &      175.18121        &      -26.47711        & $4.9 \pm 3.0$ & $34.1 \pm 6.6$ & $ <1.5 $ & $ 14.5 \pm 4.0$ \\
57       &      175.15547        &      -26.50483        & $41.2 \pm 7.1$ & $26.6 \pm 6.4$ & $ 5.6 \pm 1.1 $ & $ 10.8 \pm 3.0$ \\
58       &      175.20148        &      -26.48577        & $5355.1 \pm 73.4$ & $4033.0 \pm 63.7$ & $ 689.6 \pm 52.0 $ & $ 1575.0 \pm 208.8$ \\
59       &      175.27609        &      -26.48332        & $4.0 \pm 4.3$ & $20.6 \pm 6.9$ & $ <2.0 $ & $ 8.9 \pm 3.2$ \\
60       &      175.22218        &      -26.43210        & $9.9 \pm 4.6$ & $17.5 \pm 6.4$ & $ 1.43 \pm 0.68 $ & $ 7.1 \pm 2.7$ \\
61       &      175.20068        &      -26.43743        & $8.2 \pm 4.1$ & $23.8 \pm 6.7$ & $ 1.15 \pm 0.58 $ & $ 9.7 \pm 3.0$ \\
62       &      175.19254        &      -26.43634        & $7.9 \pm 4.0$ & $7.9 \pm 4.6$ & $ 1.12 \pm 0.55 $ & $ 3.2 \pm 1.9$ \\
63       &      175.22252        &      -26.45065        & $26.4 \pm 5.9$ & $49.4 \pm 8.0$ & $ 3.6 \pm 0.8 $ & $ 19.7 \pm 4.1$ \\
64       &      175.25787        &      -26.44586        & $30.0 \pm 6.8$ & $14.0 \pm 6.7$ & $ 4.7 \pm 1.1 $ & $ 5.8 \pm 2.9$ \\
65       &      175.27296        &      -26.44957        & $21.9 \pm 6.0$ & $22.6 \pm 7.2$ & $ 4.2 \pm 1.2 $ & $ 11.2 \pm 3.9$ \\
66       &      175.27232        &      -26.45238        & $9.7 \pm 4.5$ & $23.9 \pm 7.8$ & $ 1.73 \pm 0.81 $ & $ 11.1 \pm 3.9$ \\
67       &      175.13544        &      -26.52808        & $53.3 \pm 8.3$ & $53.8 \pm 9.2$ & $ 10.2 \pm 1.8 $ & $ 27.9 \pm 6.0$ \\
68       &      175.14703        &      -26.47453        & $3.7 \pm 3.9$ & $18.9 \pm 6.4$ & $ <1.6 $ & $ 7.7 \pm 2.8$ \\
69       &      175.16040        &      -26.46140        & $5.3 \pm 3.9$ & $29.6 \pm 6.8$ & $ <1.8 $ & $ 12.0 \pm 3.2$ \\
70       &      175.16412        &      -26.46660        & $11.2 \pm 4.3$ & $13.3 \pm 5.0$ & $ 1.50 \pm 0.59 $ & $ 5.4 \pm 2.1$ \\
71       &      175.16164        &      -26.48617        & $73.7 \pm 9.0$ & $17.4 \pm 5.3$ & $ 9.9 \pm 1.4 $ & $ 7.1 \pm 2.4$ \\
72       &      175.18102        &      -26.46208        & $20.4 \pm 5.1$ & $11.2 \pm 4.6$ & $ 2.8 \pm 0.7 $ & $ 4.6 \pm 2.0$ \\
73       &      175.18962        &      -26.46955        & $12.5 \pm 4.2$ & $24.9 \pm 5.7$ & $ 1.68 \pm 0.57 $ & $ 10.3 \pm 2.7$ \\
74       &      175.18440        &      -26.48541        & $60.7 \pm 8.0$ & $39.7 \pm 6.8$ & $ 8.5 \pm 1.3 $ & $ 17.0 \pm 3.7$ \\
75       &      175.19153        &      -26.48805        & $495.0 \pm 22.4$ & $334.8 \pm 18.7$ & $ 64.2 \pm 5.6 $ & $ 133.0 \pm 19.0$ \\
76       &      175.20314        &      -26.46463        & $1.1 \pm 2.5$ & $11.8 \pm 4.5$ & $ <0.8 $ & $ 4.8 \pm 1.9$ \\
77       &      175.22797        &      -26.45959        & $5.0 \pm 3.4$ & $42.3 \pm 7.4$ & $ <1.6 $ & $ 17.2 \pm 3.8$ \\
78       &      175.23540        &      -26.46467        & $14.4 \pm 4.6$ & $9.8 \pm 4.7$ & $ 1.90 \pm 0.63 $ & $ 3.8 \pm 1.9$ \\
79       &      175.25610        &      -26.46475        & $0.5 \pm 3.2$ & $24.4 \pm 6.6$ & $ <1.1 $ & $ 10.9 \pm 3.3$ \\
80       &      175.25993        &      -26.46278        & $248.4 \pm 16.1$ & $314.3 \pm 18.6$ & $ 41.2 \pm 4.1 $ & $ 144.2 \pm 20.8$ \\
81       &      175.20267        &      -26.48068        & $58.9 \pm 8.0$ & $43.6 \pm 7.2$ & $ 7.6 \pm 1.2 $ & $ 17.1 \pm 3.6$ \\
82       &      175.27940        &      -26.47438        & $10.1 \pm 4.8$ & $29.9 \pm 7.7$ & $ 1.7 \pm 0.8 $ & $ 13.4 \pm 3.9$ \\
83       &      175.14645        &      -26.49300        & $4.4 \pm 3.6$ & $38.8 \pm 7.5$ & $ <1.6 $ & $ 16.0 \pm 3.7$ \\
84       &      175.15465        &      -26.49537        & $7.1 \pm 4.0$ & $15.2 \pm 5.4$ & $ <2.0 $ & $ 6.2 \pm 2.4$ \\
85       &      175.16492        &      -26.49530        & $56.2 \pm 8.0$ & $100.1 \pm 10.5$ & $ 7.5 \pm 1.2 $ & $ 40.7 \pm 6.8$ \\
86       &      175.18535        &      -26.48916        & $51.6 \pm 7.5$ & $52.0 \pm 7.8$ & $ 7.1 \pm 1.2 $ & $ 21.8 \pm 4.4$ \\
87       &      175.19443        &      -26.48622        & $24.2 \pm 5.4$ & $6.3 \pm 3.8$ & $ 3.1 \pm 0.7 $ & $ 2.5 \pm 1.5$ \\
88       &      175.20360        &      -26.48595        & $37.7 \pm 8.0$ & $23.5 \pm 6.2$ & $ 4.9 \pm 1.1 $ & $ 9.2 \pm 2.7$ \\
89       &      175.21483        &      -26.48368        & $7.6 \pm 3.5$ & $42.7 \pm 7.2$ & $ 1.00 \pm 0.46 $ & $ 16.9 \pm 3.6$ \\
90       &      175.21136        &      -26.49230        & $25.2 \pm 5.5$ & $23.3 \pm 5.6$ & $ 3.25 \pm 0.75 $ & $ 9.0 \pm 2.5$ \\
91       &      175.22507        &      -26.49026        & $17.8 \pm 4.8$ & $20.5 \pm 5.4$ & $ 2.4 \pm 0.7 $ & $ 8.1 \pm 2.4$ \\
92       &      175.22772        &      -26.49112        & $89.5 \pm 9.7$ & $14.6 \pm 4.9$ & $ 11.8 \pm 1.5 $ & $ 5.7 \pm 2.1$ \\
93       &      175.23019        &      -26.49082        & $66.7 \pm 8.5$ & $34.6 \pm 6.7$ & $ 8.7 \pm 1.3 $ & $ 13.5 \pm 3.1$ \\
94$^{r1}$        &      175.23413        &      -26.48683        & $5.2 \pm 2.9$ & $8.7 \pm 3.8$ & $ <1.3 $ & $ 3.4 \pm 1.5$ \\
95       &      175.24232        &      -26.50758        & $47.2 \pm 7.5$ & $54.0 \pm 8.5$ & $ 6.3 \pm 1.1 $ & $ 21.1 \pm 4.3$ \\
96       &      175.26640        &      -26.51311        & $116.2 \pm 11.3$ & $344.1 \pm 19.3$ & $ 18.4 \pm 2.3 $ & $ 146.2 \pm 20.9$ \\
97       &      175.26569        &      -26.51449        & $13.4 \pm 4.4$ & $41.4 \pm 7.4$ & $ 2.1 \pm 0.7 $ & $ 17.4 \pm 3.9$ \\
98       &      175.24670        &      -26.53411        & $77.1 \pm 10.0$ & $99.3 \pm 11.4$ & $ 12.9 \pm 1.9 $ & $ 43.1 \pm 7.5$ \\
99       &      175.24873        &      -26.53539        & $53.5 \pm 8.5$ & $52.1 \pm 8.9$ & $ 8.8 \pm 1.5 $ & $ 21.8 \pm 4.7$ \\
100      &      175.22601        &      -26.56204        & $40.9 \pm 7.2$ & $36.2 \pm 7.8$ & $ 11.3 \pm 2.2 $ & $ 25.8 \pm 6.5$ \\
101      &      175.18367        &      -26.51827        & $14.0 \pm 4.7$ & $3.5 \pm 3.9$ & $ 1.9 \pm 0.6 $ & $ <4.0$ \\
102      &      175.14857        &      -26.54114        & $163.1 \pm 13.8$ & $135.8 \pm 13.8$ & $ 31.4 \pm 3.5 $ & $ 70.2 \pm 11.7$ \\
103      &      175.16844        &      -26.55954        & $6.7 \pm 5.1$ & $61.5 \pm 9.8$ & $ <3.6 $ & $ 35.3 \pm 7.3$ \\
104$^{r3}$       &      175.20998        &      -26.56691        & $-1.3 \pm 1.8$ & $9.0 \pm 4.2$ & $ <0.7 $ & $ 6.8 \pm 3.3$ \\
105      &      175.13370        &      -26.49614        & $-2.1 \pm 1.5$ & $7.5 \pm 3.6$ & $ <0.2 $ & $ 3.2 \pm 1.6$ \\
106$^{r3}$       &      175.26247        &      -26.43484        & $8.3 \pm 3.5$ & $3.7 \pm 3.6$ & $ 1.61 \pm 0.70 $ & $ <5.5 $ \\
107      &      175.26004        &      -26.49588        & $2.0 \pm 3.5$ & $13.8 \pm 5.6$ & $ <1.3 $ & $ 5.9 \pm 2.5$ \\
   &    &     &              &              &   &   \\
\hline
\end{longtable}
\label{source_list}
\tablefoot{The RA and DEC positions refer to the 
centroid of the X-ray detected emission.  If a source does not satisfy $S/N>2$ in one band, 
we report the $2\, \sigma$ upper limit.  Sources with apex $^{r1}$ and $^{r3}$ have an extraction 
radius 1 and 3 pixels, respectively, smaller than $r_{ext}$ as defined in Section 5.1.}
% \end{table}
\FloatBarrier

\newpage

\section{Spectra of X-ray members}

In Figure \ref{spectra}, we show the folded spectra of the eight X-ray members of the 
Spiderweb protocluster that have at least 40 net counts in the soft (0.5-2 keV) or
hard (2-7 keV) band, along with the best-fit spectral model.  Spectra have 
been rebinned with a minimum of 20 counts per energy bin for display purpose.
The spectrum of the Spiderweb Galaxy (XID 58) will be discussed in much greater
detail in a companion paper (Tozzi et. al. in preparation).  The spectral model used in all 
the fits consists in a fixed modelization of the Galactic absorption with the {\sl Xspec} model 
{\tt tbabs}, with $N_{\rm H\,gal} = 3.18\times 10^{20}$ cm$^{-2}$, an intrinsic power
law with a free slope $\Gamma$, an intrinsic absorption modeled with {\tt zwabs}, and 
a Gaussian line component that is allowed to vary in a range $1.9-2.1$ keV, approximately
centered on the energy 2.03 keV corresponding to the neutral iron K$_\alpha$ line at 
a rest frame energy of 6.4 keV. 

For completeness, in Figure \ref{spectra2}, we show the folded spectra of the six X-ray members of the 
Spiderweb protocluster that do not reach 40 in the soft and in the hard band.  Spectra have 
been rebinned with a minimum of only ten counts per energy bin for display purposes.
The spectral model used in the fits is significantly simpler than the previous one, 
and consists in a fixed modelization of the Galactic absorption with the {\sl Xspec} model 
{\tt tbabs}, with $N_{\rm H\,gal} = 3.18\times 10^{20}$ cm$^{-2}$, an intrinsic power
law with a slope frozen to $\Gamma=1.8$, and an intrinsic absorption modeled with {\tt zwabs}. 
We note that for sources XID 34 and XID 87, the spectral slope is left free instead of the 
intrinsic absorption, because the simple Compton-thin model is not able to provide acceptable fits. 
Best-fit parameters for all the sources are shown in Table \ref{src_Xrayfit}.

\begin{figure}[h]
\begin{center}
\includegraphics[width=0.49\textwidth]{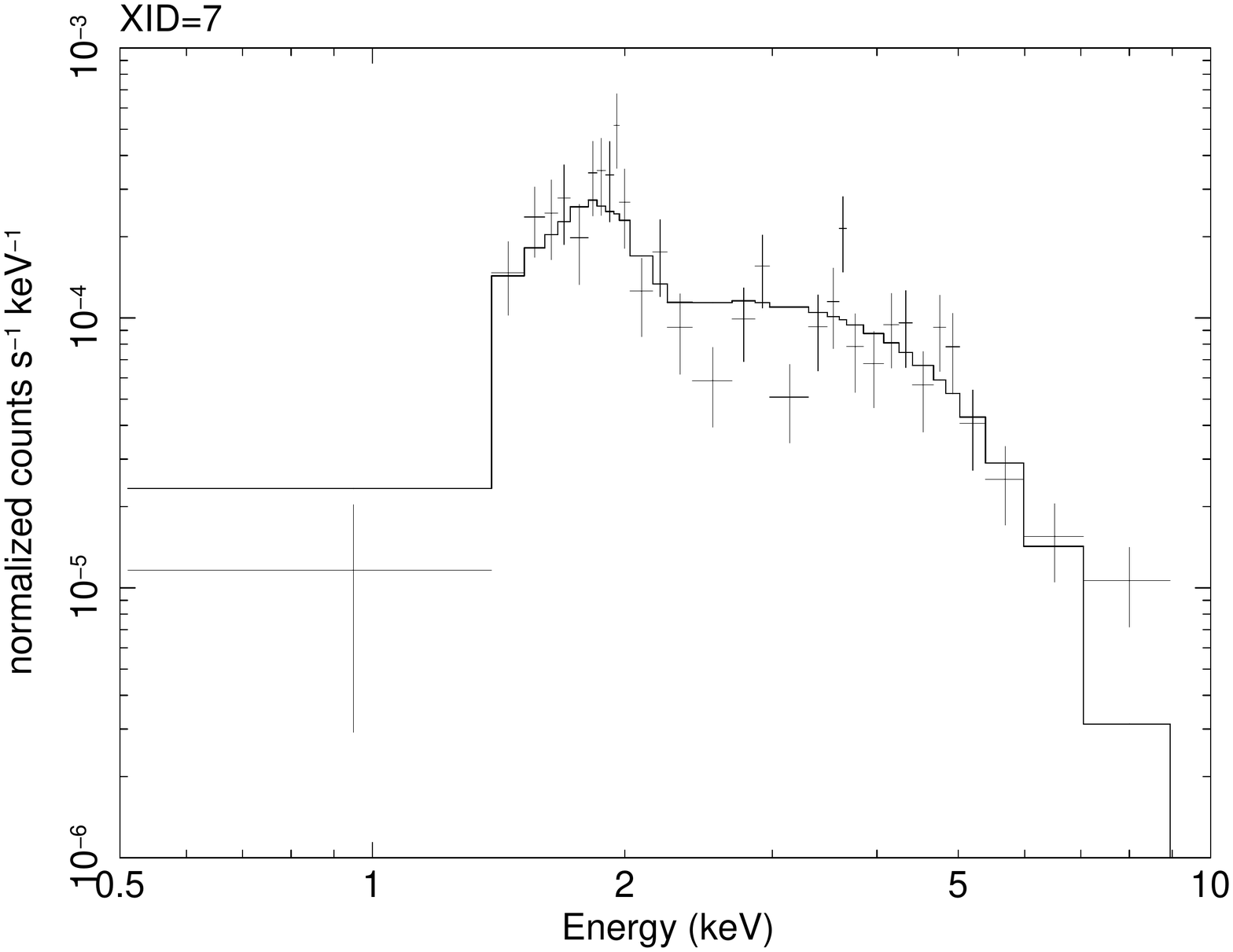}
\includegraphics[width=0.49\textwidth]{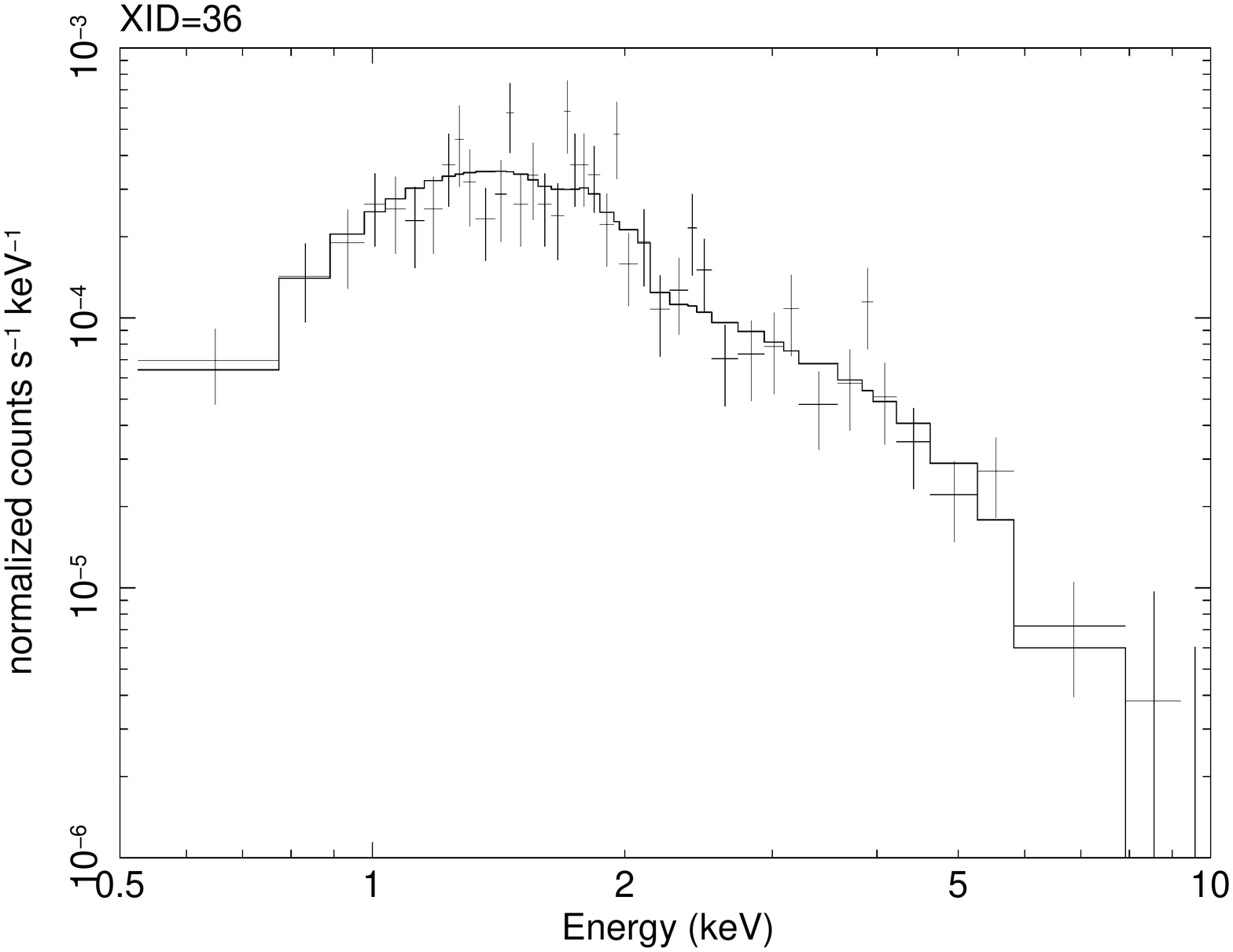}
\includegraphics[width=0.49\textwidth]{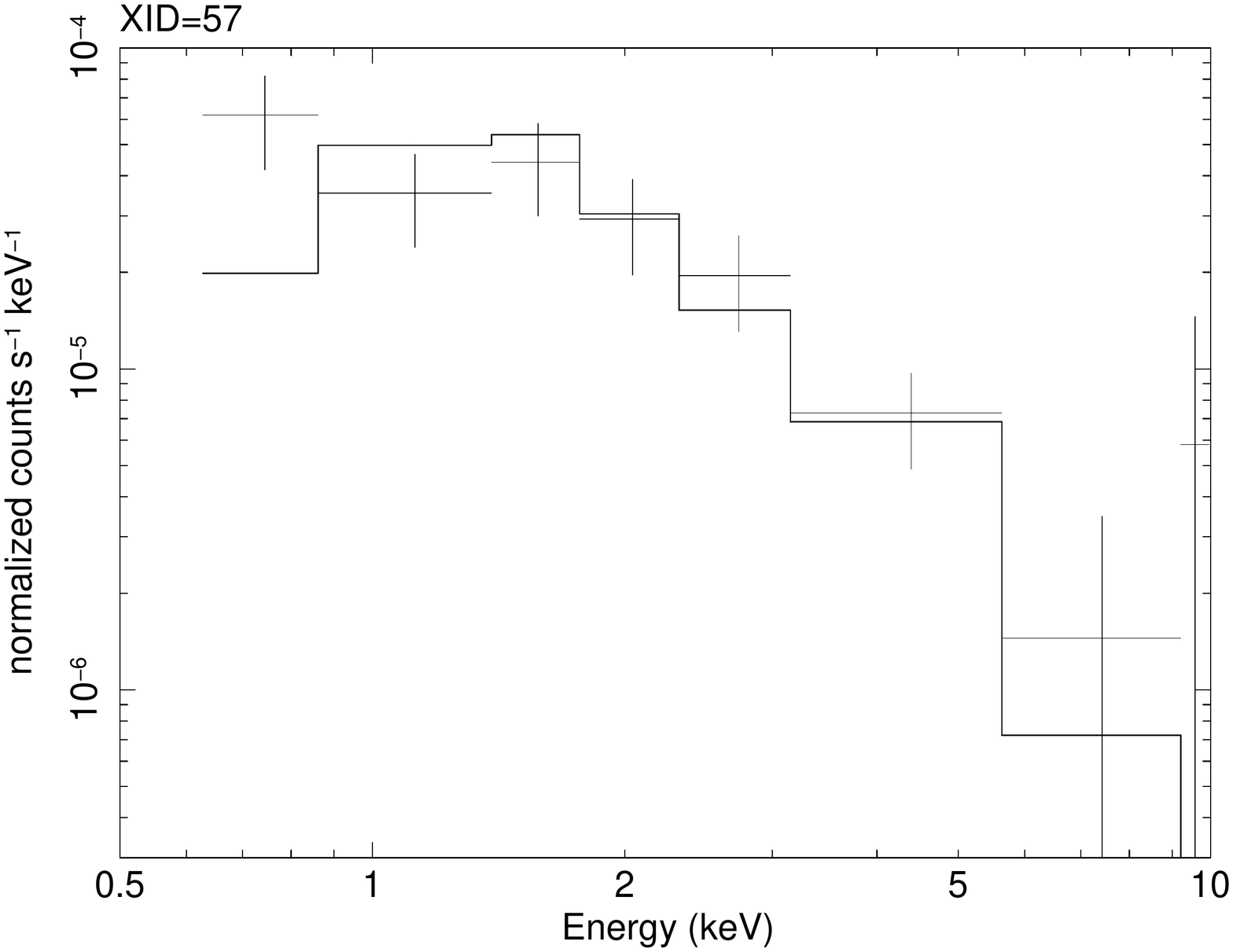}
\includegraphics[width=0.49\textwidth]{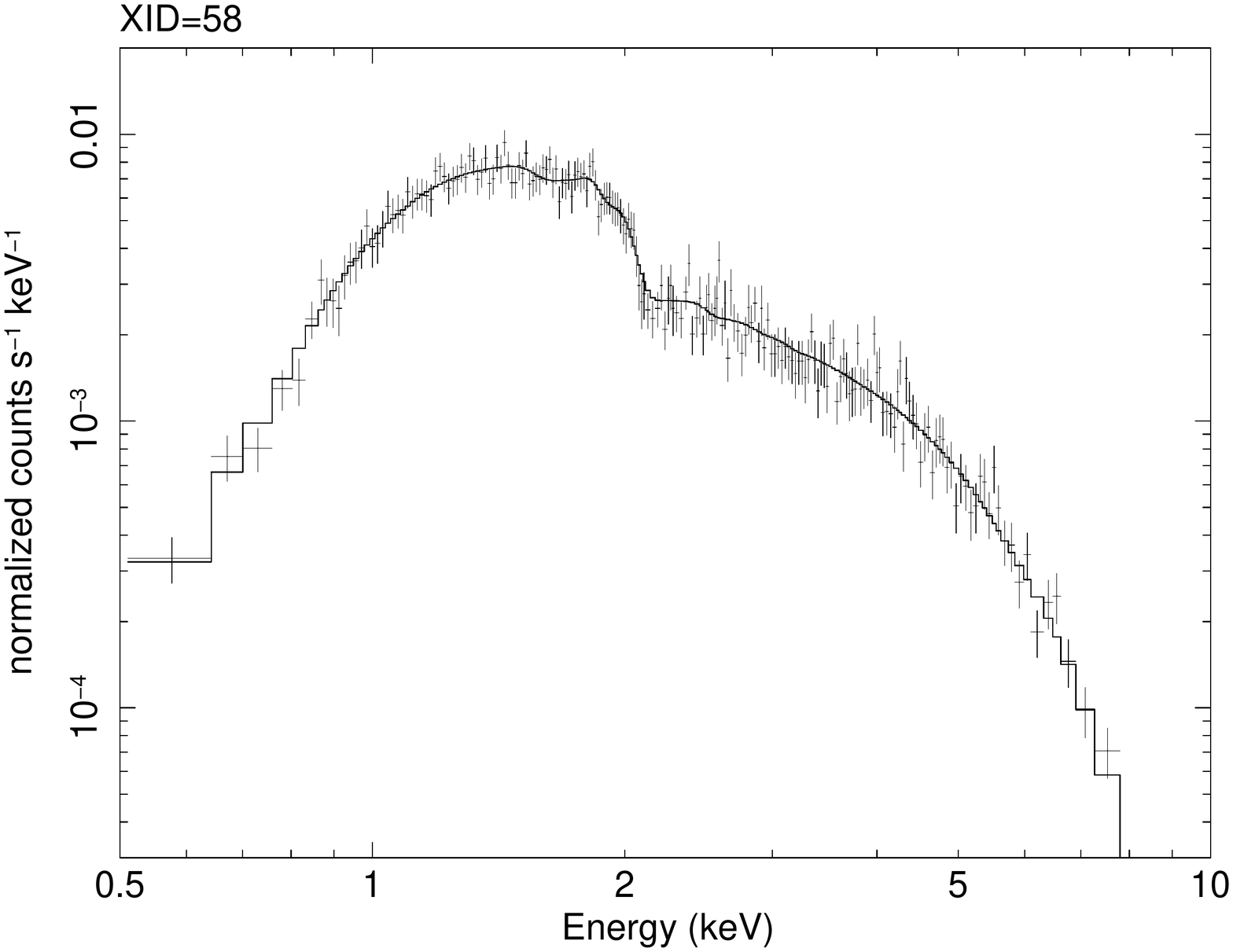}
\caption{Folded spectra of the X-ray members of the Spiderweb protocluster
with more than 40 net counts in the soft or in the hard band, including the
Spiderweb Galaxy itself (XID 58).}
\label{spectra}
\end{center}
\end{figure}
\FloatBarrier

\addtocounter{figure}{-1}
\begin{figure*}
\begin{center}
\includegraphics[width=0.49\textwidth]{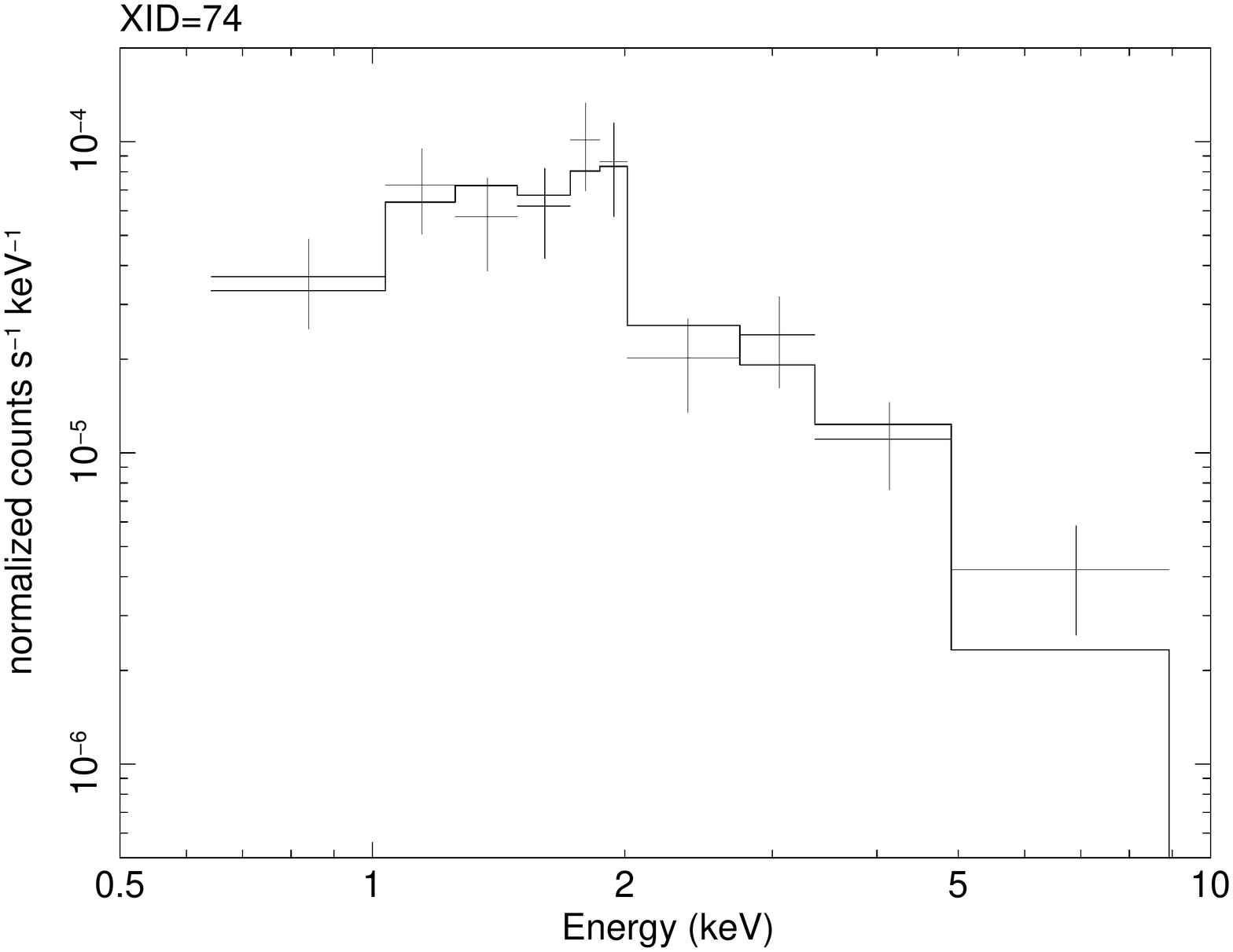}
\includegraphics[width=0.49\textwidth]{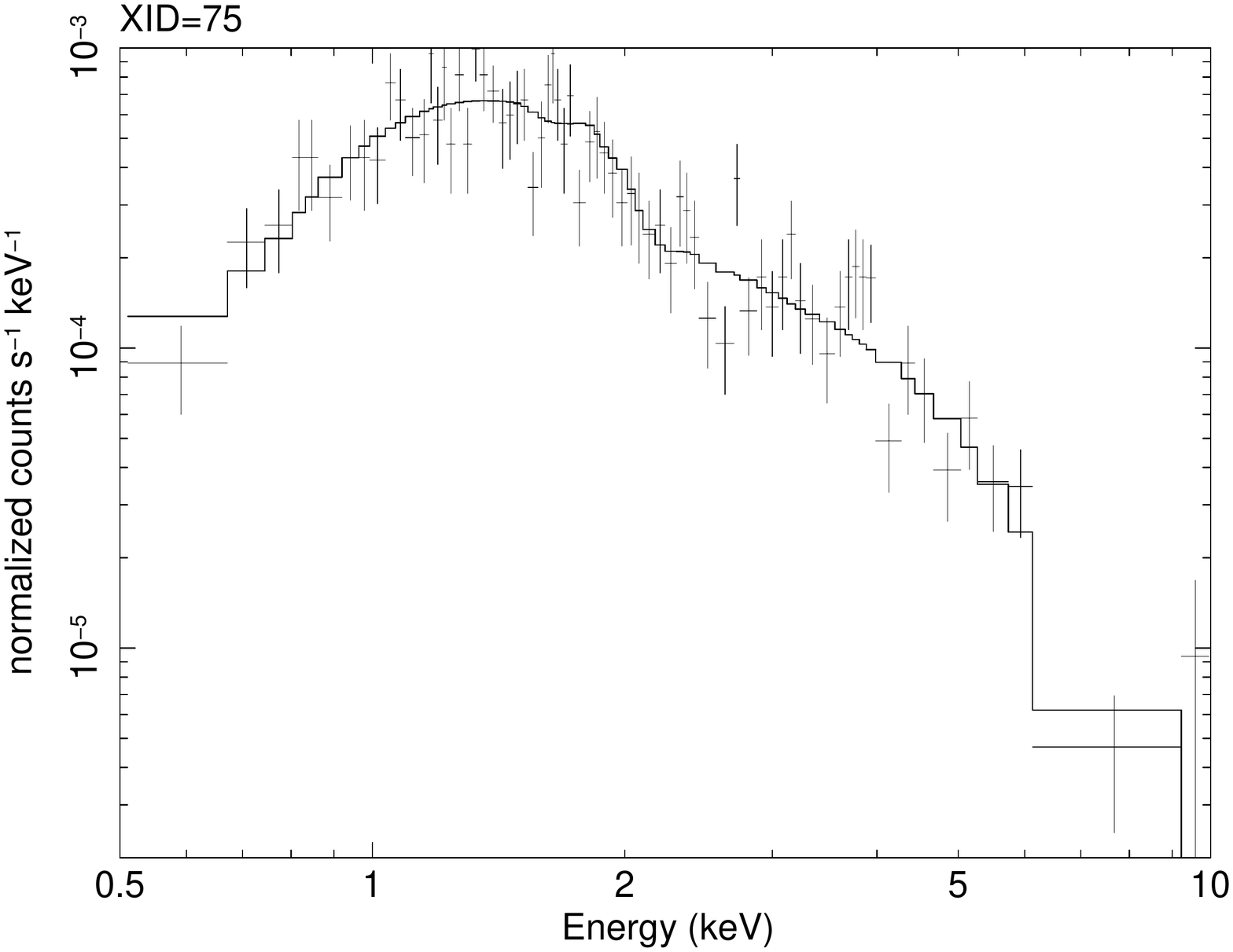}
\includegraphics[width=0.49\textwidth]{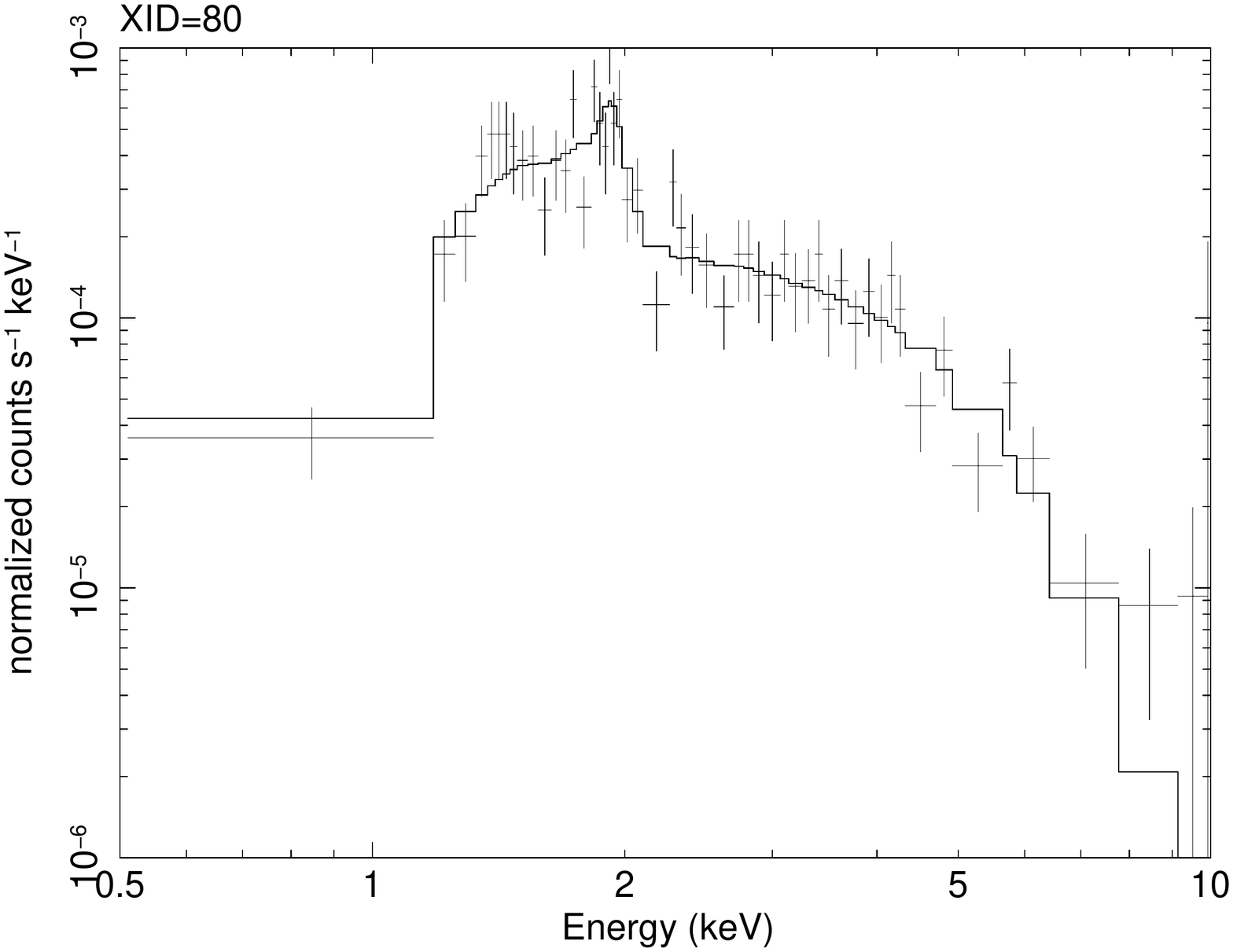}
\includegraphics[width=0.49\textwidth]{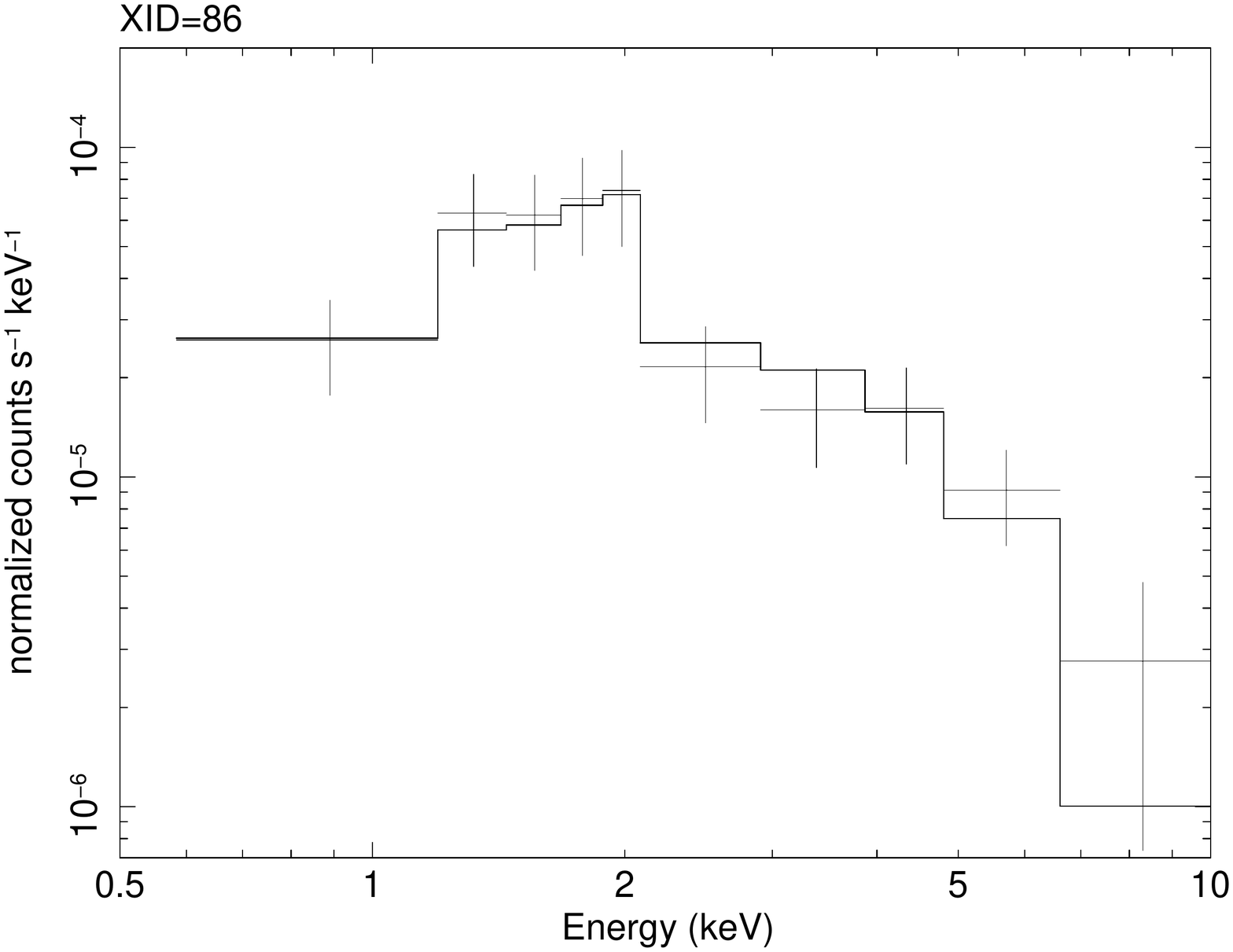}
     \caption{ - continued .}
%\label{spectra}
\end{center}
\end{figure*}
\FloatBarrier

\begin{figure*}[h]
\begin{center}
\includegraphics[width=0.49\textwidth]{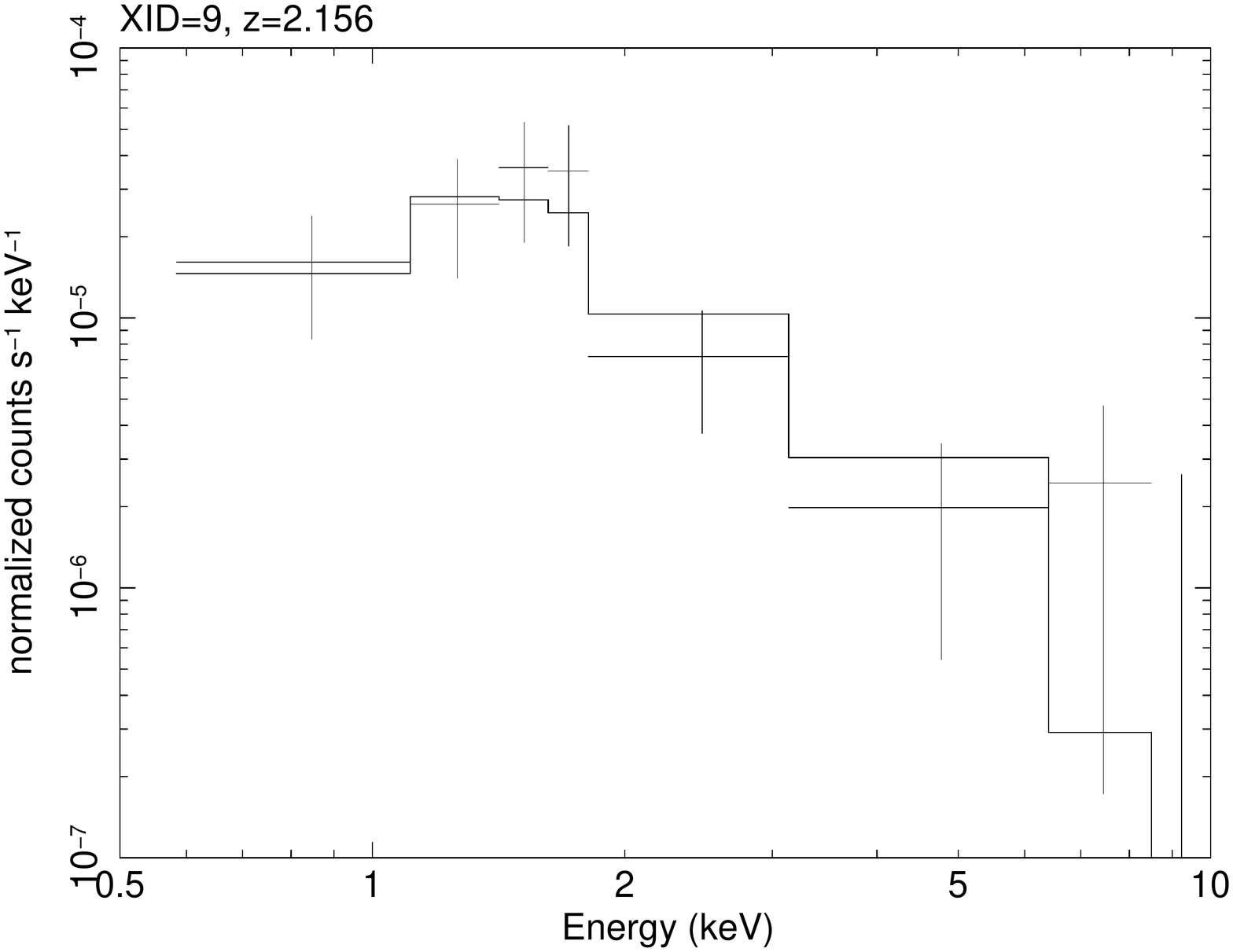}
\includegraphics[width=0.49\textwidth]{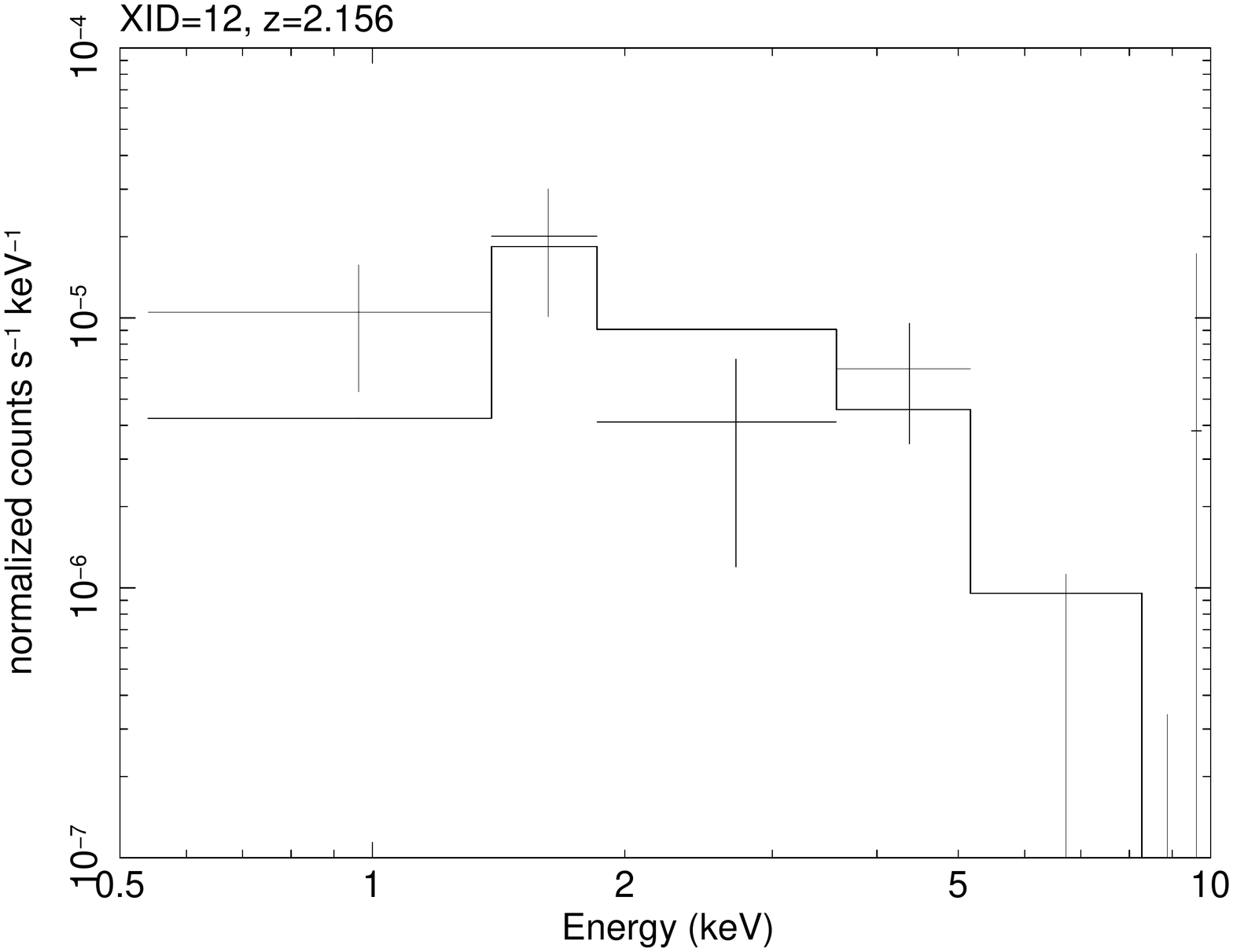}
\includegraphics[width=0.49\textwidth]{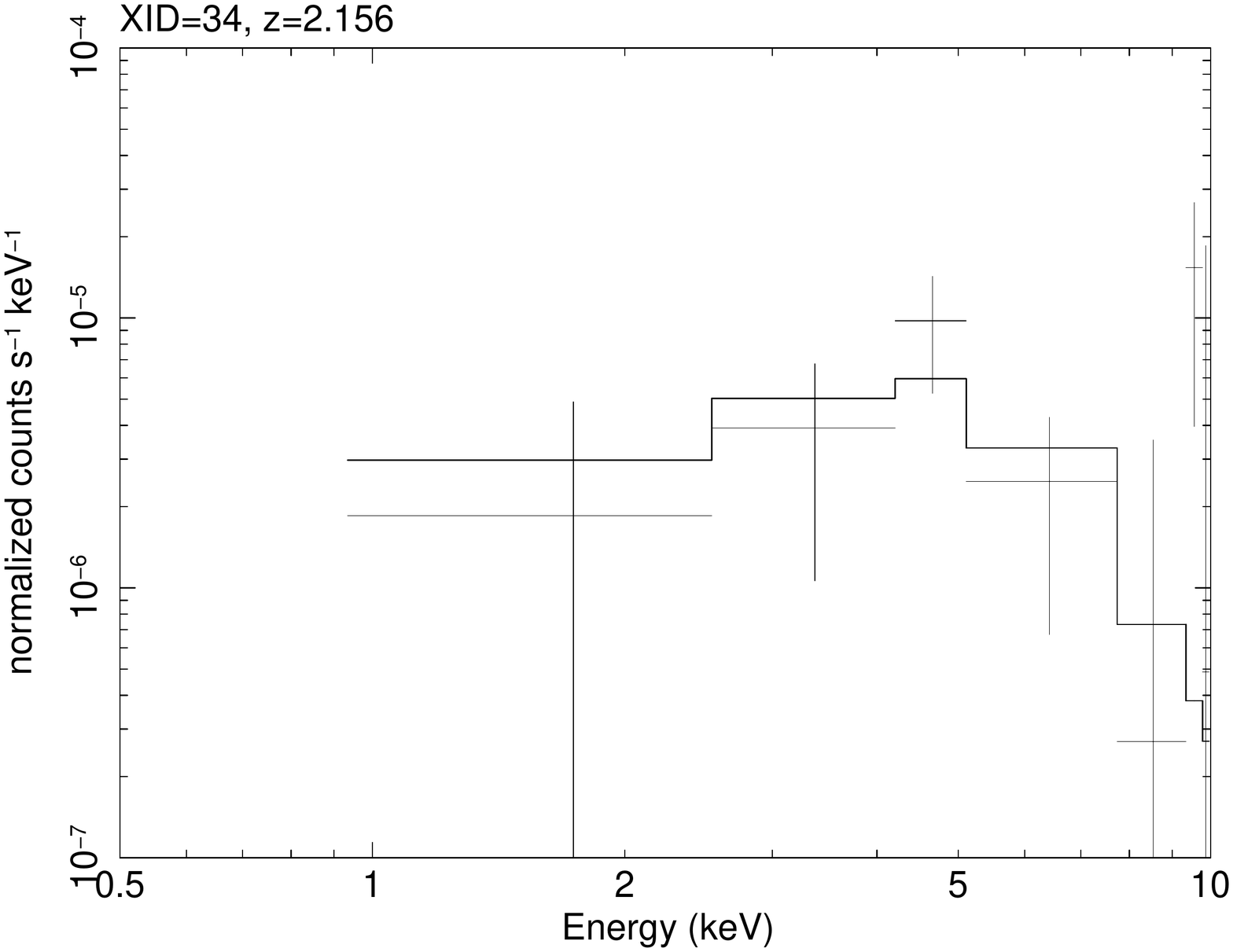}
\includegraphics[width=0.49\textwidth]{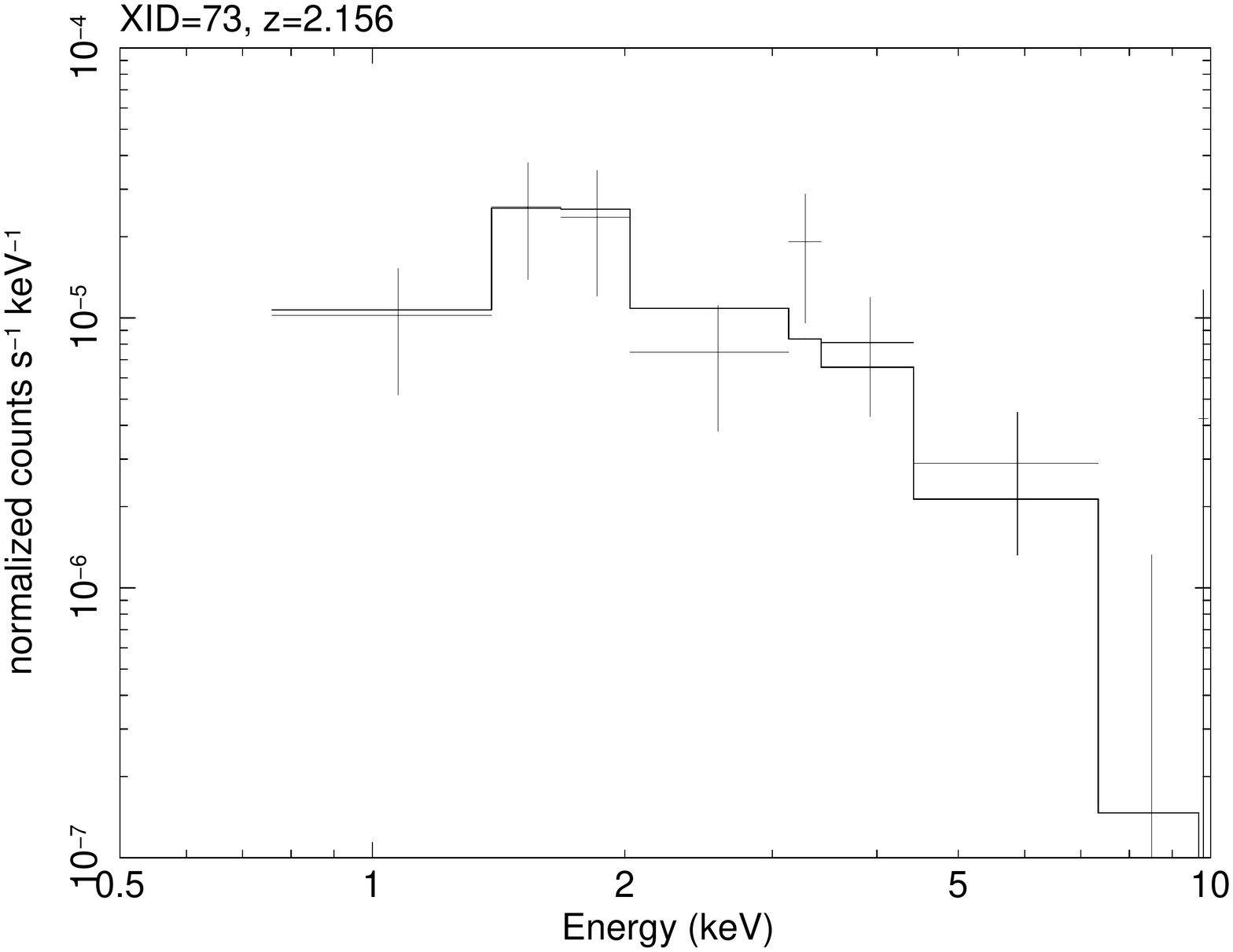}
\includegraphics[width=0.49\textwidth]{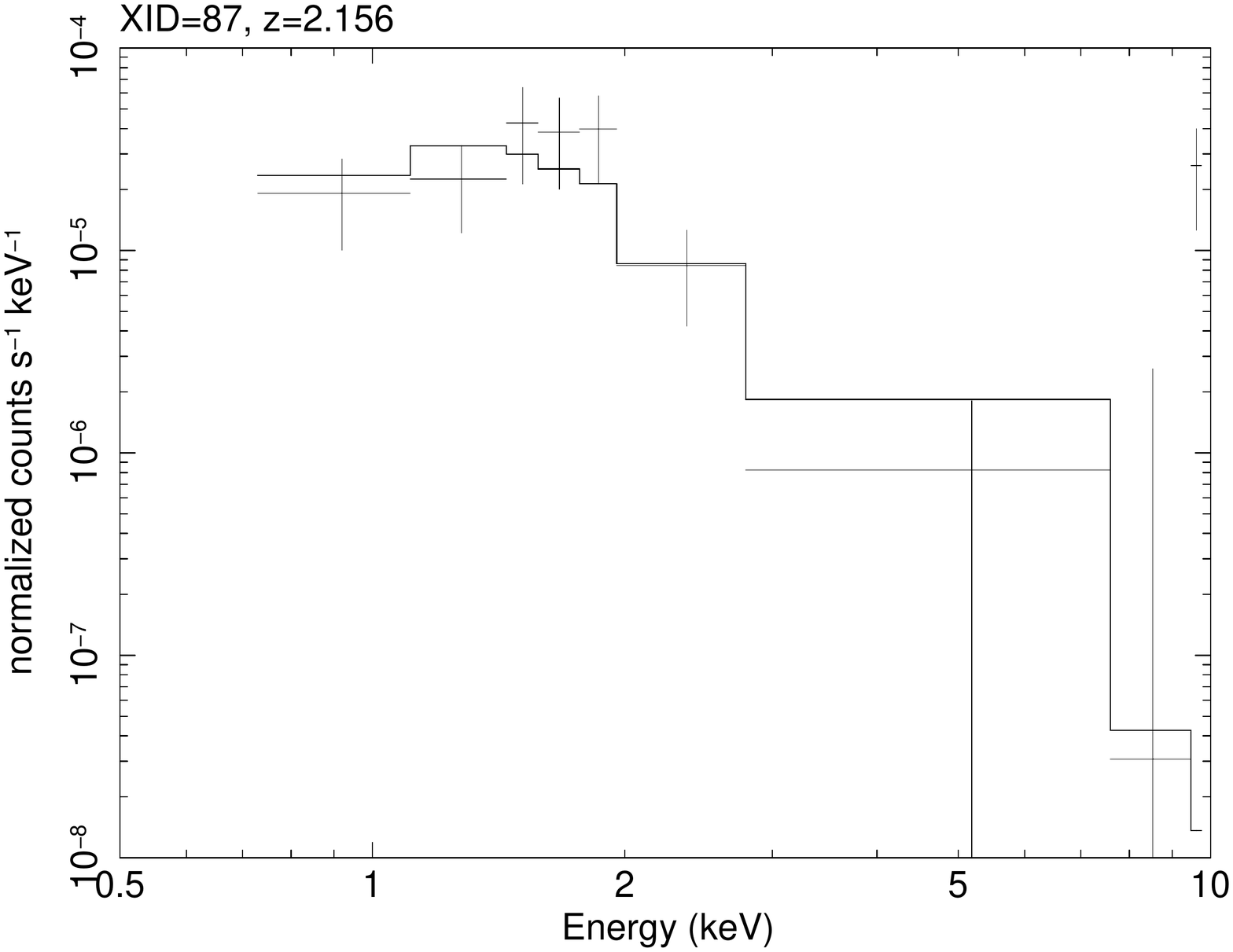}
\includegraphics[width=0.49\textwidth]{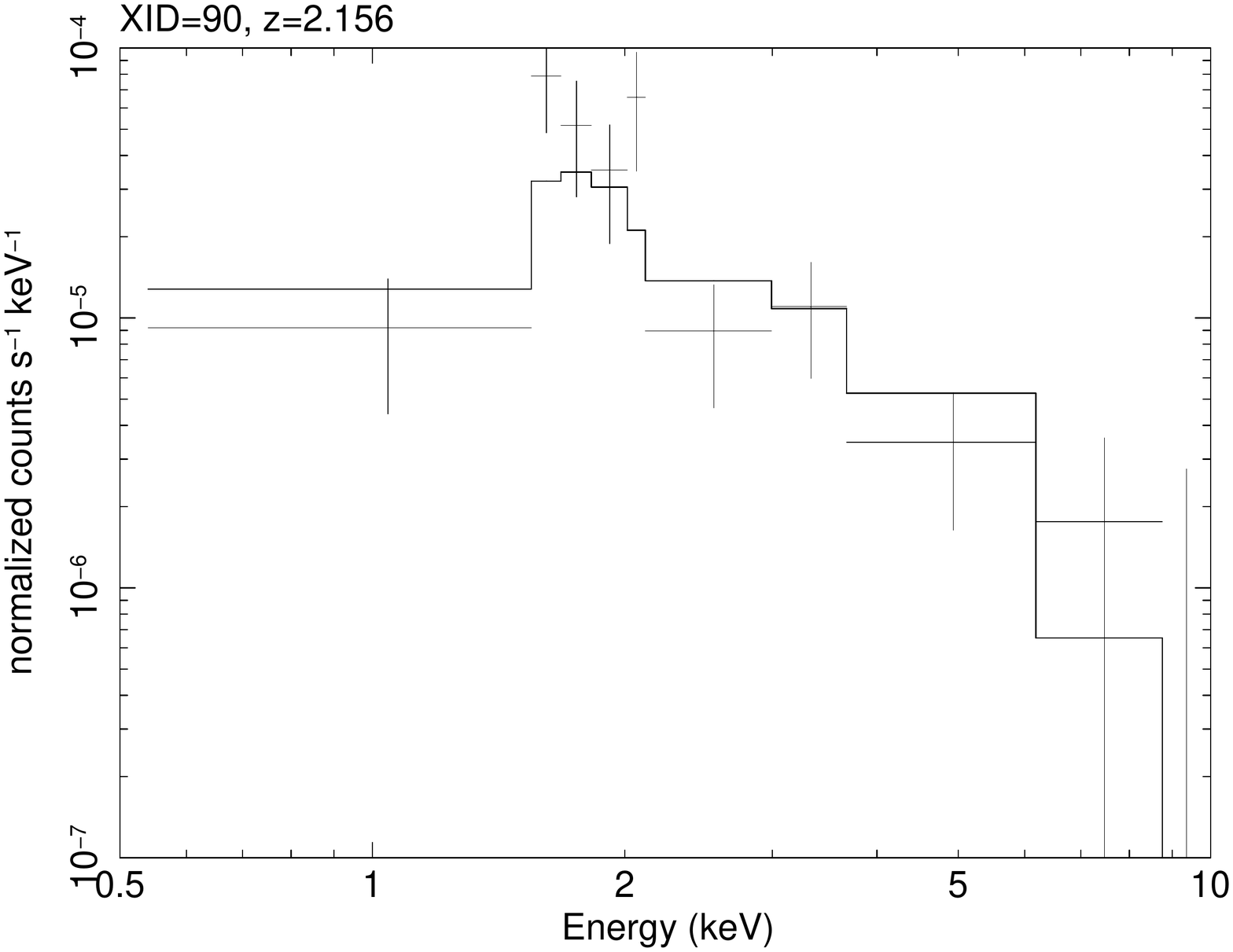}
\caption{Folded spectra of the X-ray members of the Spiderweb protocluster
with less than 40 net counts in the soft and in the hard band.}
\label{spectra2}
\end{center}
\end{figure*}
\FloatBarrier

%%%%%%%%%%%%%%%%%%%%%%%%%%%%%%%%%%%%%%%%%%
% Notes ON SINGLE SOURCES
%%%%%%%%%%%%%%%%%%%%%%%%%%%%%%%%%%%%%%%%%%

\section{Notes on single sources}

In this Appendix we briefly discuss specific features of all the X-ray-detected members
of the Spiderweb protocluster, except the Spiderweb Galaxy (XID 58). 

\subsection{XID 7}

This source has been identified as a candidate 
H$\alpha$ emitter in \citet{2004aKurk} (ID=272), and a photometric redshift of
$z=2.18^{+0.08}_{-0.05}$ (ID=493) has been measured by \citet{2013Tanaka}.  
\citet{2011Kuiper} investigated the NIR spectrum with SINFONI (see their ID 12), and 
found a spectroscopic redshift of $z=2.1447\pm 0.0023$ from the O[III] line, 
observed with a flux of $8.2\pm 1.5 \times 10^{-17}$ erg/s/cm$^2$/$\AA$.
% A redshift of  $z=2.147$ has been measured with Subaru in \citet{2013Tanaka}.
% (ID=58).
% This source has also been identified with an H$\alpha$ emitter in \citet{2019Tadaki} (ID=1138.58).  
In the X-ray, this source shows a Compton thin spectrum with the highest
intrinsic absorption in the sample ($\sim 4\times 10^{23}$ cm$^{-2}$), and a luminosity in the 
QSO range ($L_{0.5-10 keV} \sim 7.5 \times 10^{44}$ erg/s).  It is the X-ray source
closest to the Spiderweb Galaxy.

\subsection{XID 9}

The optical counterpart of source XID 9 has been originally selected as a 
candidate Ly$\alpha$ emitter.
% with $B=27.4$.  
The optical spectrum is shown in \citet{2002Pentericci} (ID=1612), where 
the Ly$\alpha$ line has been measured to have a flux of $3.9\times 10^{-17}$ erg/s/cm$^2$, 
with an $EW>30$ $\AA$ and a $FWHM = 490 $ km/s
% Also Selected as an Ly$\alpha$ emitter candidate in \citet{2004aKurk} with ID=675
This source has been confirmed also with Keck spectroscopy by \citet{2005Croft}
(ID=L675) and the Ly$\alpha$ has been revised to have $FWHM = 200 \pm 30$ km/s
with a flux of  $6.4\times 10^{-17}$ erg/s/cm$^2$.  The NV line is also
detected with $FWHM = 200 \pm 30$ km/s for a flux of  $4.1\times 10^{-17}$ erg/s/cm$^2$.
% Identified as an H$\alpha$ emitter in \citet{2018Shimakawa} (ID=77).

\subsection{XID 12}

The redshift of XID 12 (z=2.1628) has been obtained with Subaru in \citet{2014Shimakawa} (ID=71). 
The X-ray spectrum is consistent with negligible intrinsic absorption, while the 
total luminosity $L_{0.5-10 keV} \sim 2.2 \times 10^{43}$ erg/s is the lowest 
among the X-ray protocluster member, but still in the Seyfert-like regime.
This source has the largest offset from its optical counterpart, due
to the extended nature of its emission as noted in the text. Possible association 
of the extended emission with radio jets will be investigated when the wide-field, high-resolution
radio images are available.  
% The spectrum is not discussed in detail in the paper.

\subsection{XID 34}

This source has been identified with APEX LABOCA in \citet{2014Dannerbauer} (ID=DKB12a)
with a flux $S_{870\mu}=5.0 \pm 1.4$ mJy for a $S/N= 3.6$. 
It also has a strong
detection at 24$\mu$ with $S_{24\mu} = 303.4 \pm 5.0$ $\mu$Jy and significant detection 
in the range 160 - 500 $\mu$Jy, but it only have an upper limit at 1.4 GHz with JVLA.
There are four H$\alpha$ emitters consistent with the LABOCA position, however DKB12a is the closest 
to the X-ray position. The H$\alpha$ and FIR-SED derived SFR are 
240 and 860 $M_\odot$/yr, respectively, suggesting the presence of dust. 
The total FIR luminosity is $5.0\times 10^{12} \, L_\odot$. 
Eventually, a redshift of  $z=2.1690$ was measured with Subaru in \citet{2014Shimakawa} (ID 29).
\citet{2018Shimakawa} identified a possible subgroup at the position of this source.
The source was also detected in the CO line in the COALAS survey
\citep[][see their Table 3]{2021Jin}, with an offset of 2 arcsec between the ATCA and
X-ray centroids.  % ID=04
The high FIR luminosity is consistent with the X-ray tentative classification 
as a Compton-thick AGN, due to an extremely flat (if not inverted) power-law spectrum.  
Despite the very low flux, the estimated intrinsic luminosity would be
$L_{2-10 keV} \sim 2.6 \times 10^{43}$ erg/s in our simplistic assumption 
that the observed (reflected) luminosity is $\sim 10$\% of the intrinsic emission.

\subsection{XID 36}

The optical counterpart of source XID 36 was originally selected as a Ly$\alpha$ 
emitter with $B=24.8$.  The optical spectrum is shown in \citet{2002Pentericci}, where 
the source was classified as a QSO (ID=1687). Its spectrum shows Ly$\alpha$, SiIV and CIV 
emission lines.  The Ly$\alpha$ line was measured to have a flux of 
$56\times 10^{-17}$ erg/s/cm$^2$, with an $EW=164$ $\AA$ and a $FWHM = 5800 $ km/s.
The Ly$\alpha$ and CIV lines have a double-peaked profile due to absorption by neutral, 
metal-enriched gas. The presence of dust is suggested by the low Ly$\alpha$/CIV line
ratio compared to normal values for QSOs 
\citep[Ly$\alpha$/CIV $\sim 2.5$ from composite spectra of QSO BLRs][]{1989Osterbrock}. 
% Also Selected as an Ly$\alpha$ emitter candidate in \citet{2004aKurk} with ID=c968

The spectrum (ID=X3/L968d) obtained by \citet{2005Croft} found only 
moderate absorption  blueward of Ly$\alpha$, and corrected the 
redshift from $z=2.183$ \citet{2002Pentericci} to $z=2.162$.
This AGN shows broad and narrow Ly$\alpha$, broad and narrow CIV
$\lambda$1549, and broad NV. The CIV emission was fitted with two
Gaussian components, and the broad and narrow Ly$\alpha$ and NV
were fitted with three Gaussian components.  For the
Ly$\alpha,$ these latter authors found $FWHM = 4790 \pm 40$ km/s
% for a flux of $36.9 \pm 0.2 \times 10^{-17}$ erg/s/cm$^2$ EW = 157
and  $FWHM = 122 \pm 20$ km/s for the broad and narrow component, respectively.
% for a flux of $118.4 \pm 0.8 \times 10^{-17}$ erg/s/cm$^2$, EW=49
For the CIV they found $FWHM = 12180 \pm 280$ km/s
% for a flux of $48.2 \pm 0.8 \times 10^{-17}$ erg/s/cm$^2$ 
and  $FWHM = 1070 \pm 50$ km/s for the broad and narrow component, respectively.
% for a flux of $5.7\pm 0.3 \times 10^{-17}$ erg/s/cm$^2$
The NV line has a measured $FWHM = 4041 \pm 40$ km/s.
% The spectrum is also presented in \citet{2013Tanaka} (ID=399).
% The spectrum is not discussed in detail in the paper.
The QSO nature of this source is confirmed by the high X-ray luminosity 
($L_{0.5-10 keV} \sim 3.1\times 10^{44}$ erg/s). The source shows a moderate 
absorption in the X-ray at the level of $\sim 10^{22}$ cm$^{-2}$, but also shows 
hints of a strong Fe line with rest-frame EW $\sim 0.22$ keV, despite with 
large uncertainties.

\subsection{XID 57}

The redshift of XID 57 (z=2.1426) was obtained with Subaru in \citet{2014Shimakawa} (ID=14). 
% The spectrum is not discussed in detail in the paper.
The X-ray spectrum is typical of a Type I AGN with typical Seyfert luminosity
$L_{0.5-10 keV} \sim 5.0\times 10^{43}$ erg/s.

\subsection{XID 73}

The source was identified with an ERO candidate in \citet{2004aKurk} with ID=463.
It is the only X-ray-emitting protocluster member in our catalog 
that has not been spectroscopically confirmed. In the X-ray, it appears as an obscured
Seyfert galaxy.

\subsection{XID 74}

% Also Selected as an ERO candidate in \citet{2004aKurk} with ID=270
% Identified as an H$\alpha$ emitter in \citet{2018Shimakawa} (ID=55).

% The source is also presented in \citet{2013Tanaka} with ID=491
This source is classified as a quiescent galaxy despite it being detected in H$\alpha$ 
probably due to the presence of the 
AGN that nevertheless does not contribute significantly to the overall SED.
This source is investigated in detail in \citet{2010Doherty} 
where they measure it to have $z=2.172$ (ID 456) and to be consistent with 
being formed in an intense burst of star 
formation, with a derived age of $1.6^{+1.1}_{-0.7}$ Gyr.
% an e-folding time scale $\tau = 0.1^{+0.4}_{-0.1}$ Gyr, and dust
% extinction of $\tau_V = 0.4^{+1.4}_{-0.4}$, where $\tau_V$ is the optical depth in the Vband.
% The photometric stellar mass is measured to be $2.8^{+1.5}_{-1.0}\times  10^{11}$ $M_\odot$
% The low value of $\tau$   
The low current star formation rate $<1.0\, M_\odot$/yr derived from the 
SED fit confirms that the galaxy is now in a quiescent phase.
% The measured H$\alpha$ line flux of $6.2 \pm 2 \times 10^{-17}$  erg/s/cm$^2$ is
% \citet{2004bKurk}. 
The X-ray spectrum is consistent with a Type I or 
moderately obscured Seyfert galaxy. Also, the neutral Fe emission line is detected
with $EW\sim 0.5$ keV, despite with significant uncertainties.

\subsection{XID 75}

% ID=6 IN TABLE 2 OF PENTERICCI ET AL. 2002
% This source is identified in \citet{2002Pentericci} with an H$\alpha$ emitter with mag I=20.8 
% It is identified with another faint quasar at $z =2.157 \pm 0.002$ 
% showing very broad H$\alpha$ emission in the near
% infrared spectrum (Kurk et al. 2002b).
This source was selected as an H$\alpha$ emitter candidate in \citet{2004aKurk} and eventually
spectroscopically confirmed \citep{2004bKurk} (ID=215).
Its spectrum shows continuum emission plus a very broad line which almost 
covers the full spectral range, and it is identified with H$\alpha$.
For this AGN the N[II] line is blended with the very broad H$\alpha$ line
and impossible to discern.  The measured FWHM of the line is $5300 \pm  800$ km/s
for a flux of $46.2 \pm  8.8 \times  10^{-17}$ erg/s/cm$^2$ and an $EW=150 \pm 30 \AA$.
The H$\alpha$ emission is therefore not considered to be associated to star formation rate.  
% Also continuum emission is detected \citep{2004bKurk}.
The X-ray spectrum is typical of an unobscured QSO, with $L_{0.5-10 keV} 
\sim 5.2\times 10^{44}$ erg/s, consistent with the presence of the broad optical line.

% The source is also presented in \citet{2013Tanaka}, with ID=537. 
% however they have a different redshift
% (a zspecphot of 0.1, and a redshift of 1.5!) but they claim their redshift is not secure due to
% the AGN contamination of the SED.

% In Tadaki 2013 (where it has KID 1592) 
% it has a wrong zphot, but the reason is that the SED is contaminated by the AGN.

\subsection{XID 80}

Source XID 80 has been identified with source ID 16 in \citet{2002Pentericci} as a 
candidate Ly$\alpha$ emitter encompassed by a very strong and extended
Ly$\alpha$ emission, and therefore classified as an AGN.  
% Because this object is at the very edge of the I band image, it was not included
% in the original sample of candidate Ly$\alpha$ emitters (Kurk
% et al. 2000) and an optical spectrum was not obtained.
% Also Selected as an Ly$\alpha$ emitter candidate in \citet{2004aKurk} with ID=b778
Eventually its spectrum was obtained by \citet{2005Croft} 
at $z=2.149$ (ID X16/L778d). The spectrum shows a broad Ly$\alpha$ (rest-frame deconvolved FWHM
890 km/s), and several emission lines such as NV ($\lambda$1240), CIV ($\lambda$1549), 
HeII ($\lambda$1640), CIII] ($\lambda$1909),[CII] ($\lambda$2326), and 
MgII ($\lambda$2798), all with $FWHM$ ranging from 1200 to 2000 km/s 
\citep[see Table 2 of][]{2005Croft}.
% Also identified as an H$\alpha$ emitter in \citet{2018Shimakawa} (ID=95).

This source was also identified with APEX LABOCA at 870 $\mu m$ with 
$S_{870} \mu m =4.2 \pm 1.4$ mJy for a $S/N= 2.9$
in \citet{2014Dannerbauer} (ID=DKB16). Despite the faint signal, this source is
clearly detected in IR and radio bands (from $S_{24\mu}=572.1 \pm 5.0$ $\mu$Jy 
to $S_{1.4GHz}=76.2 \pm 19.0$ $\mu$Jy with the VLA (see their Table 2).
% QUOTED FROM DANNERBAUER 14: The velocity offset between
% the Lyα and Hα line is +476 km s−1 which is typical for LAEs
% and LBGs (Shapley et al. 2003) indicating gas outflow from this
% source.
The estimated SFR is $\sim 1140 M_\odot$/yr and  $\sim 830 M_\odot$/yr from the 
H$\alpha$ and FIR SED, respectively, and the total FIR luminosity is measured to be
$4.8\times 10^{12} \, L_\odot$.  
% Interestingly, this is the only LABOCA source
% in \citet{2014Dannerbauer} that has a consistent H$\alpha$ and FIR derived SFR, 
% indicating that the other sources are probably significantly dust obscured.  
The X-ray spectrum shows that this source is an obscured Type II QSO 
with with an unobscured luminosity $L_{0.5-10 keV} \sim 7.4\times 10^{44}$ erg/s.

Finally, this source has the second-largest offset from its optical counterpart due
to a possible extended nature of the emission as noted in the text. Possible association 
of the extended emission with radio jets will be investigated when the wide-field, high-resolution
radio images are available.

\subsection{XID 86}

Source XID 86 is identified in \citet{2002Pentericci} (Source ID=5)
with a faint extended ERO with $R-K = 5.8$ ($I-K = 5.3$) consistent
with having a redshift close to the Spiderweb Galaxy.
Eventually, the source was spectroscopically confirmed by \citet{2005Croft} 
at $z=2.162$ (ID 5Xd).
Faint but significant Ly$\alpha$ emission (rest-frame deconvolved
$FWHM \sim 400$ km/s) has been detected.  It was not possible to classify this source as 
a starburst or an AGN, but the X-ray luminosity provides strong evidence 
for the presence of an AGN, as already noted in the literature 
\citep{2002Pentericci,2005Croft}. 
% \citet{2013Tanaka} failed to
% obtain consistent zspecphot due probably to the AGN contamination
% to the overall SEDs for this source (ID=559).
In the X-ray, this source 86 shows a flat spectrum ($\Gamma \sim 1$), 
with hardness ratio $HR\sim 0$. We consider it a candidate Compton-thick 
source, with the X-ray spectrum dominated by a reflection component, 
and an estimated intrinsic luminosity $L_{2-10 keV} \sim 3.4\times 10^{44}$ erg/s.
The detection of the neutral Fe emission line with $EW\sim 0.50$ keV 
strengthen the Compton-thick hypothesis.
% Also Selected as an ERO candidate in \citet{2004aKurk} with ID=226

\subsection{XID 87}

Source XID 87 was selected as an ERO candidate in \citet{2004aKurk} with ID=284
and identified as an H$\alpha$ emitter in \citet{2018Shimakawa} (ID=48).
The redshift was measured thanks to a ALMA CO(3–2) observation presented in 
\citet{2019Tadaki}.  The CO(3-2) line has a redshift of 2.157 with a $FWHM=232$ km/s.
% This source (ID=1138.48) 
% has been shown to have $log(M_*/M\odot) = 10.65 \pm 0.13$, 
% $SFR = 144 \pm 66 M_\odot$/yr, $log(M_{gas}/M_\odot)= 10.99\pm 0.69$.
This source was also detected in the CO(1-0) line in the COALAS survey
\citep[][see their Table 3]{2021Jin}.  % ID=15
The X-ray spectrum shows the steepest slope in the sample, and is
typical of a Type I AGN, with  $L_{0.5-10 keV}  \sim 3.3\times 10^{43}$ erg/s.

\subsection{XID 90}

This source was originally selected as an H$\alpha$ emitter candidate 
in \citet{2004aKurk} with ID=154.
% note: probably a wrong association with 2004aKurk.
% Close to an H$\alpha$ emitter in \citet{2018Shimakawa} (ID=28).
% In both cases the displacement is of 1.7". 
It was also detected in the CO line in the COALAS survey
\citep[][see their Table 3]{2021Jin}.  % ID=12
We note that the X-ray centroid is closer to 
a nearby source that nevertheless has similar color and preliminary photo-z consistent
with being in the protocluster. 
% However, also in this case there is an offset, still within the uncertainty of the ATCA beam.
This is the only X-ray source in our sample that has two possible counterparts, both consistent
with being in the protocluster, but with only one spectroscopically confirmed.
The X-ray spectrum is typical of Type II AGN, 
with a Seyfert-like unabsorbed luminosity of  $L_{0.5-10 keV}  \sim 4.6\times 10^{43}$ erg/s.

%%%%%%%%%%%%%%%%%%%%%%%%%%%%%%%%%%%%%%%%%%%%%%%%%%%%%%%%%%%%%%%%%%%%%%%%%%%%%%%%%%%%%%%%%%%%%%%%%
% PRELIMINARY DISCUSSION ON THE LX-M*-PLANE
%%%%%%%%%%%%%%%%%%%%%%%%%%%%%%%%%%%%%%%%%%%%%%%%%%%%%%%%%%%%%%%%%%%%%%%%%%%%%%%%%%%%%%%%%%%%%%%%%

\newpage

\section{$L_X$-$M_*$ relation\label{LXMstar}}

\begin{figure}[h]
\begin{center}
\includegraphics[width=0.49\textwidth]{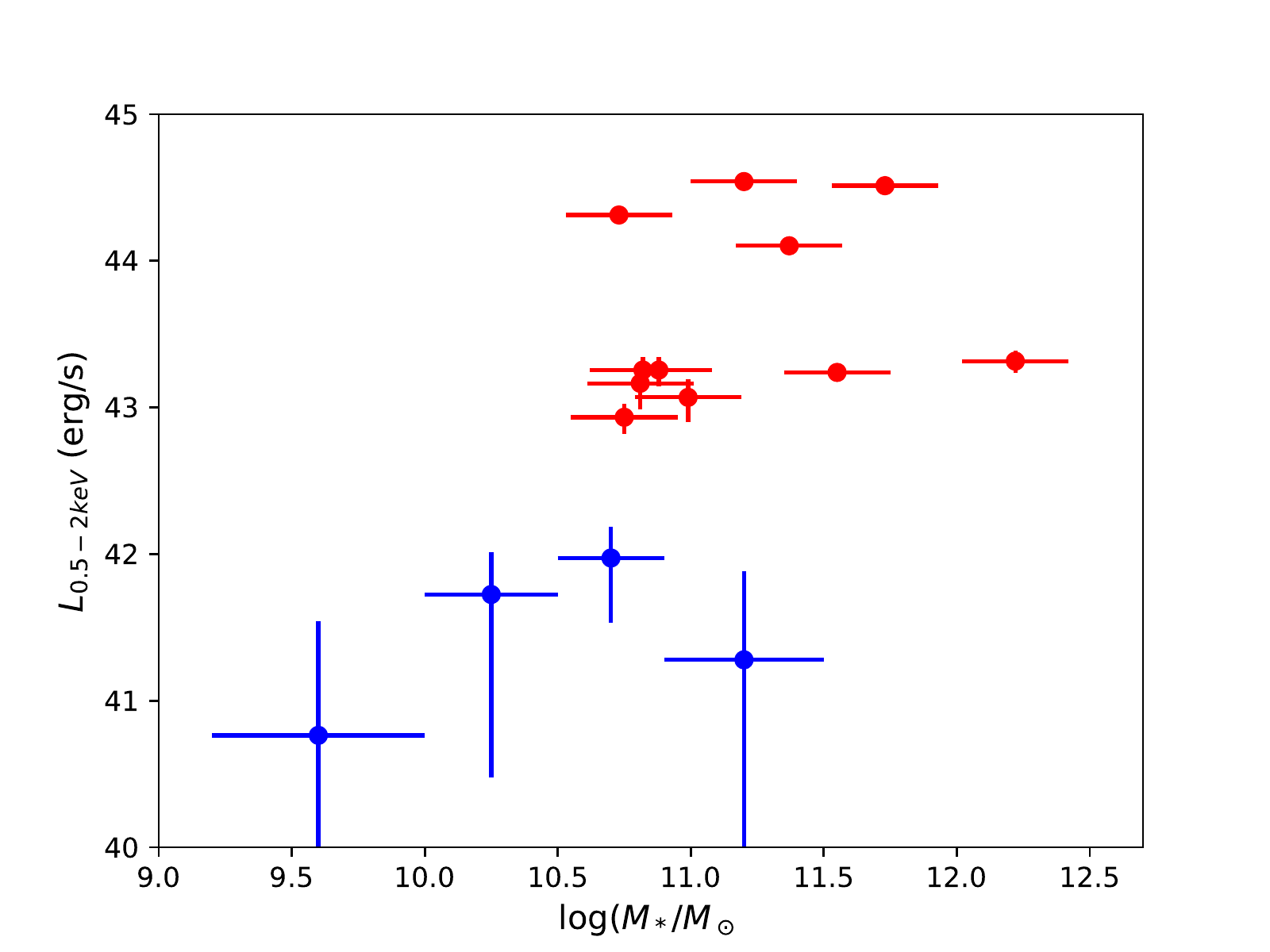}
\includegraphics[width=0.49\textwidth]{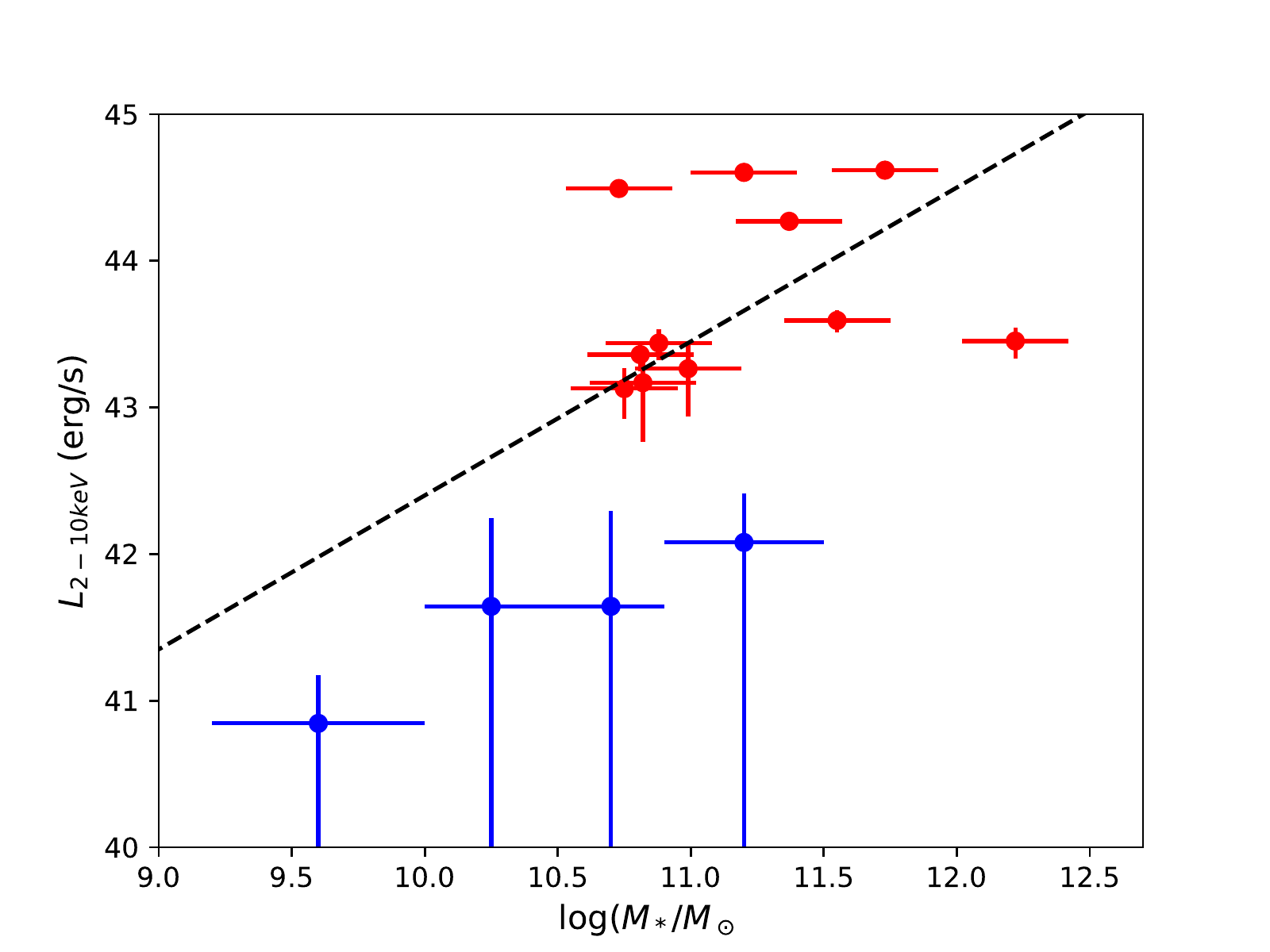}
\caption{X-ray luminosity and stellar mass relations.  
Left panel: Red points show the distribution of 12 protocluster members (including the 
Spiderweb Galaxy but excluding the two Compton-thick candidates) in the 
soft-band, intrinsic luminosity  and stellar mass plane. Blue points are
obtained stacking the protocluster members without X-ray detection 
in bins of stellar mass.  We note that stacked $L_X$ values should be considered
upper limits, because, despite the formal positive detection, all
the stacked values are consistent with noise within 1 or 1.5 $\sigma$.
Right panel: Same as the left panel but in the rest-frame hard band. } 
\label{LX_mstar_plane}
\end{center}
\end{figure}

In Figure \ref{LX_mstar_plane} we show the $L_X-M_*$ plane for all the protocluster
members with spectroscopic redshift.  We plot 11 sources with X-ray emission, excluding the
spiderweb and excluding also the two Compton-thick candidates, whose intrinsic luminosity cannot be
estimated accurately. In addition, we also show the average X-ray luminosity per source
that we obtain by stacking the sources without X-ray detection in four stellar mass bins.  
The mass bins have been chosen in order to have a similar number of sources ($\geq 10$)
in each of them, except for the first bin ($ 9.5<{\rm log}(M_*)<10.0$) where we include all
the sources without a $K_{\rm S}$-band counterpart.  Before stacking the X-ray images at the source position, 
we removed the sources closer than 12 arcsec to the Spiderweb Galaxy in order to avoid spurious signal due to the
irregular diffuse emission. 
We have 38 sources in the first bin, then 13 sources in the $ 10.0<{\rm log}(M_*)<10.5$ bin, 
and 11 and 12 sources in the $ 10.5<log(M_*)<10.9$ and $ 10.9<{\rm log}(M_*)<11.6$ bin, respectively. 
Despite in all the cases we formally have a positive photometry, most of them 
are very close to zero, and, in all the cases, consistent with zero whitin 1 or 1.5
$\sigma$.  Therefore, we must consider the blue points and the corresponding error bars as upper limits.
The error bars on ${\rm log}(M_*)$ are $\sim 0.2$ for the X-ray detected sources,
while corresponds to the bin width for the stacked points.

Apart from a very scattered, global trend of having higher intrinsic $L_X$ at higher $M_*$, 
we note that the sources not detected in X-ray do not have luminosity close to our
detection limit, but at least an order of magnitude lower, implying that 
they are not hosting AGN comparable to those hosted by the X-ray detected 
protocluster members. In the right panel of Figure \ref{LX_mstar_plane} 
we also plot, as a reference, the average $L_X$-$M_*$ relation obtained 
in the field at $z\sim 2$ in \citet{2012bMullaney} for the rest-frame 
2-10 keV band. Apart from the expected scatter at high masses 
\citep[see][]{2012aMullaney}, which is entirely expected because the relation holds for average values of the X-ray luminosity, we notice 
that the stacked values of the protocluster galaxies without X-ray detection 
appear to be lower than the relation measured in the field.  This comparison 
is preliminary, because only star forming galaxies should be included.  However, 
this is a hint that, at low mass, the accretion onto the SMBH may have 
been suppressed among the protocluster members, while it is enhanced at high masses. 
In other words, the average $L_X$--$M_*$ relation may be significantly steeper within the
protocluster.  This issue will be explored in Pannella et al. (in preparation).

\end{appendix}

\end{document}